 \documentclass[twocolumn]{aastex6}

\graphicspath{{imagens_arxiv/}}

\newlength{\picwd}
\newlength{\thinpic}
\usepackage[utf8]{inputenc}
\usepackage{natbib}
\usepackage{xcolor}

\usepackage{gensymb}

\newcommand{\e}[1]{\ensuremath{\times 10^{#1}}} %
\newcommand{\un}[1]{\ensuremath{\ \mathrm{#1}}}
\newcommand{\uns}[1]{\ensuremath{\ \mathsf{#1}}}
\newcommand{\rsun}{\ensuremath{\ R_\odot}}
\newcommand{\prop}[1]{}

\newcommand{\rev}[1]{#1}
\begin{document}

\title{A multiple flux-tube solar wind model}

%% Use \author, \affil, plus the \and command to format author and affiliation 
%% information.  If done correctly the peer review system will be able to
%% automatically put the author and affiliation information from the manuscript
%% and save the corresponding author the trouble of entering it by hand.
%%
%% The \affil should be used to document primary affiliations and the
%% \altaffil should be used for secondary affiliations, titles, or email.

%% Authors with the same affiliation can be grouped in a single
\author{Rui F. Pinto\altaffilmark{1} and Alexis P. Rouillard}
\affil{
  Université de Toulouse; UPS-OMP; IRAP;  Toulouse, France \\
  CNRS; IRAP; 9 Av. colonel Roche, BP 44346, F-31028 Toulouse cedex 4, France
}

% \author{Yi-Ming Wang}
% \affil{Other affiliations}
  
% %% Use the \and command so offset the last author.
% \and

% \author{Roland Grappin}
% \affil{Other affiliations}


\altaffiltext{1}{rui.pinto@irap.omp.eu}

\begin{abstract}

  We present a new model, MULTI-VP, that computes the three-dimensional structure of the solar wind which includes the chromosphere, the transition region, and the corona and low heliosphere.
  MULTI-VP calculates a large ensemble of wind profiles flowing along open magnetic field-lines which sample the whole three-dimensional atmosphere or, alternatively, on a given region of interest.
  The radial domain starts from the photosphere and extends, typically, to about $30\rsun$.
  The elementary uni-dimensional wind solutions are based on a mature numerical scheme which was adapted in order to accept any flux-tube geometry.
  We discuss here the first results obtained with this model.
  We use Potential Field Source-Surface (PFSS) extrapolations of magnetograms from the Wilcox Solar Observatory (WSO) to determine the structure of the background magnetic field.
  Our results support the hypothesis that the geometry of the magnetic flux-tubes in the lower corona controls the distribution of slow and fast wind flows.
  The inverse correlation between density and speed faraway from the Sun is a global effect resulting from small readjustments of the flux-tube cross-sections in the high corona (necessary to achieve global pressure balance and a uniform open flux distribution).
  In comparison to current global MHD models, MULTI-VP performs much faster and does not suffer from spurious cross-field diffusion effects.
  We show that MULTI-VP has the capability to predict correctly the dynamical and thermal properties of the background solar wind (wind speed, density, temperature, magnetic field amplitude and other derived quantities) and to approach real-time operation requirements.

\end{abstract}
\keywords{Sun, Solar wind}

%%%%%%%%%%%%%%%%%%%%%%%%%%%%%%%%%%%%%%%%%
\section{Introduction}
\label{sec:intro}

\begin{figure*}
  \centering
  \textsf{\textbf{CR 2056}} \\
  \includegraphics[width=0.31\linewidth]{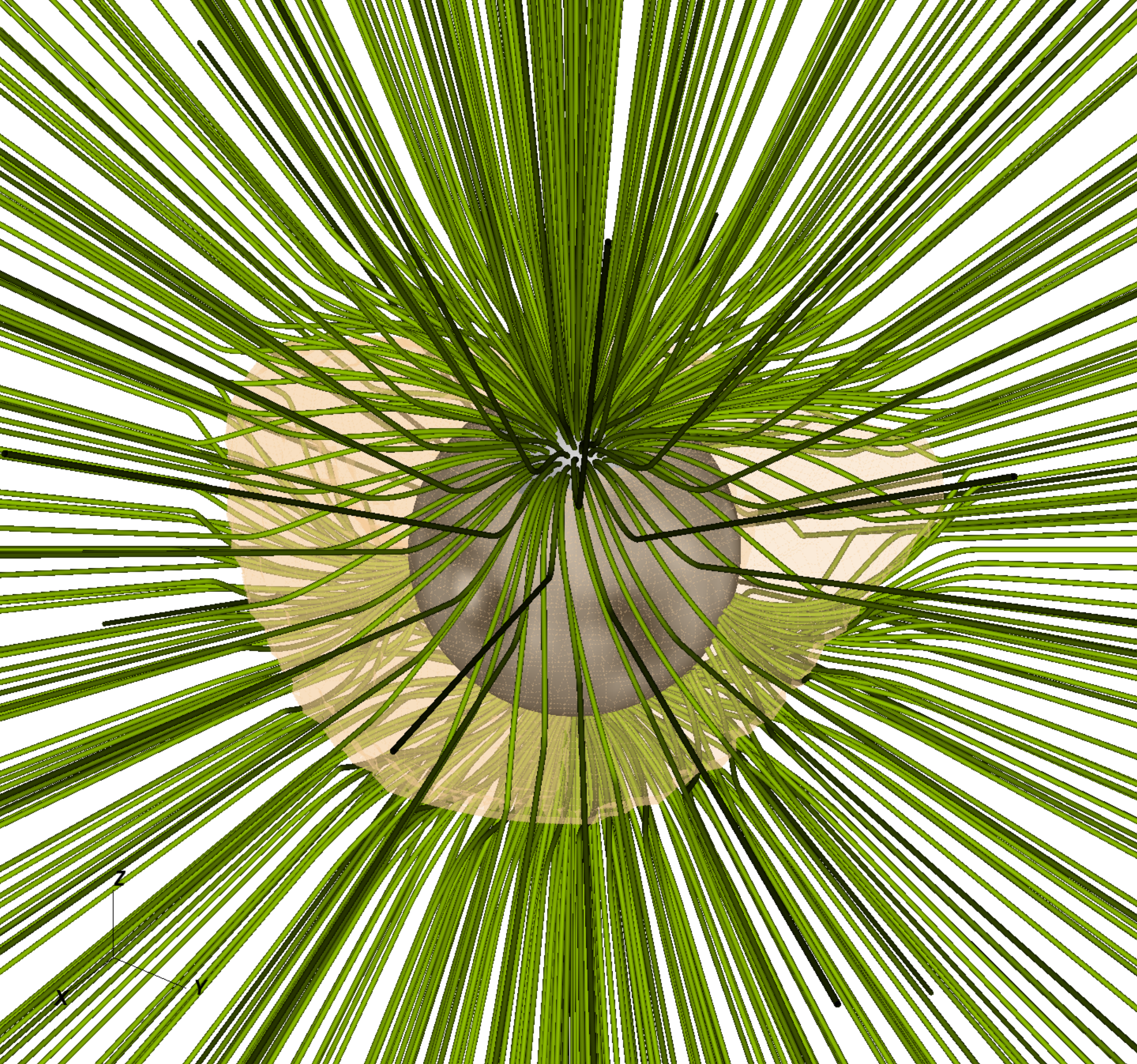}
  \includegraphics[width=0.31\linewidth]{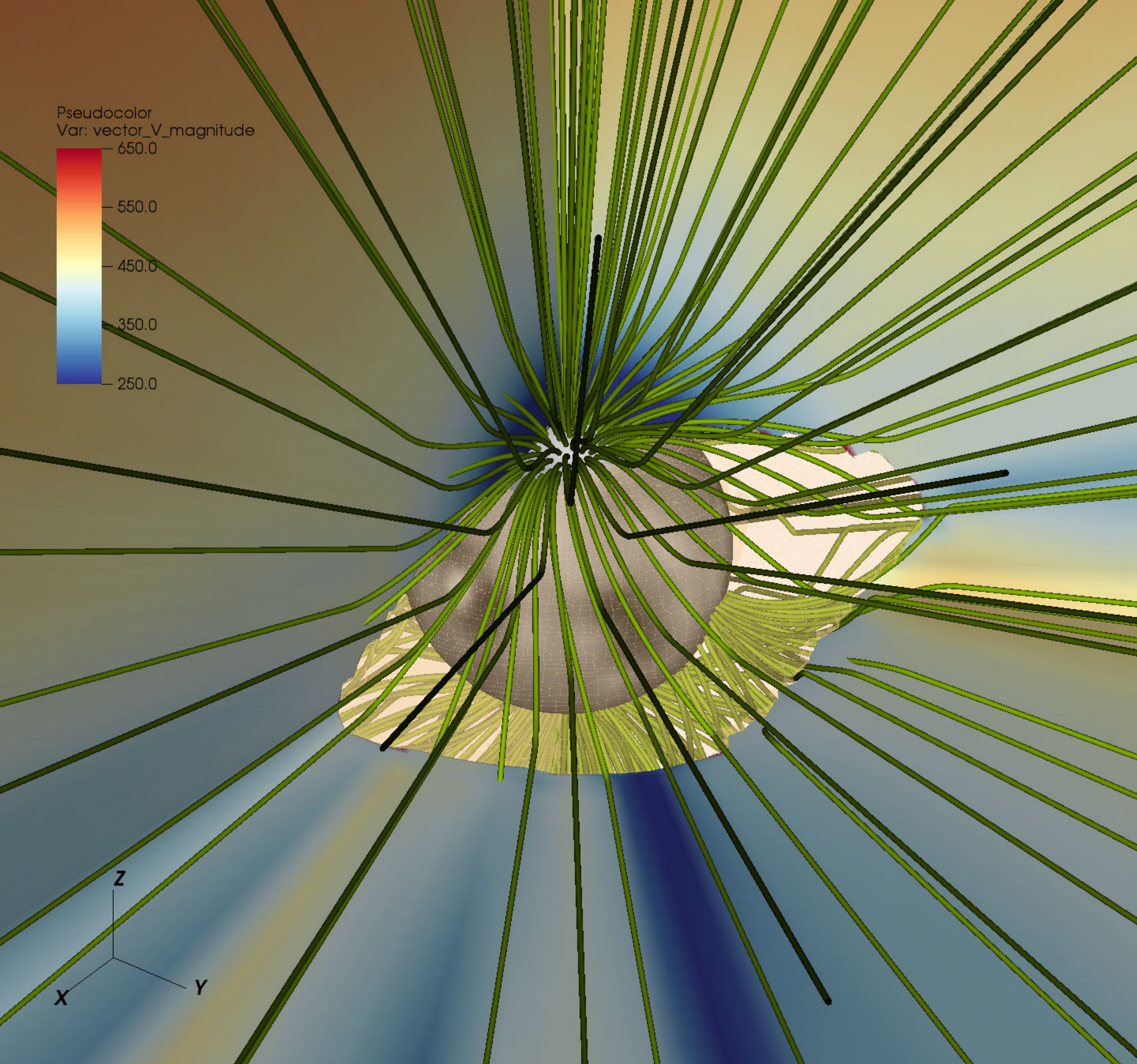}
  \includegraphics[width=0.31\linewidth]{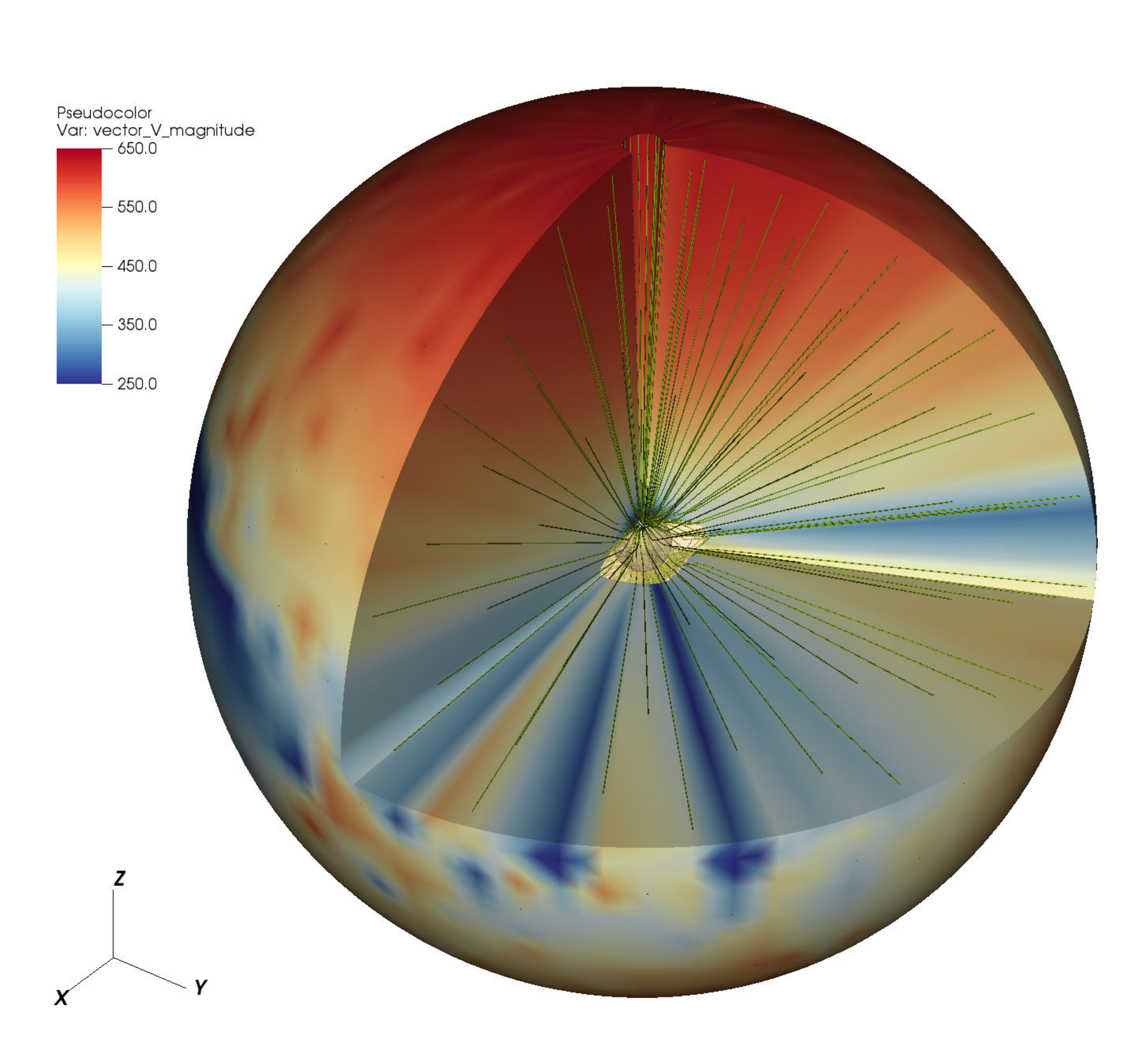} \\ \vspace{0.05\linewidth}

  \textsf{\textbf{CR 2121}} \\
  \includegraphics[width=0.31\linewidth]{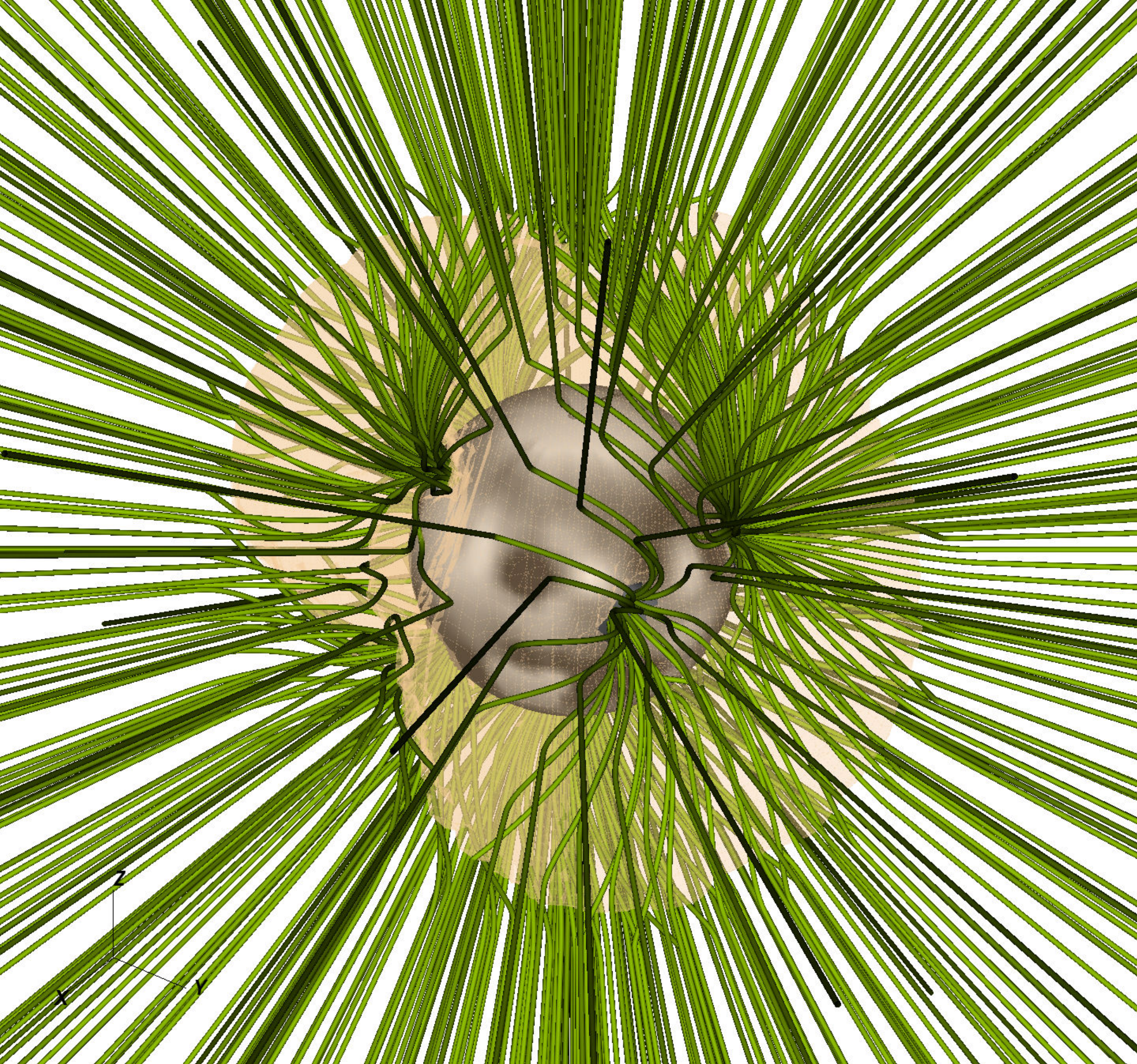}
  \includegraphics[width=0.31\linewidth]{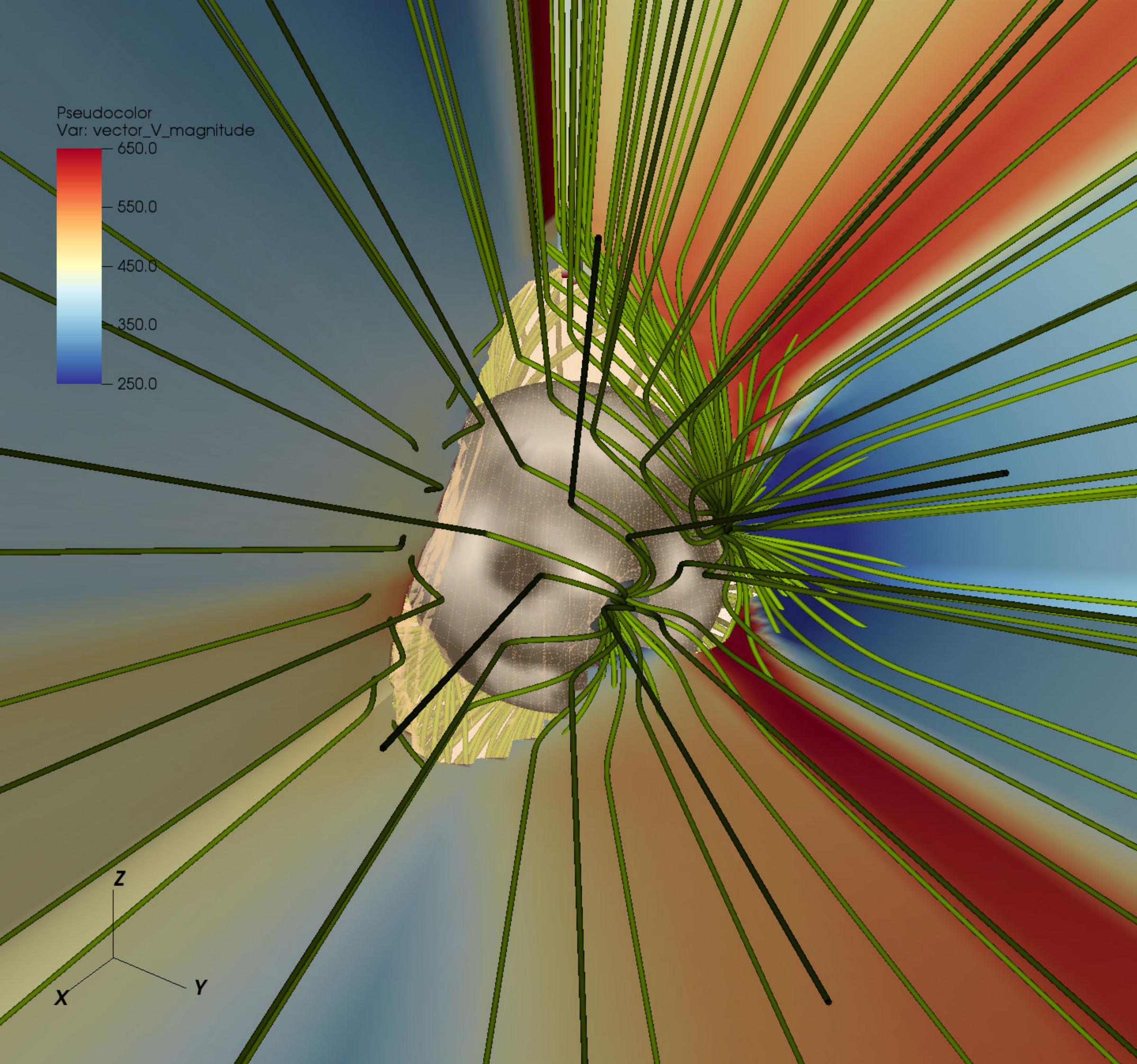}
  \includegraphics[width=0.31\linewidth]{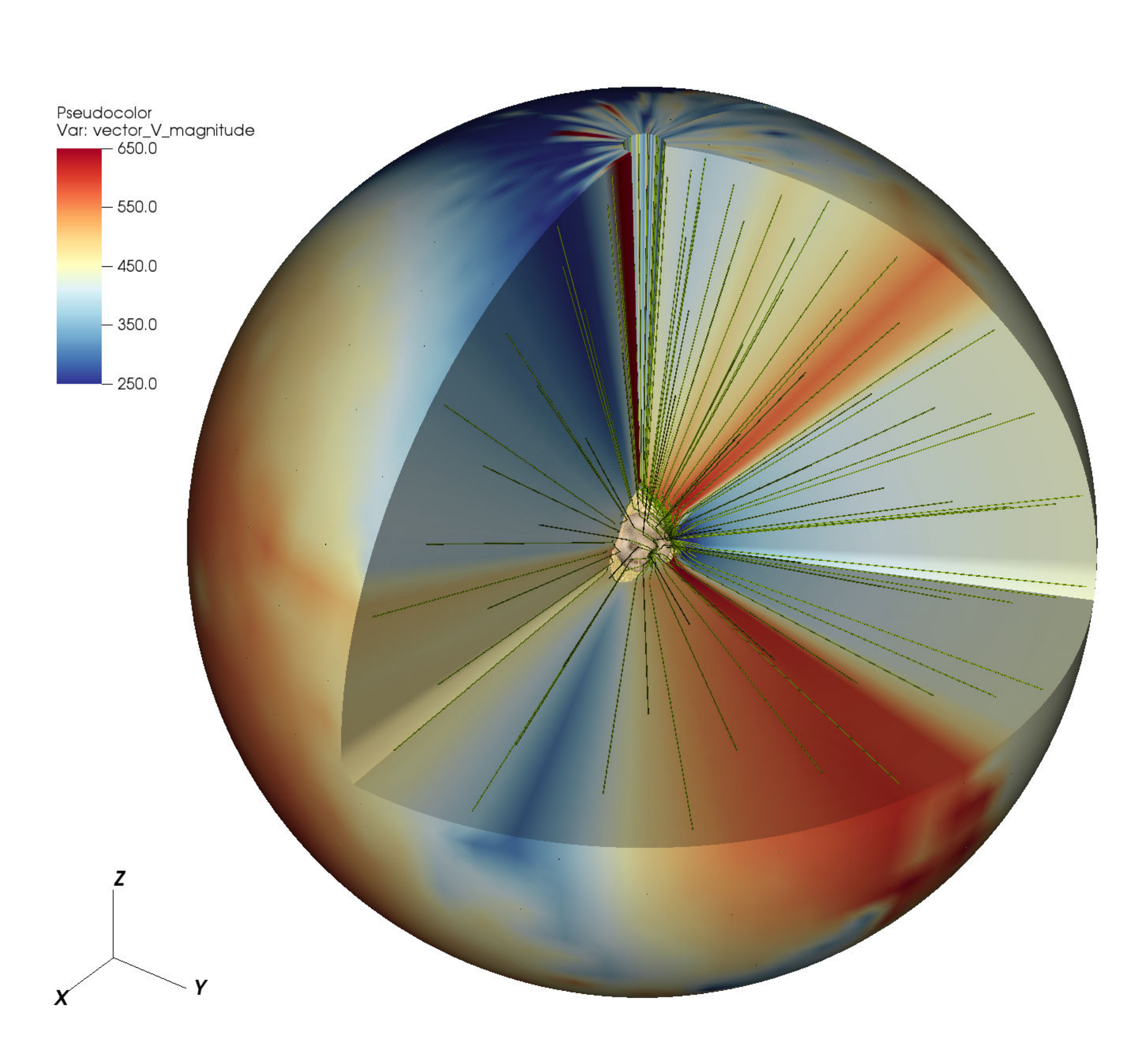} \\

  \caption{
    Illustration of the operation of the model.
    The first row corresponds to CR 2056 (April -- May 2007, close to solar minimum) and the second one to CR 2121 (Mars 2012, close to solar maximum).
    The first column shows the input WSO magnetogram rendered in gray-scale over the surface of the Sun and a sample of the magnetic field lines obtained via PFSS extrapolation used to initiate the model.
    The transparent yellow surface indicates the coronal hole boundaries (the closed-field regions are excluded from the domain).
    The second column shows a close-up of the wind speeds in the low corona, represented in colour-scale (from dark blue at $250\un{km/s}$ to dark red at $650\un{km/s}$).
    The third column shows the same information at global scale (truncated at a radius of $15\un{\rsun}$ and with one octant removed).
  }
  \label{fig:mvp_scheme}
\end{figure*}

\prop{Solar wind properties, general intro.}
The solar wind separates into fast and slow wind streams whose large-scale distribution evolves markedly during the solar activity cycle.
Slow solar wind flows are usually found in the vicinity of streamer / coronal hole boundaries (S/CH), while the fastest wind flows stream out of coronal holes.
The distribution of fast and slow wind streams hence follows the cyclic variations of the magnetic field structure of the solar corona \citep{mccomas_weaker_2008,smith_solar_2011,richardson_solar_2008}.
In fact, slow wind flows are essentially confined to the equatorial regions of the Sun during solar minima, with the fast wind flows covering all other latitudes. 
During solar maxima, slow wind flows also occur at higher latitudes following the appearance of non-equatorial streamers and pseudo-streamers.
During the rise and decay phase of the cycle, fast wind flows also make incursions into the equatorial regions \citep{mccomas_three-dimensional_2003}.
The terminal wind speeds seem to be determined to a great extent by the geometrical properties of the magnetic flux-tubes through which the solar wind flows \citep{wang_solar_1990,pinto_flux-tube_2016}.
In fact, the properties of the surface motions (assumed as energy sources for the heating and acceleration of the wind) are more uniform across the solar disk than the variations in amplitude of the wind flows that stream out, suggesting that it is the coronal environment that causes the segregation between fast and slow solar wind flows.
Other than their asymptotic speed, slow and fast wind streams are  differ consistently in terms of density, mass flux, heavy ion composition and thermal structure. 

%\nota{(\ldots)} \bigskip

\begin{figure*}
  \centering
  \includegraphics[width=\linewidth, clip, trim = 0 20 0 0]{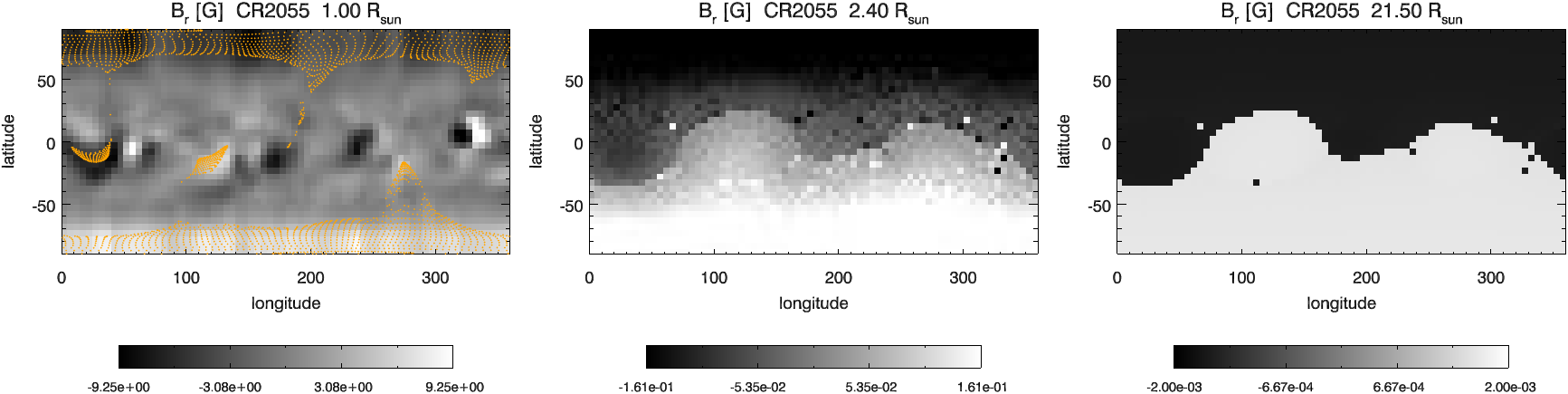} \medskip \\
  \includegraphics[width=\linewidth, clip, trim = 0 20 0 0]{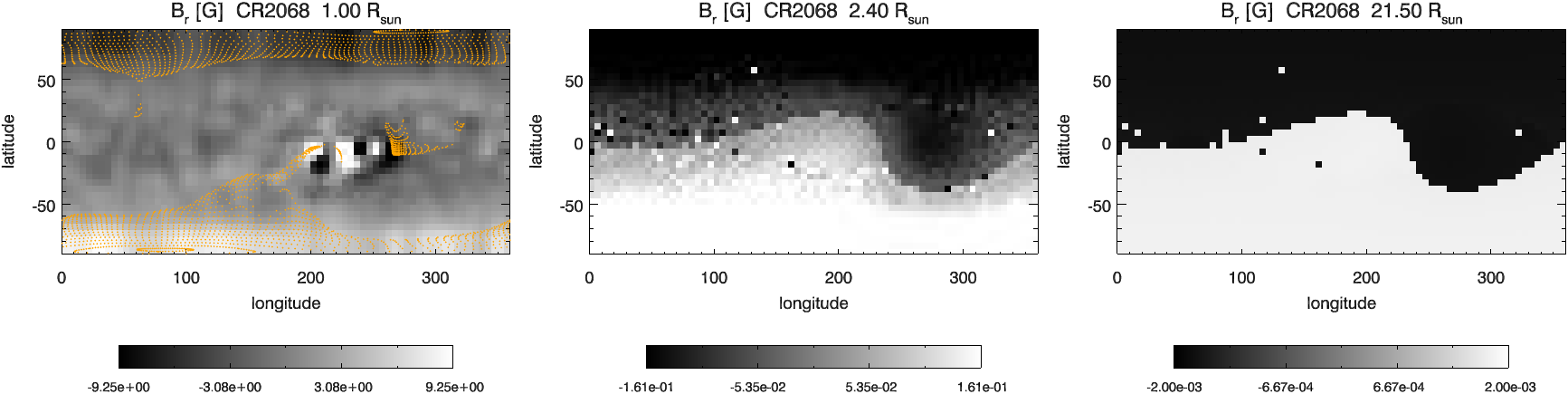} \medskip \\
  \includegraphics[width=\linewidth, clip, trim = 0 20 0 0]{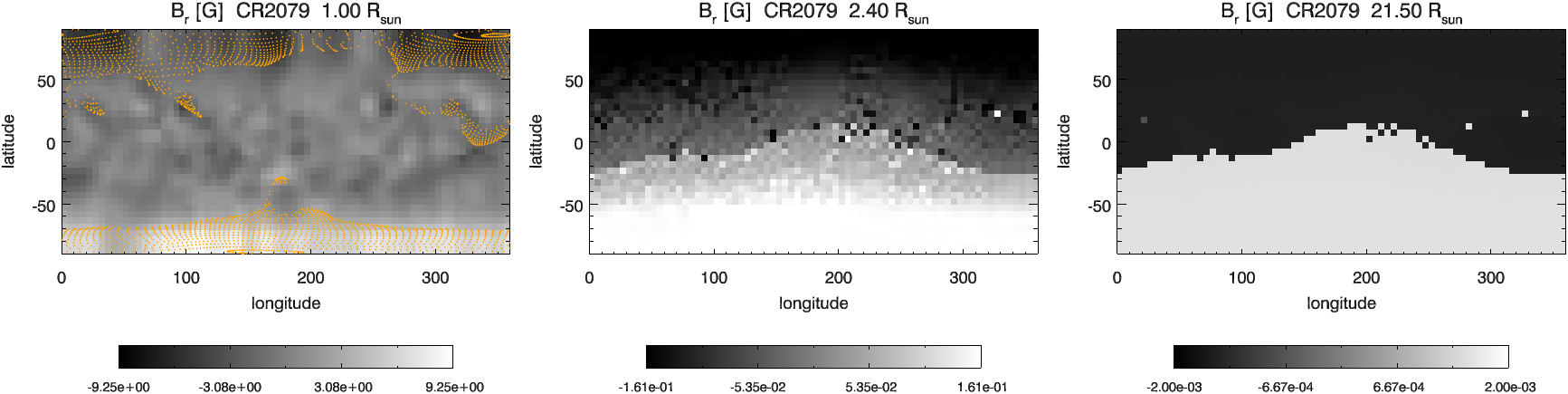} \medskip \\
  \includegraphics[width=\linewidth, clip, trim = 0 20 0 0]{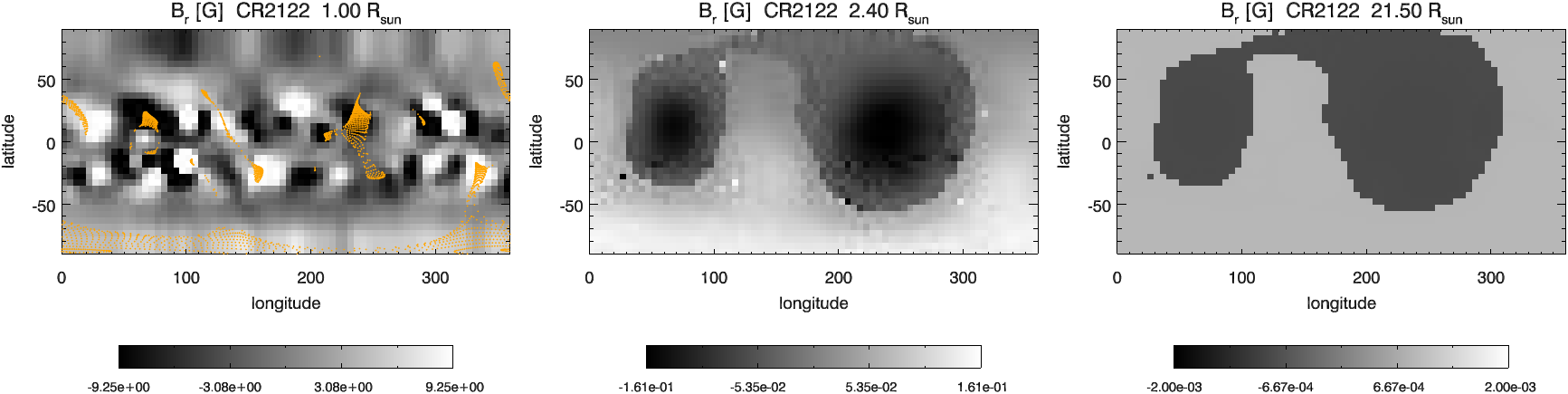} \medskip \\
  \includegraphics[width=\linewidth                       ]{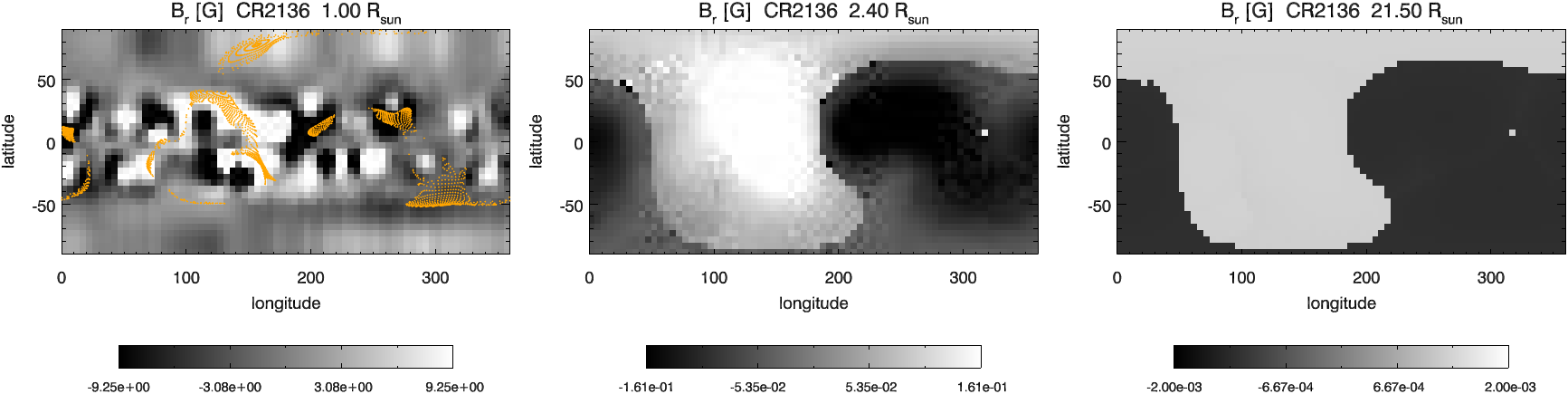} 
  \caption{
    Maps of the magnetic field at the solar surface, close to the source-surface and at $21.5\rsun$ obtained after extrapolating the WSO magnetograms via PFSS and applying the corrections discussed in the text.
    The orange dots on the first column show the foot-points of the sample of magnetic field-flux tubes used to compute the solar wind solutions. These flux-tubes cover the whole latitude -- longitude map uniformly beyond the source-surface, with and angular resolution of $5\degree$.
    The amplitude of the open magnetic field is very variable at the source-surface, but becomes nearly uniform faraway from the Sun.
  }
  \label{fig:maps_inital_bfield}
\end{figure*}

\prop{Overview of the current solar wind models (empirical, numerical, global MHD, specialized 1D models).}
Understanding solar wind acceleration and predicting the terminal solar wind speed (together with other properties) has been the subject of intense research over the last decades.
The solar wind speed is commonly associated with simple parameters describing the variations of the cross-sections of the flux-tubes as a function of height, namely the expansion factor
\begin{equation}
  \label{eq:expans_definition_general}
  f = \frac{A_1}{A_0} \left(\frac{r_0}{r_1}\right)^2,
\end{equation}
where $A_0$ and $A_1$ are the cross-section of a given elemental flux-tube respectively at the surface of the Sun ($r=r_0$) and at some point higher in the corona ($r = r_1 > r_0$) above which the flux-tubes expand radially outwards (and not super-radially).
The total expansion ratio is equal to $1$ for a flux-tube expanding radially, while a very strongly expanding flux-tube has $f >>1$.
The continued exploitation of potential field extrapolations with source-surface (PFSS) of magnetogram data lead to associating $r_1$ with the radius of the source-surface, commonly placed at a fixed height of $r_{SS} = 2.5\un{\rsun}$, and to evaluation the expansion factor at this height: $f\equiv f_{SS}$ \citep{wang_solar_1990}.
This source-surface radius is the one that produces the best matches between the geometry of the extrapolated magnetic fields and the shapes of the coronal structures observed in white-light during solar eclipses, especially the size of the streamers and coronal hole boundaries \citep{wang_formation_2010,wang_coronal_2009}.
However, matching quantities such as the open magnetic flux seems to require defining $r_{SS}$ as a function of the solar activity \citep{lee_coronal_2011,arden_breathing_2014}, or more generally as a function of the properties of the global coronal magnetic field \citep{reville_solar_2015}.
\citet{suzuki_forecasting_2006} suggested that the terminal wind speed would be better predicted by a combination of the expansion factor and the magnetic field amplitude at the foot-point of any given flux-tube or, equivalently, to the open magnetic flux \citep[see also][]{fujiki_relationship_2015}.
Other authors also invoke empirically derived parameters such as the angular distance from the foot-point of a given magnetic flux-tube to the nearest streamer / coronal-hole (S/CH) boundary \citep[parameter $\theta_b$; ][]{arge_improved_2003,arge_stream_2004,mcgregor_distribution_2011}.
Recent studies by \citet{li_solar_2011,lionello_application_2014,pinto_flux-tube_2016,peleikis_investigation_2016} indicate that field-line curvature and inclination also have an impact on the wind speed.
These results altogether motivate the formulation and adoption of semi-empirical strategies for predicting the state of the solar wind.
The so-called Wang-Sheeley-Arge (WSA) model is the most successful and the most widely used relation of this kind.
It provides predictions of the solar wind speed  close to the Earth's environment at any time, and it does so very fast.
But it requires empirical calibration, and cannot be guaranteed to work properly outside its narrow range of validity (typically restricted to the ecliptic plane near $1\un{AU}$).

\begin{figure}
  \centering
  \includegraphics[width=0.85\linewidth]{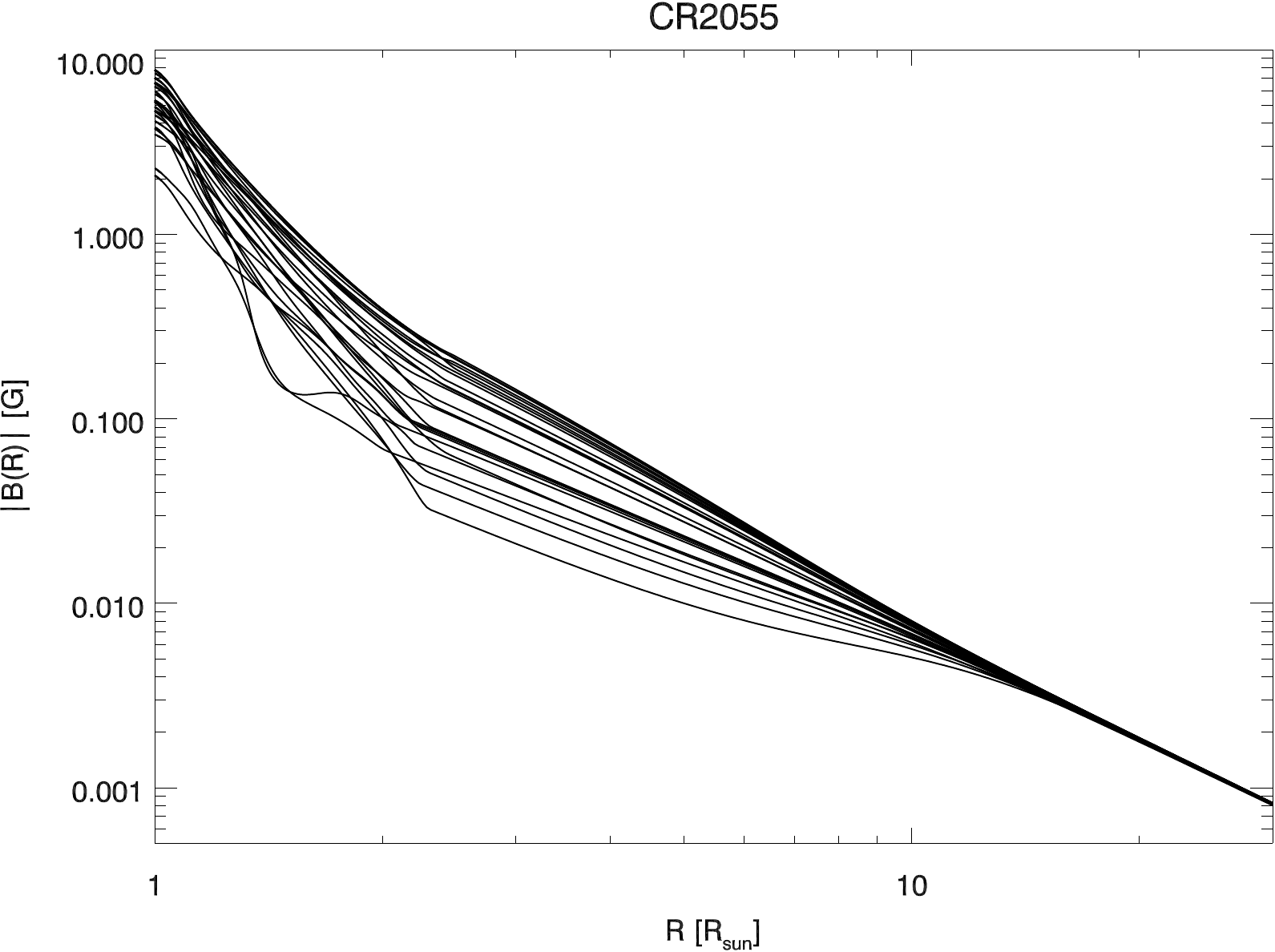} \medskip \\
  \includegraphics[width=0.85\linewidth]{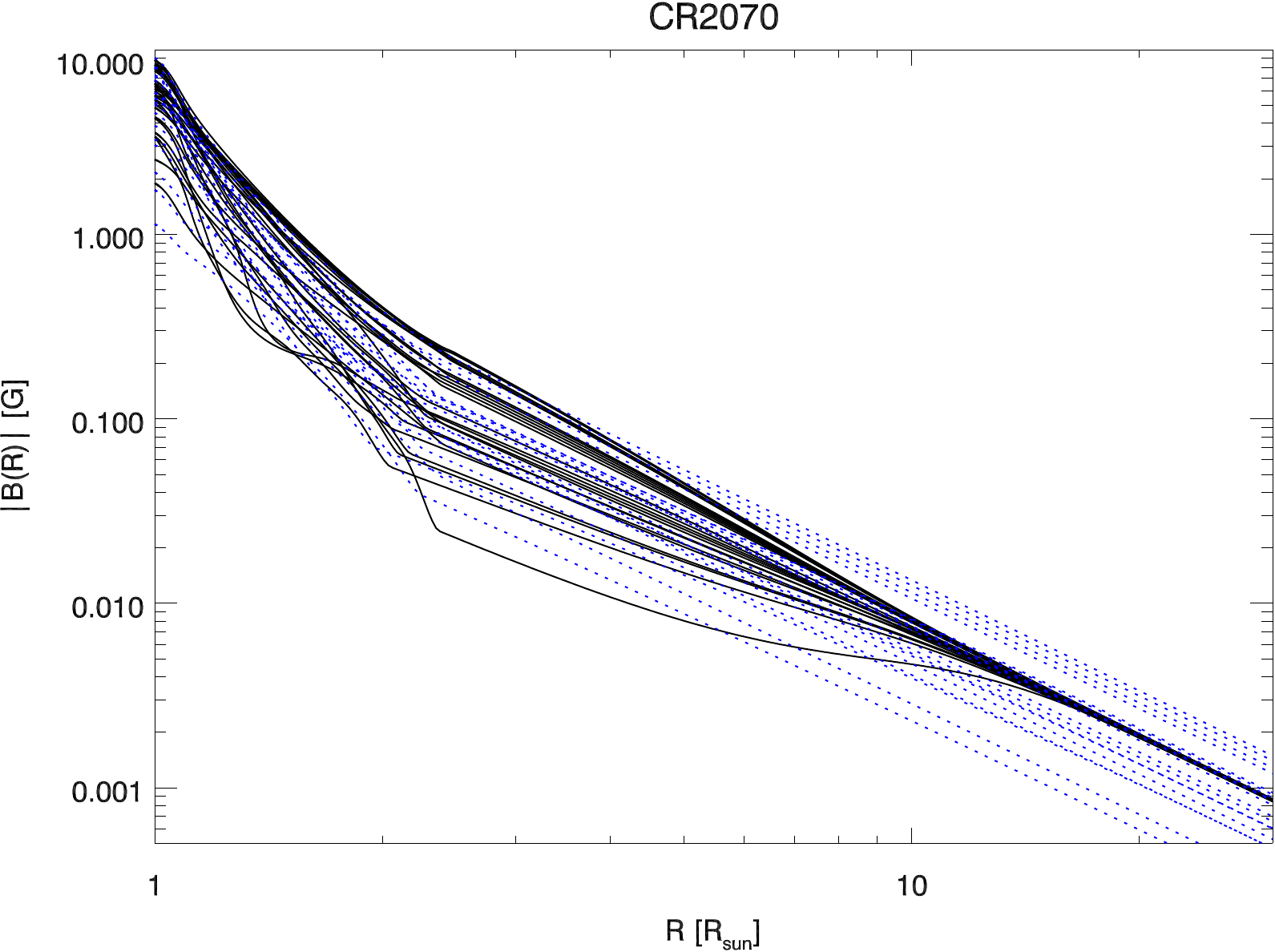} 
  \caption{
    Radial profiles of the magnetic field amplitude for a small subset of flux-tubes, at two different Carrington rotations (2055 and 2070).
    The black continuous lines represent the amplitudes after the correction to the high coronal field was applied to the background PFSS field (which makes the open magnetic field amplitude uniform above $12\un{\rsun}$).
    For comparison, the blue dotted lines on the second panel represent the same flux-tubes before the correction was performed.
  }
  \label{fig:profiles_inital_bfield}
\end{figure}
\begin{figure}
  \centering
  \includegraphics[width=0.85\linewidth]{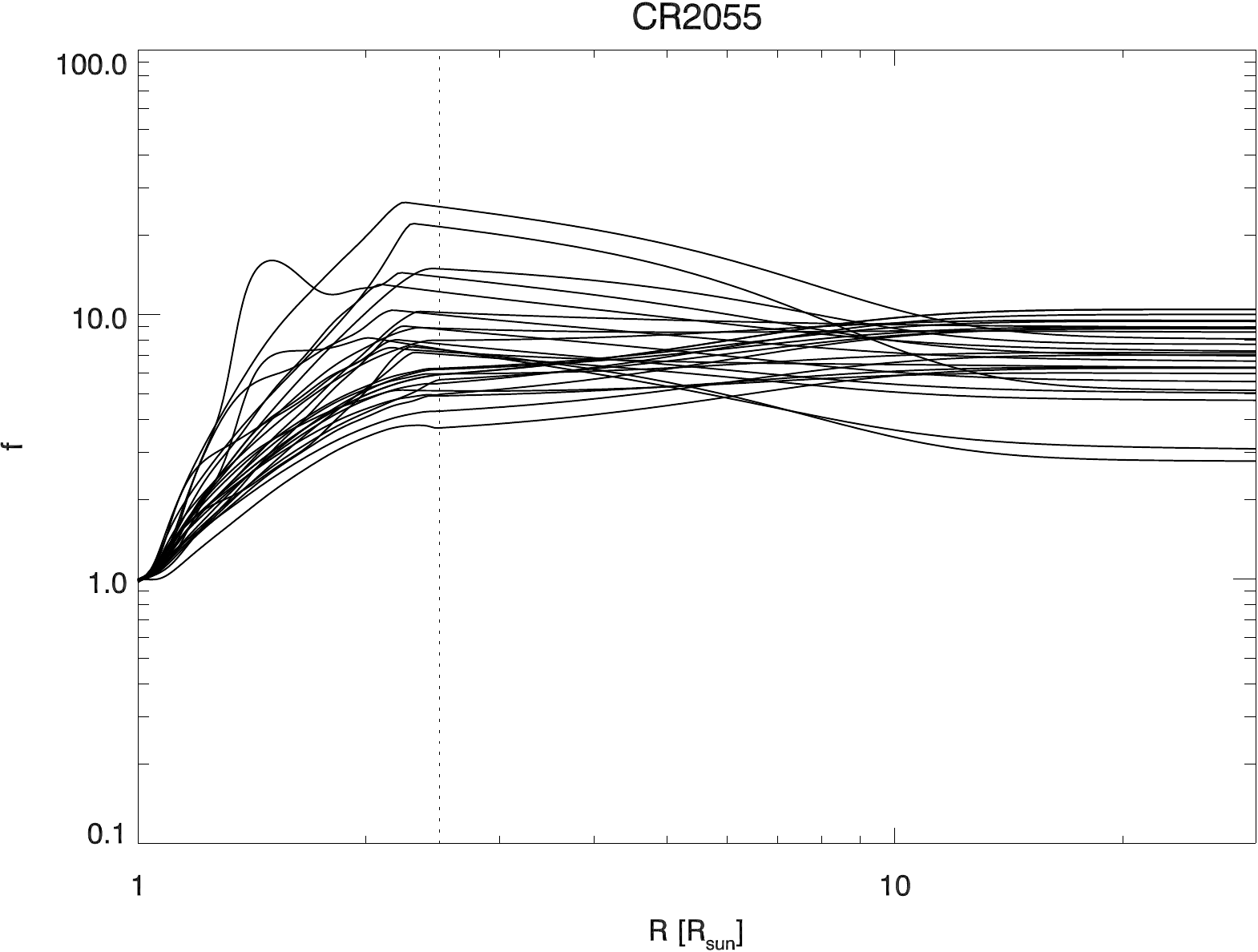} \medskip \\
  \includegraphics[width=0.85\linewidth]{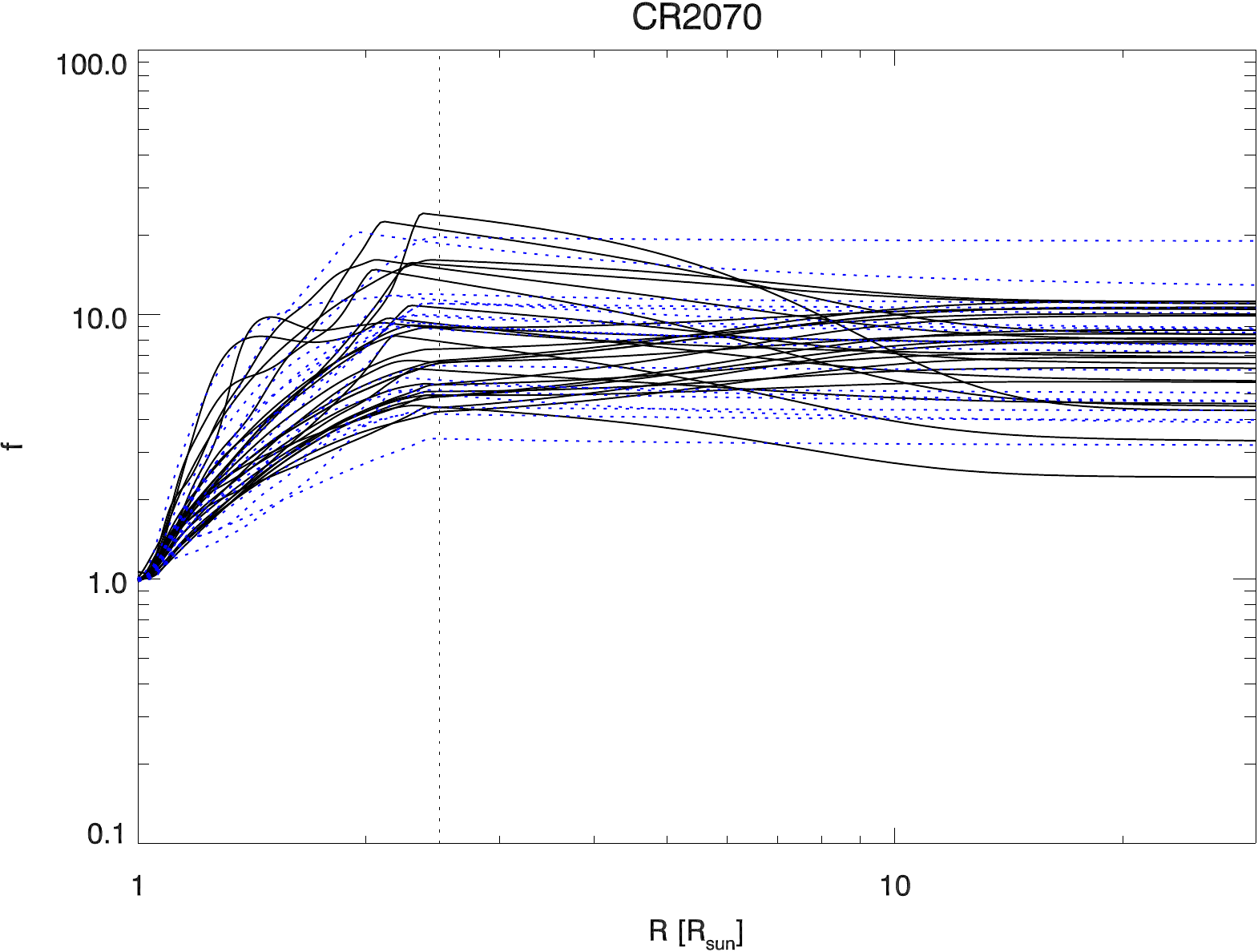} 
  \caption{
    Radial profiles of the expansion factor for the same subset of flux-tubes and for the same Carrington rotations as in Fig. \ref{fig:profiles_inital_bfield}.
    Continuous black lines and dotted blue lines have the same meaning.
  }
  \label{fig:profiles_inital_expans}
\end{figure}
\begin{figure}
  \centering
  \includegraphics[width=0.85\linewidth]{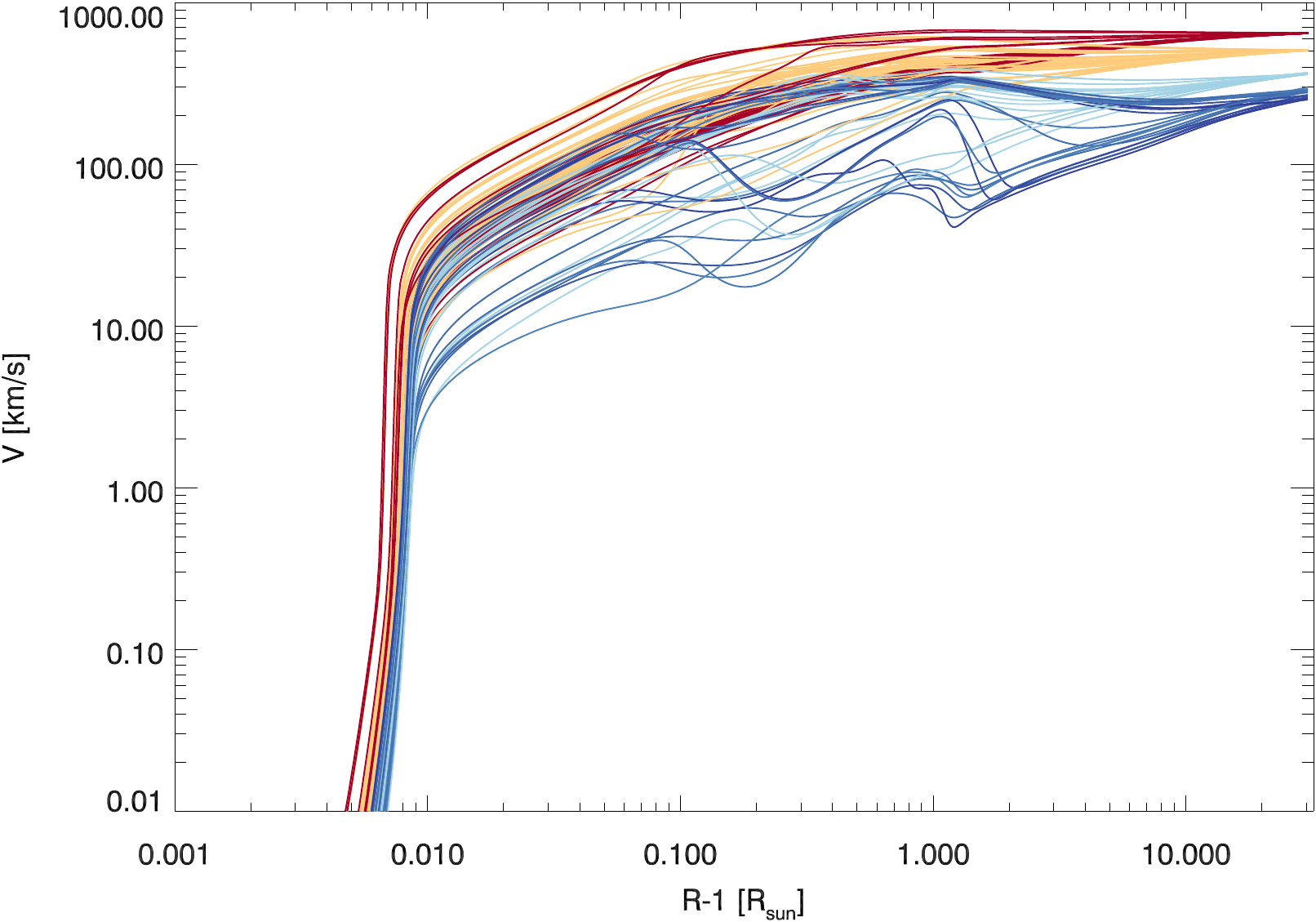} 
  \includegraphics[width=0.85\linewidth]{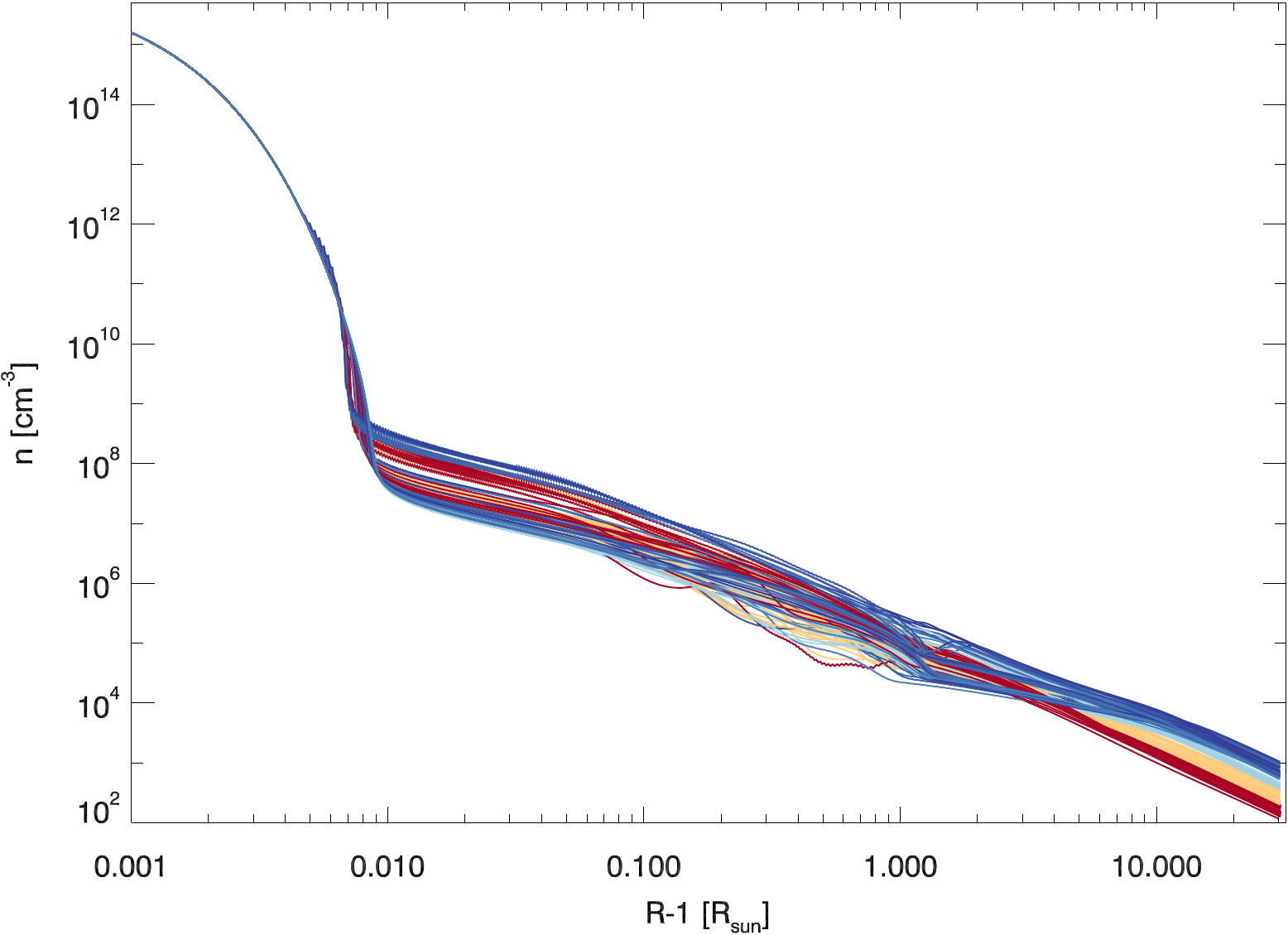} 
  \includegraphics[width=0.85\linewidth]{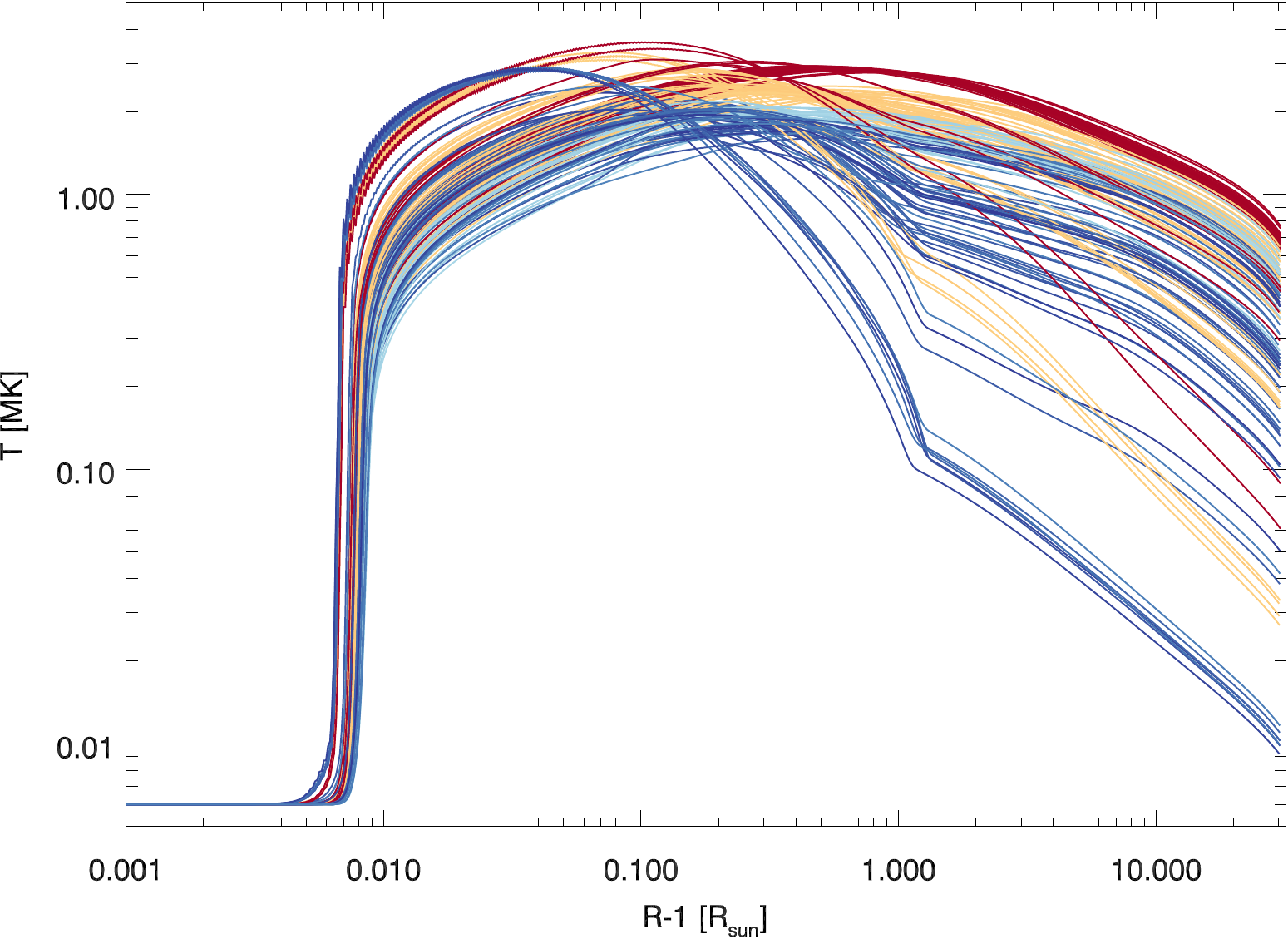} 
  \caption{
      Wind speed, plasma temperature and density as a function of height ($r-\rsun$) across the whole domain for a random subset of flux-tubes with asymptotic speeds falling close to $270,\ 360,\ 500$ and $650\un{km/s}$.
      The lines are coloured according to the asymptotic wind speed values using the colour-scheme used in Figs. \ref{fig:mvp_scheme} (second and third columns) and \ref{fig:crmaps_21rsun} (first column).
  }
  \label{fig:wind_profiles}
\end{figure}

\prop{The case for a fast and robust yet thermally adequate model.}
An alternative approach consists of modelling the solar wind acceleration and propagation based solely on physical principles.
Comprehensive modelling of the solar wind at a global scale is, however, a daunting challenge.
%Different numerical approaches undertaken so far assume essentially two different strategies.
Fully three-dimensional MHD wind models take into account the whole magnetic topology of the corona at the expense of important simplifications to the solar wind thermodynamics and of a large CPU time consumption \citep[][and many others]{gressl_comparative_2013,lee_solar_2009,yang_time-dependent_2012}.
Some uni-dimensional solar wind models consider much more sophisticated heat transport models, but over-simplify the geometry of the magnetised corona \citep{pinto_time-dependent_2009,woolsey_turbulence-driven_2014}.
%\nota{(Could be more thorough on this paragraph)}%: there are semi-empirical models, global numerical models, and 1D specialised models, each one with its pros and cons. Our approach is a synthesis of sorts.)}

\prop{Our purpose and goals: research model which can be used for real-time space weather applications.}
We propose a new solar wind model which assumes a new approach, in between those of the traditional MHD global-scale models and of the more specialized uni-dimensional models.
The model consists of computing many 1D wind solutions which sample the whole solar atmosphere (or any smaller sub-domain of interest). 
The individual 1D solutions are based on a previous uni-dimensional wind model \citep{pinto_time-dependent_2009,grappin_search_2010}, modified in order to take the full magnetic flux-tube geometry (expansion, inclination and amplitude of the field) and different heating functions (see Sect. \ref{sec:methods}).
The background magnetic field geometry is currently obtained via potential field source-surface extrapolation (PFSS) from publicly available magnetogram data, but the model is ready to use any other data source (real data, coronal reconstructions or modeled data).

\prop{Expected achievements: }
We expect this approach to bring many benefits over the current solar wind models.
Our method lets us map the wind speed, density, temperature and magnetic field amplitude across the whole atmosphere, as well as derived quantities such as the characteristic phase speeds, wind ram pressure and mass fluxes.
The model is computationally light (much lighter than full-fledged 3D MHD models) and has a perfect parallel scaling.
Another advantage of our method is that, as there is no cross-field diffusion in the model, the heliospherical current sheets remain thin and do not experience the enhanced (and spurious) resistive broadening typical of global MHD simulations (at the scale of at least a few grid cells).
The main limitations of our approach are that the realism of the background magnetic field depends on the reconstruction method used, cross stream interactions are neglected, and so are small-scale cross-field diffusive processes (the effects and strategies for addressing these caveats are discussed in Sects. \ref{sec:methods} and \ref{sec:results}).

%%%%%%%%%%%%%%%%%%%%%%%%%%%%%%%%%%%%%%%%%
\section{Methods}
\label{sec:methods}

\prop{The general philosophy of the model.}
The MULTI-VP model consists of a large ensemble of contiguous uni-dimensional wind solutions used to derive the three-dimensional structure of the solar wind.
The numerical model is therefore a multiplexed version of a mature one-dimension numerical model representing the heating and acceleration of the solar wind (called, hereafter, the baseline model).
MULTI-VP operates, typically, under the following work flow (see Fig. \ref{fig:mvp_scheme}):
\begin{enumerate}
\item Choice of a magnetogram data source.
  The source magnetograms can be full carrington maps (\emph{e.g} from Wilcox Solar Observatory -- WSO) or adaptative/forecast magnetograms (for example, from the Helioseismic and Magnetic Imager -- HMI --, or from the Air Force Data Assimilative Photospheric Flux Transport model -- ADAPT --, at much higher temporal cadences). 
\item Choice of a coronal field reconstruction method. 
We reconstruct the coronal magnetic field and sample out an ensemble of open magnetic flux-tubes extending from the surface of the Sun up to about $30\rsun$ and covering all latitudes and longitudes of interest.
\item Computation of  field-aligned wind profiles for each one of the sampled flux tubes. 
  The wind model takes full account of the magnetic field amplitude, areal expansion and inclination in respect to the vertical direction along the flux tubes.
  The model includes a simplified chromosphere, the transition region (TR) and the corona (more details are given below).
\item Assemble the flow profiles and the wind speed, temperature, density and magnetic field amplitude at all the positions desired. 
  We routinely produce maps at $21.5\rsun$ which can be used to initiate heliospheric propagation models (such as ENLIL and EUHFORIA).
\end{enumerate}
This framework was designed to be fully modular, such that any of the points above can easily be replaced by other data sources and models depending on the scientific application, and as newer methods become available.
\prop{Magnetograms and magnetic field reconstruction.}
In this manuscript, we will use Potential Field Source-Surface extrapolations from WSO synoptic maps covering several Carrington rotations both at solar minimum and at solar maximum (CR 2055 - 2079 and CR 2130 - 2149).
We set a constant source-surface radius $R_{SS} = 2.5\rsun$ for the PFSS extrapolations and follow the general method and polar-field correction of \citet{wang_potential_1992}.
We trace an ensemble of open magnetic field lines starting from the source-surface down to the solar surface with a standard angular resolution of $5\degree$.
Each field-line is, at first, assigned a purely radial expansion above the source-surface.
This leads to an interplanetary magnetic field amplitude which is very variable within each magnetic sector.
The Ulysses mission showed that the radial field component is, however, uniformly distributed with latitude
 much unlike the amplitudes measured by space probes in the interplanetary field \citep{balogh_heliospheric_1995}.
We correct for this by adding an additional flux-tube expansion profile which smoothly (and asymptotically) transforms the very non-uniform source-surface field at $r=R_{SS}$ into a uniform field at about $r\geq 12 \rsun$. 
The correction conserves the total open magnetic flux.
Its effects on the properties of the wind flow are thoroughly discussed in Sect. \ref{sec:results}.

\begin{figure*}
  \centering
  
  \includegraphics[width=0.325\linewidth]{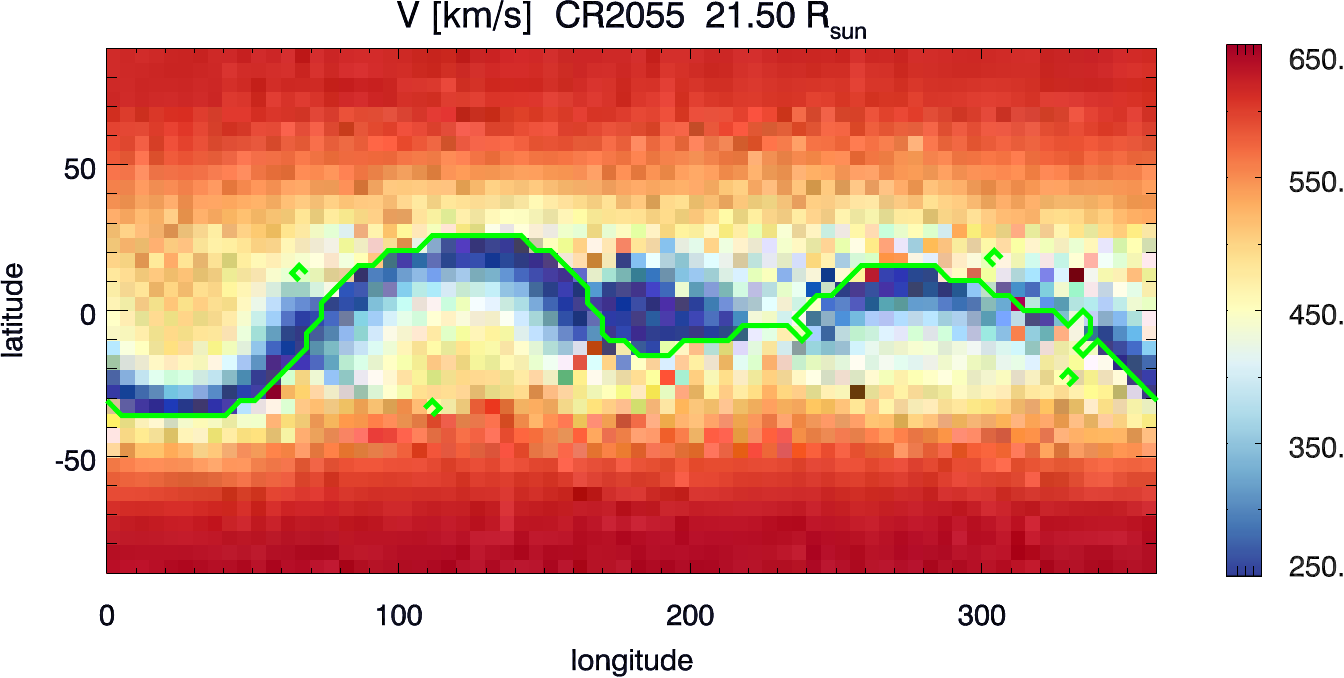} 
  \includegraphics[width=0.325\linewidth]{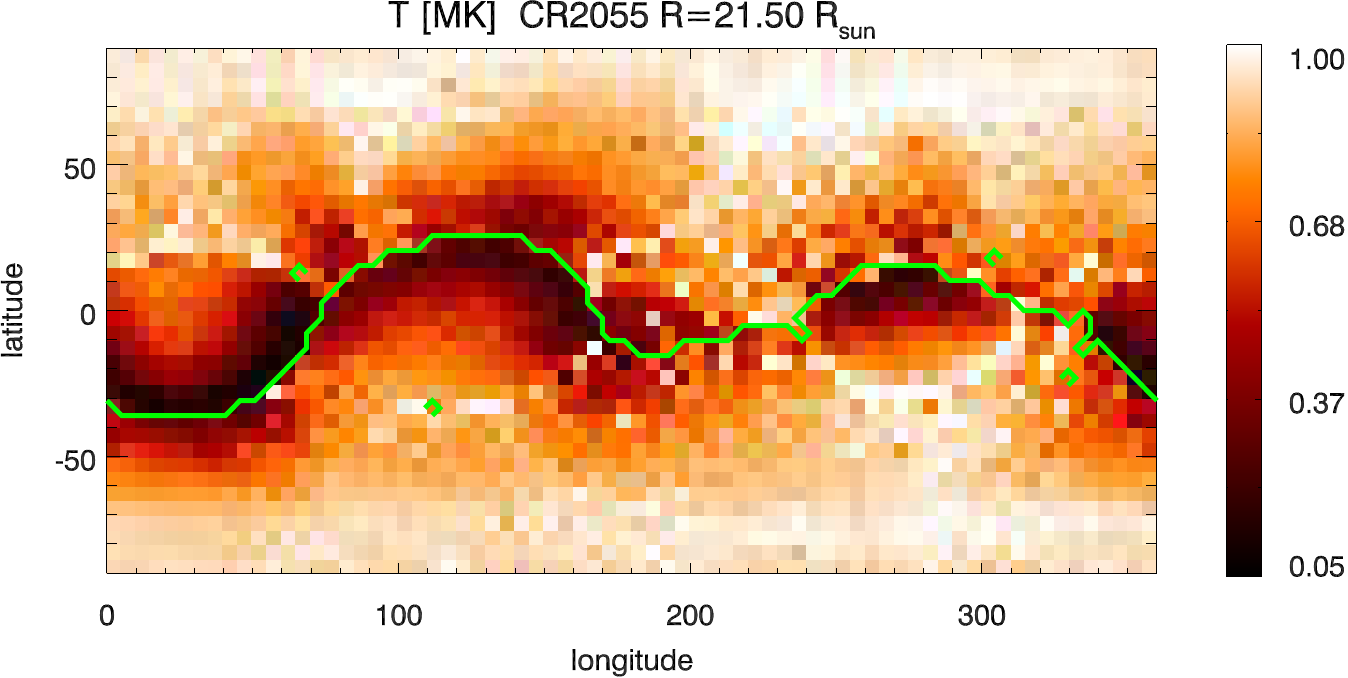} 
  \includegraphics[width=0.325\linewidth]{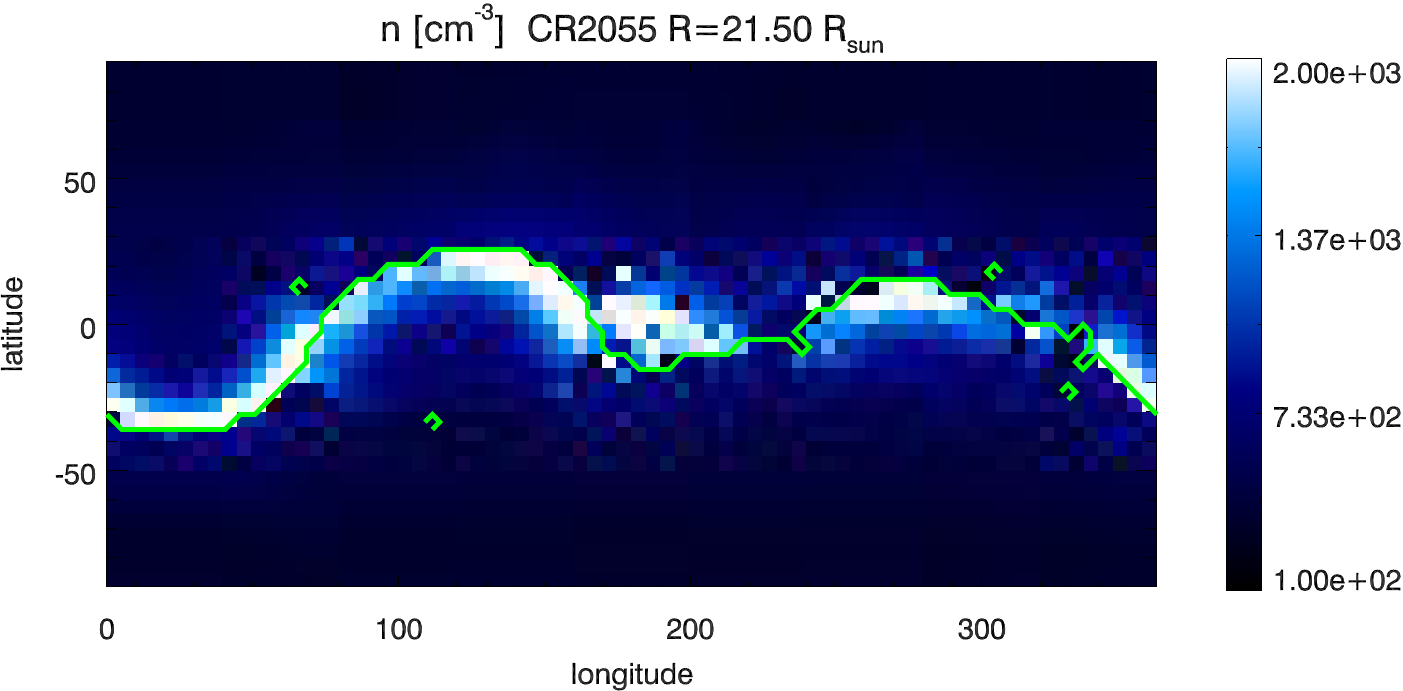}  \\ \medskip

  \includegraphics[width=0.325\linewidth]{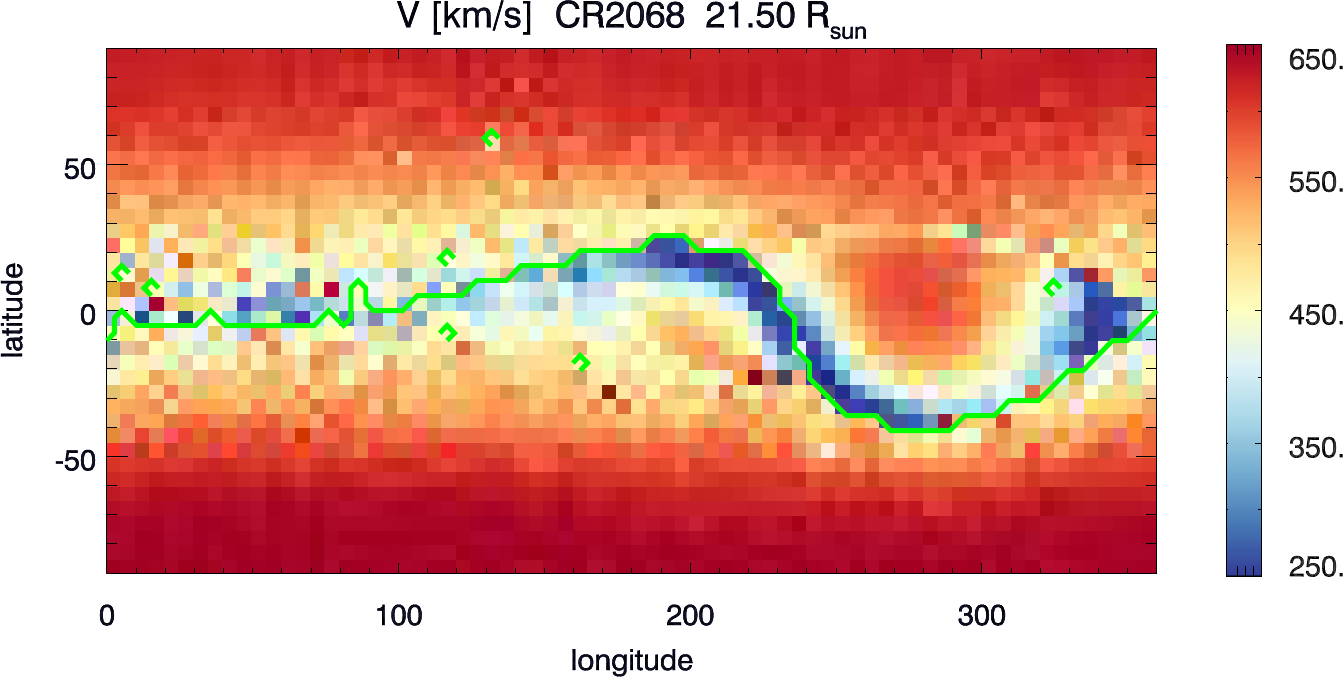} 
  \includegraphics[width=0.325\linewidth]{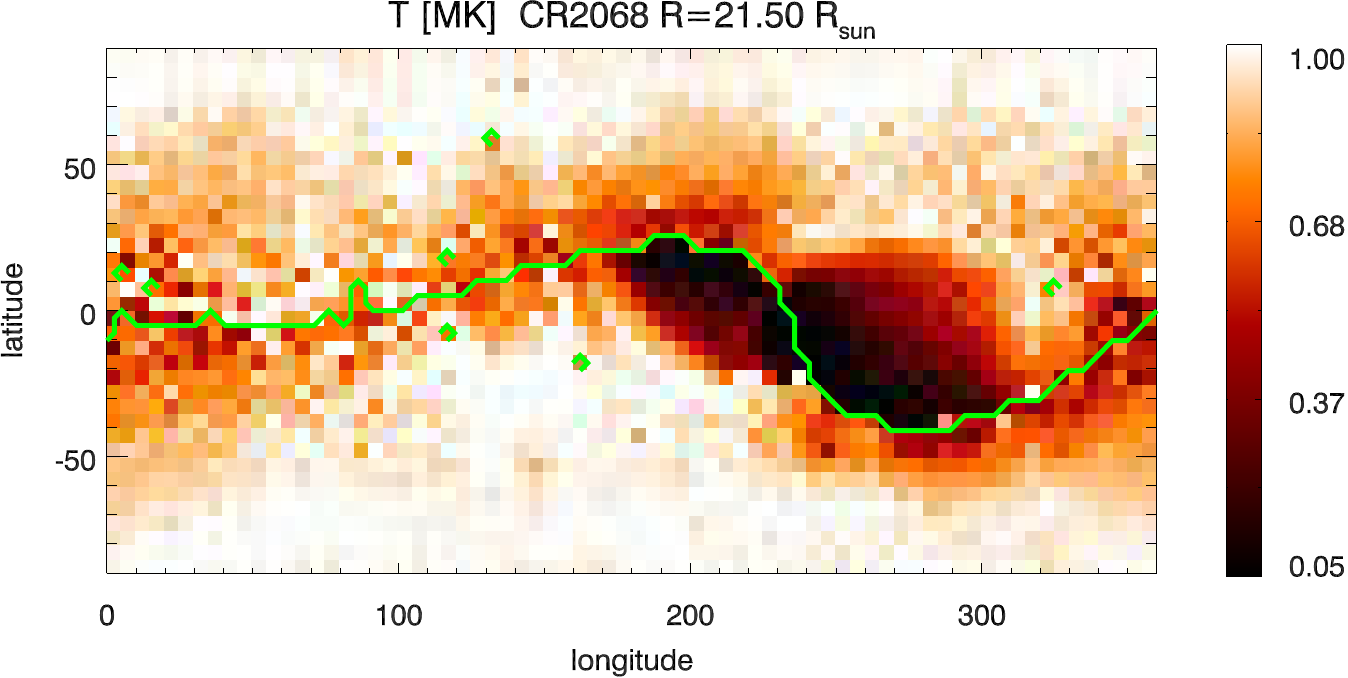} 
  \includegraphics[width=0.325\linewidth]{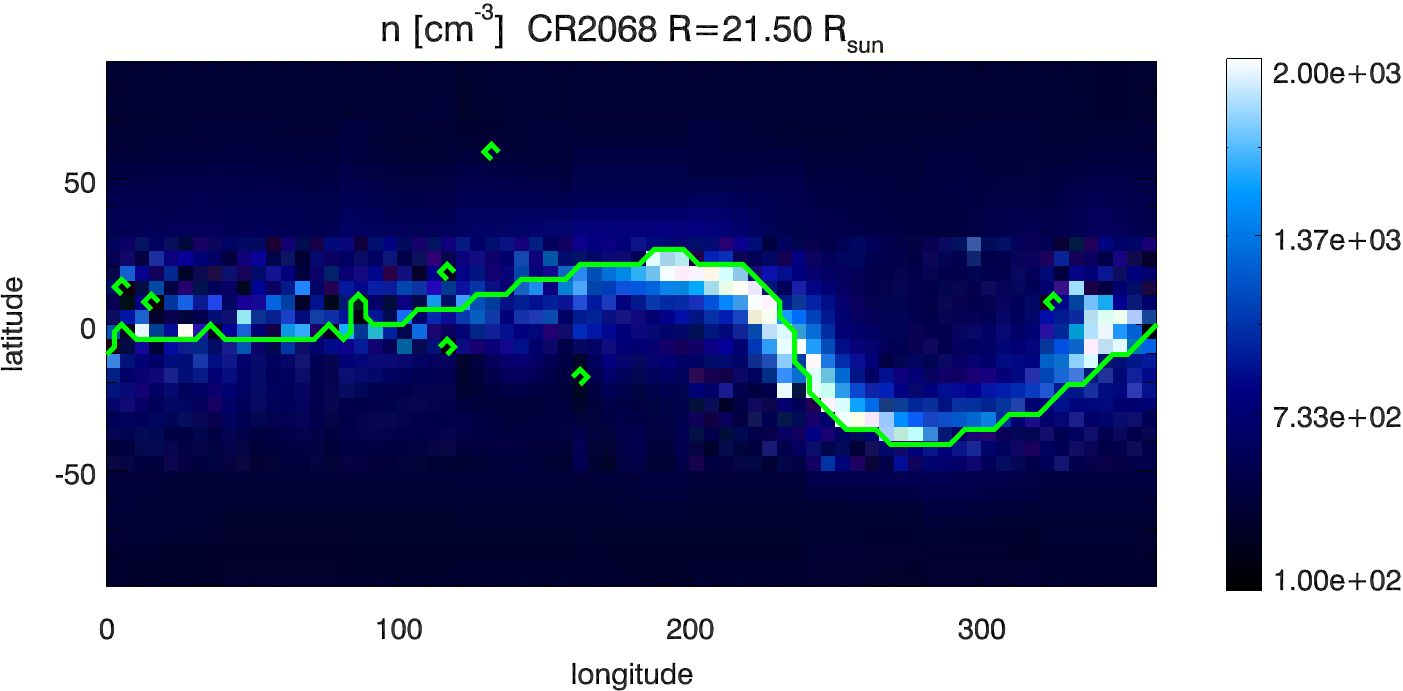}  \\ \medskip

  \includegraphics[width=0.325\linewidth]{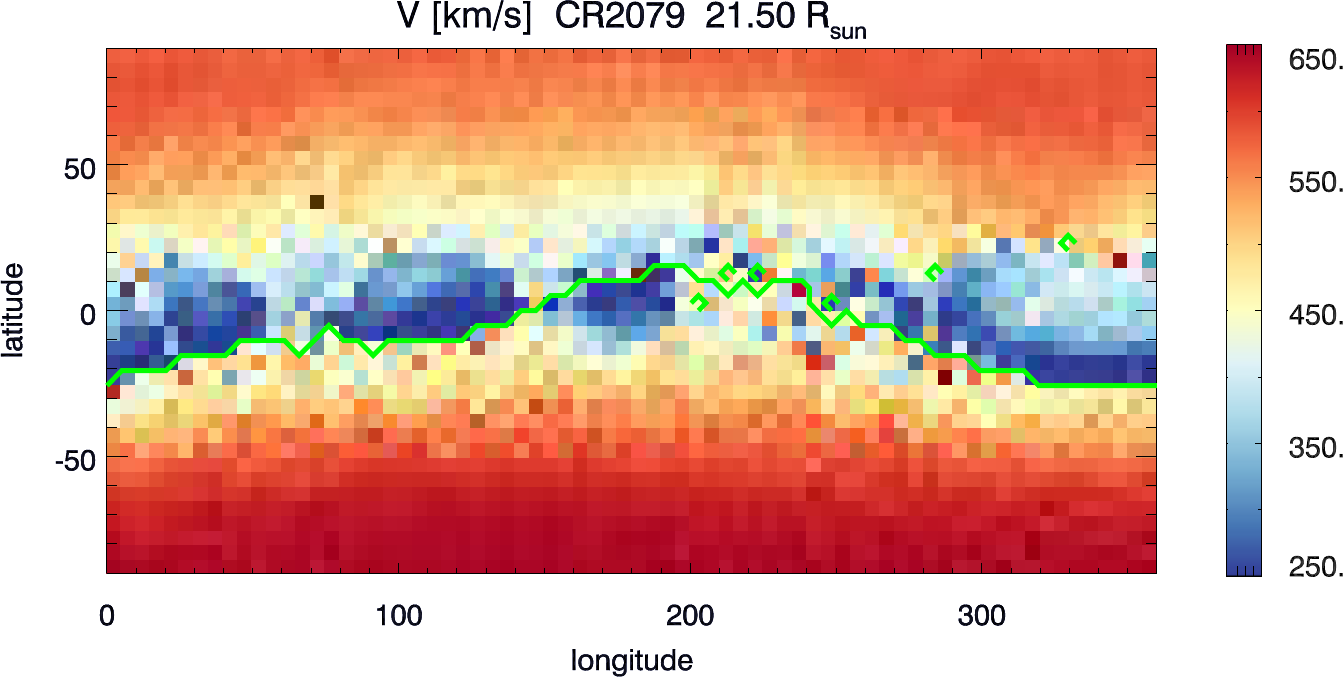} 
  \includegraphics[width=0.325\linewidth]{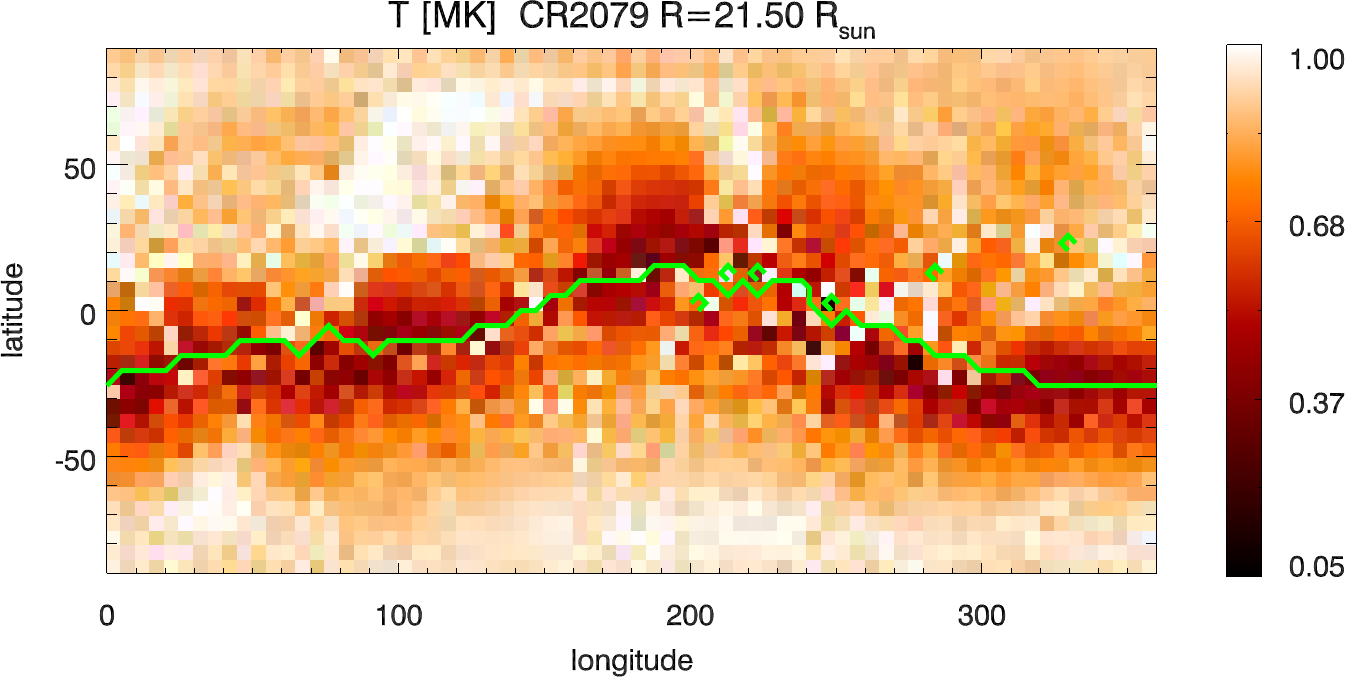} 
  \includegraphics[width=0.325\linewidth]{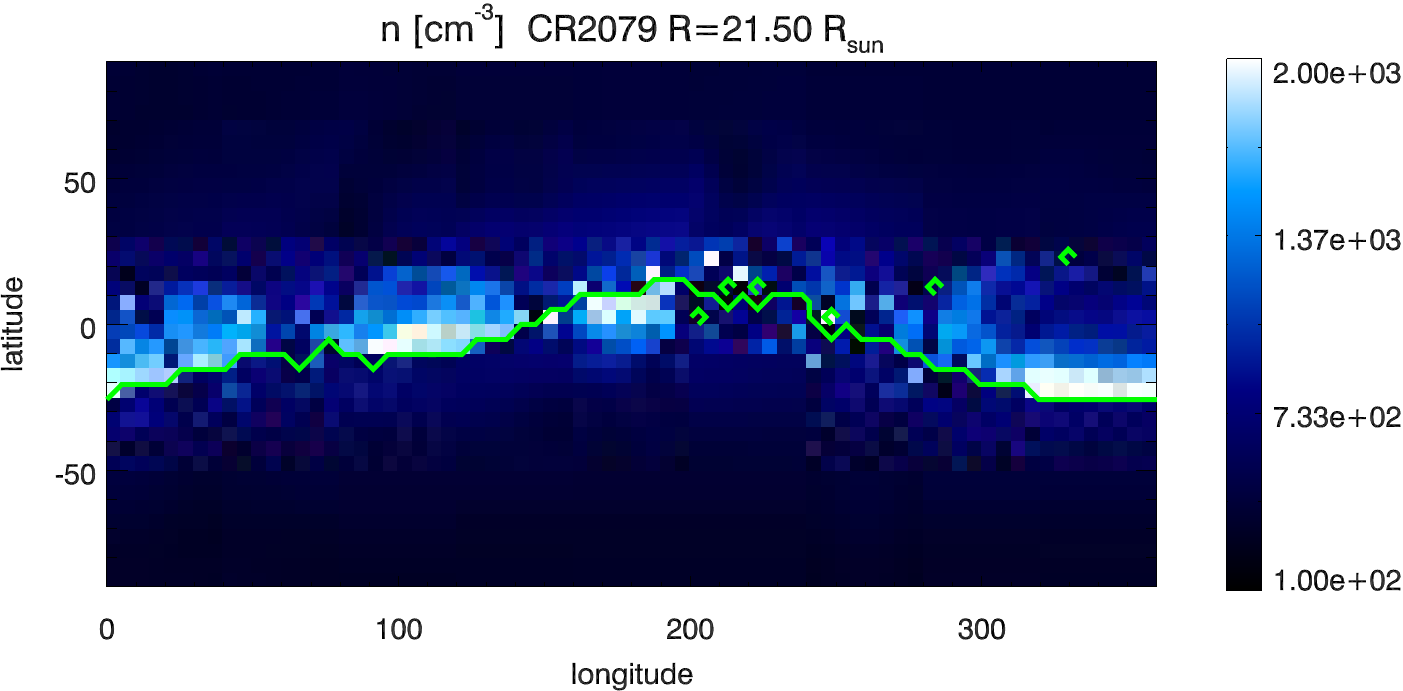}  \\ \medskip

  \includegraphics[width=0.325\linewidth]{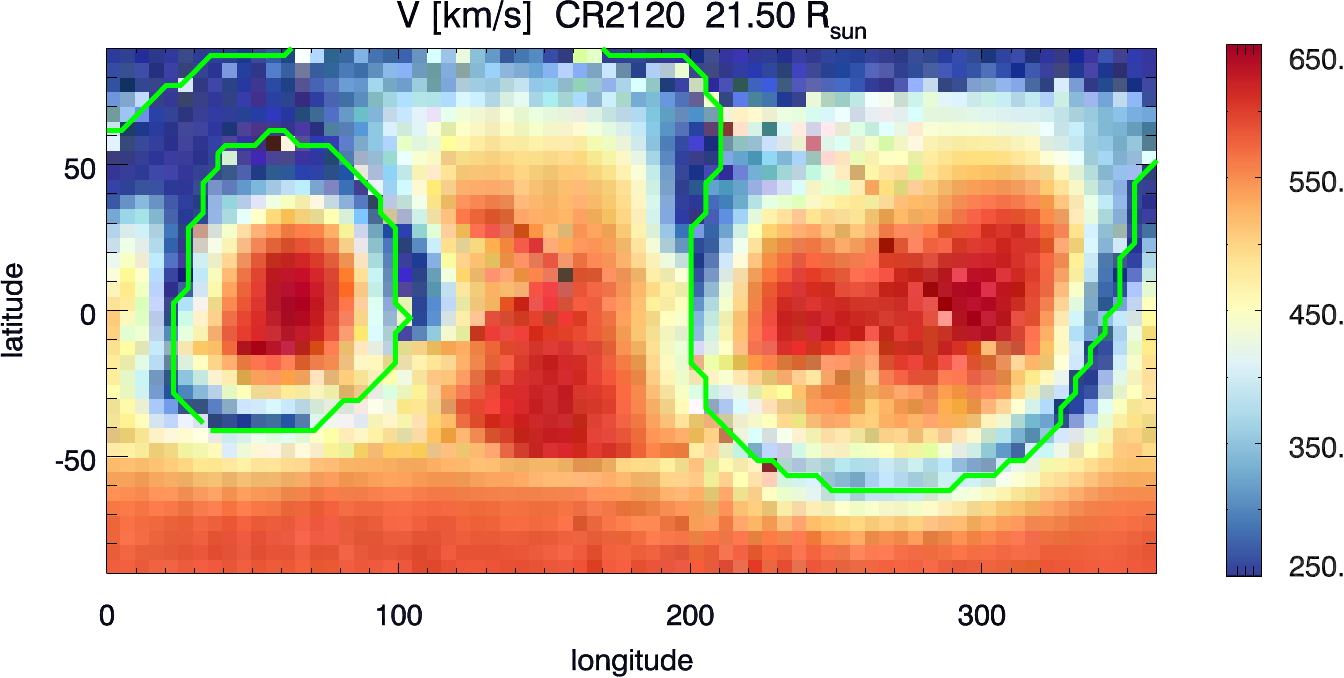} 
  \includegraphics[width=0.325\linewidth]{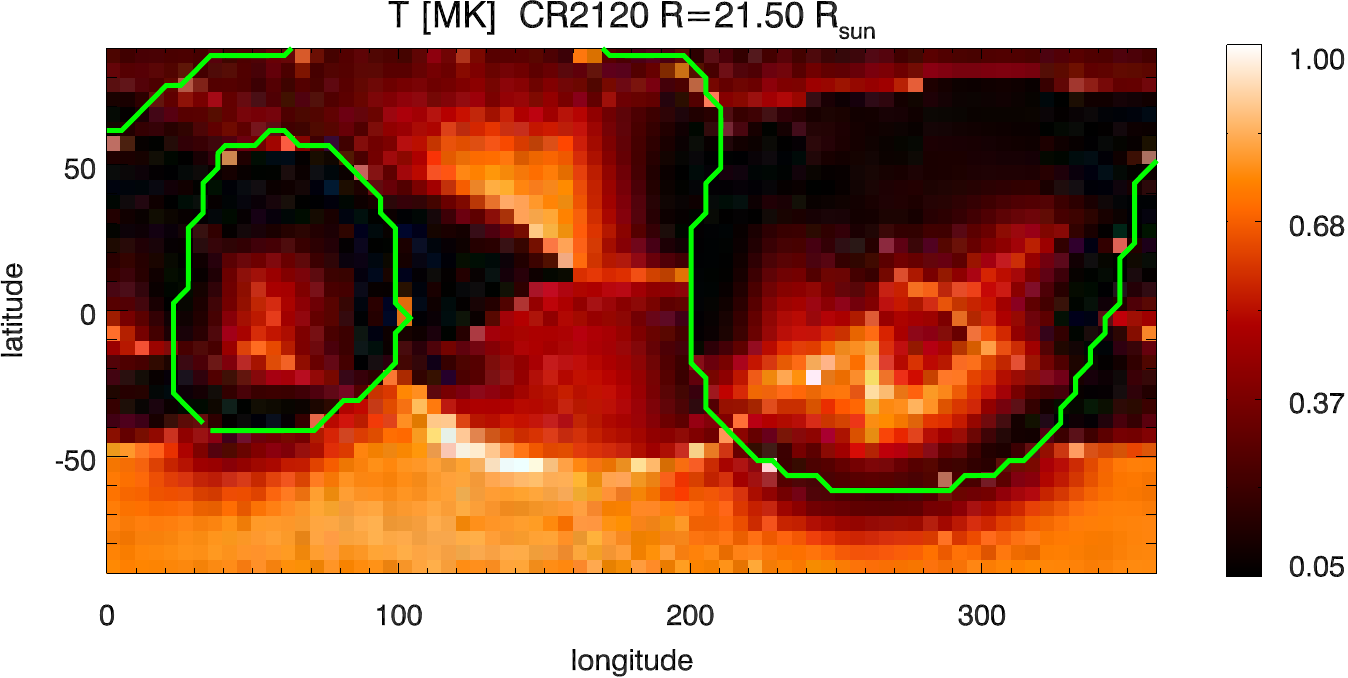} 
  \includegraphics[width=0.32\linewidth]{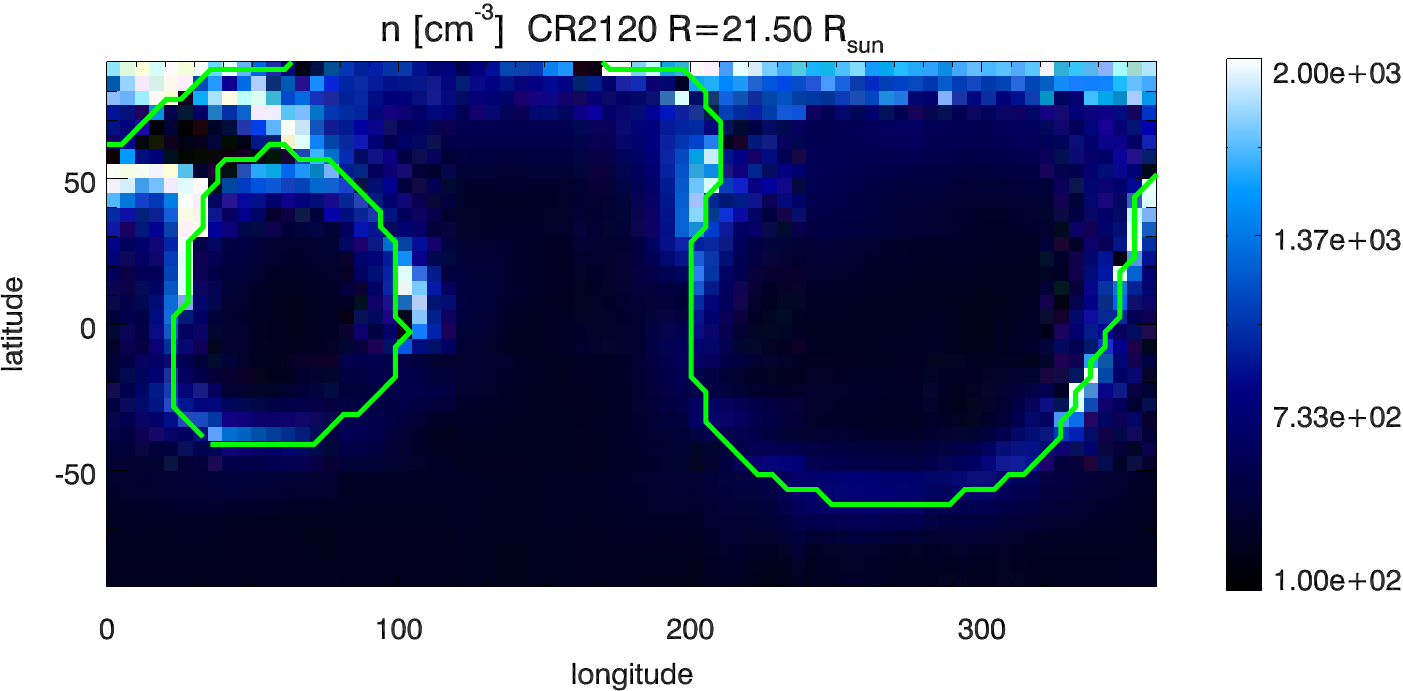}  \\ \medskip

  \includegraphics[width=0.325\linewidth]{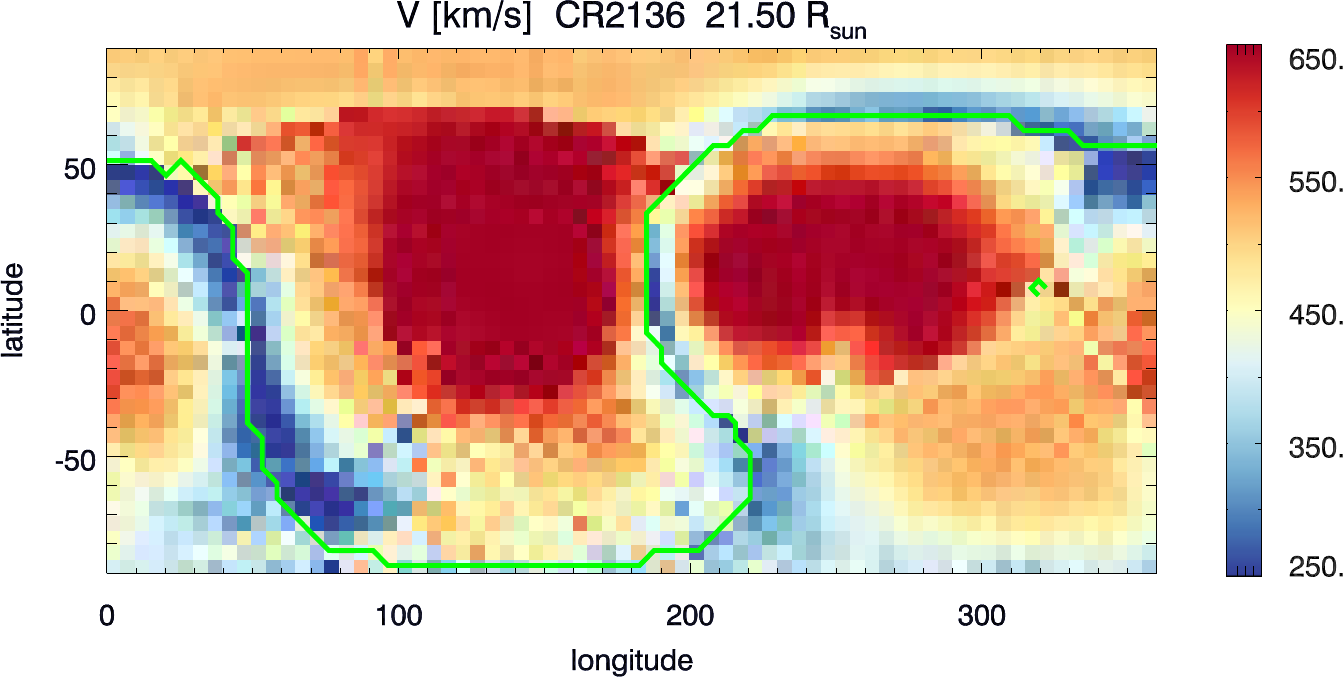} 
  \includegraphics[width=0.325\linewidth]{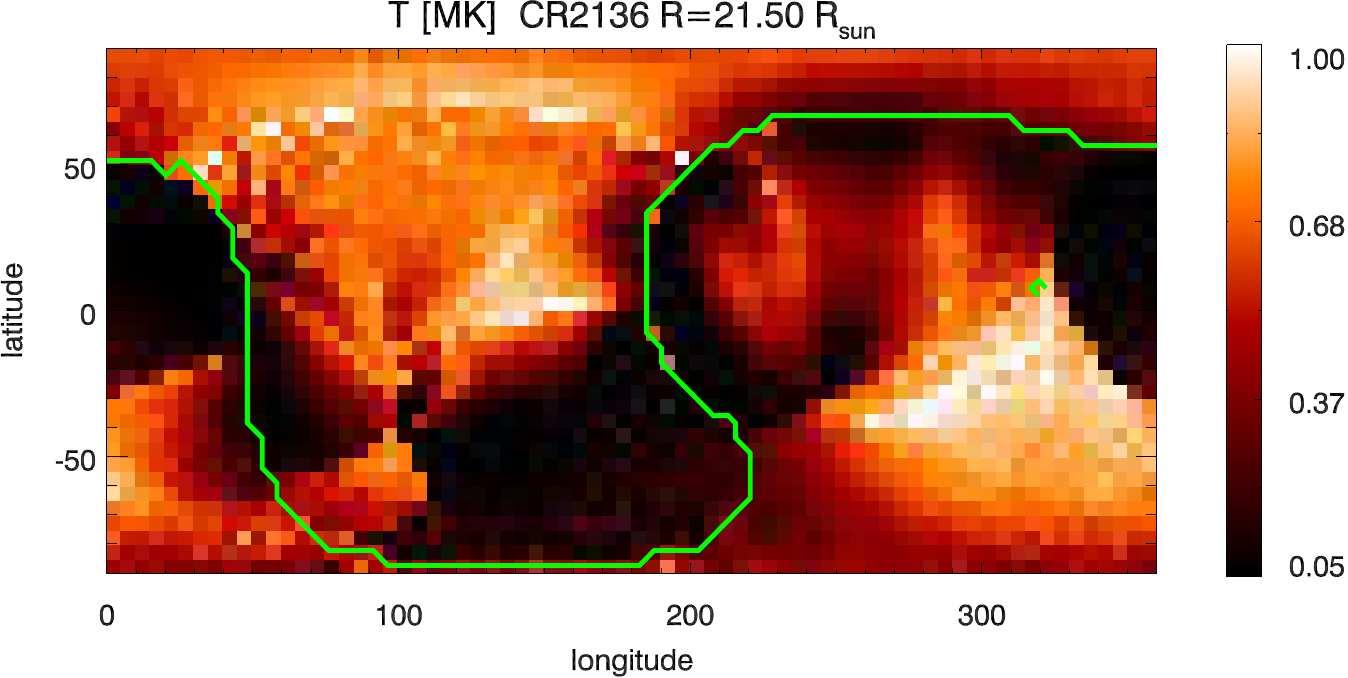} 
  \includegraphics[width=0.325\linewidth]{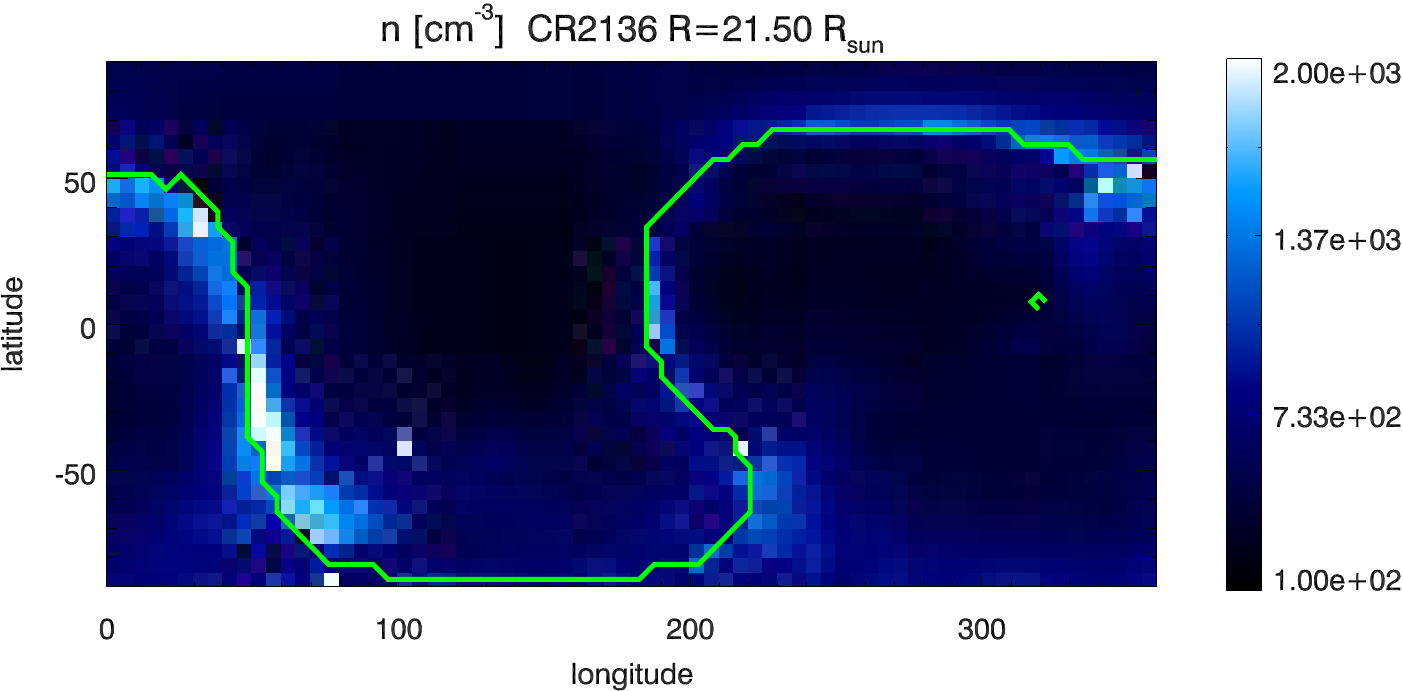} \\ \medskip

  \caption{
    Carrington maps of the computed wind speed, temperature and density for several Carrington rotations (2055, 2068, 2079, 2120 and 2136; the same as those in Fig. \ref{fig:maps_inital_bfield}) at $r\approx 21.5\un{\rsun}$.
    The green line shows the position of the heliospheric current sheet (HCS).
  }
  \label{fig:crmaps_21rsun}
\end{figure*}

% \begin{figure}
%   \centering

%   \caption{Wind speed, sound and Alfvén speed profiles.}
%   \label{fig:v_cs_ca}
% \end{figure}

\begin{figure}
  \centering
  \textsf{\textbf{CR 2055}} \\
  \includegraphics[width=0.8\linewidth]{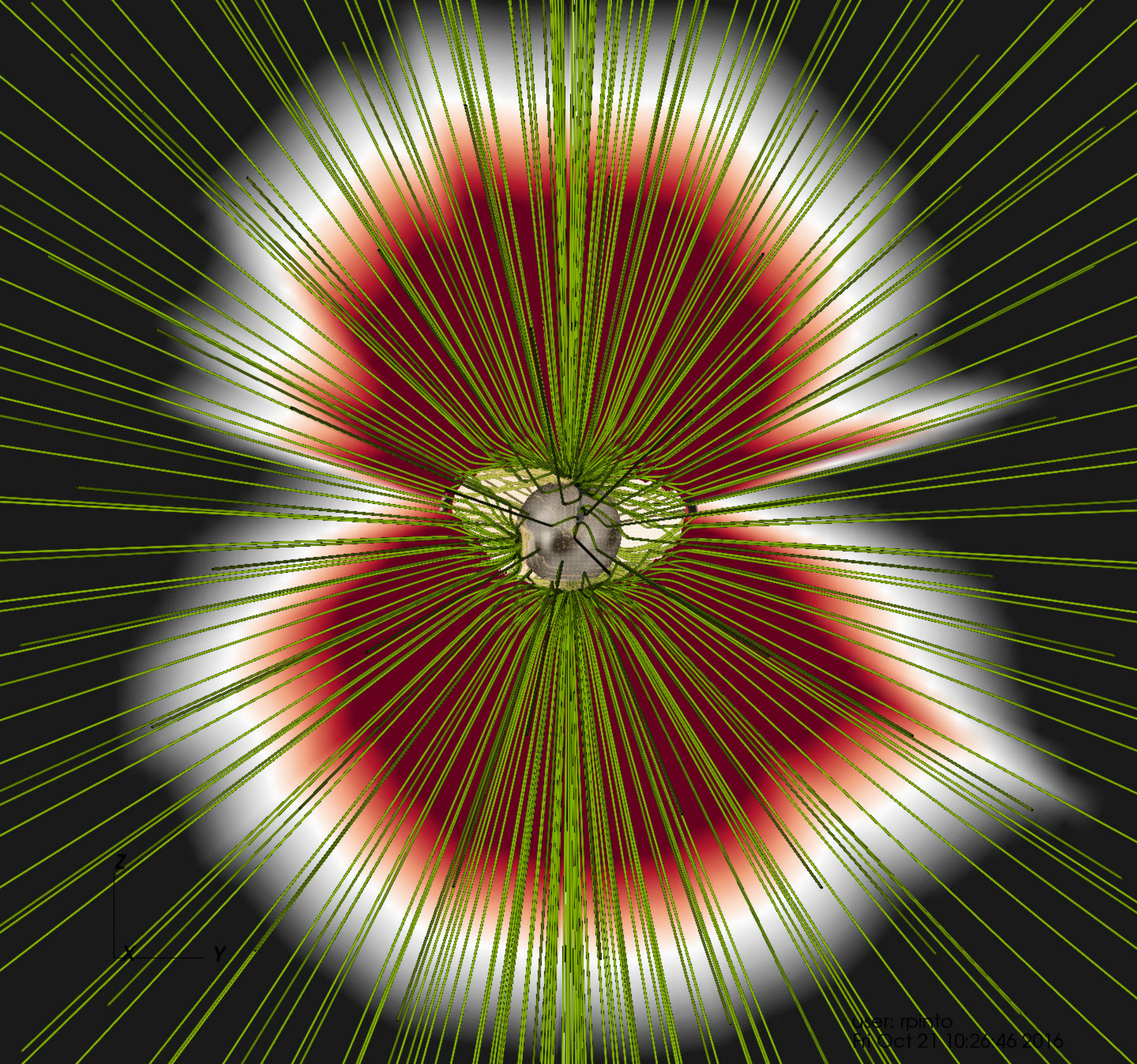}
  \\ \vspace{0.05\linewidth}
  
  \textsf{\textbf{CR 2133}} \\
  \includegraphics[width=0.8\linewidth]{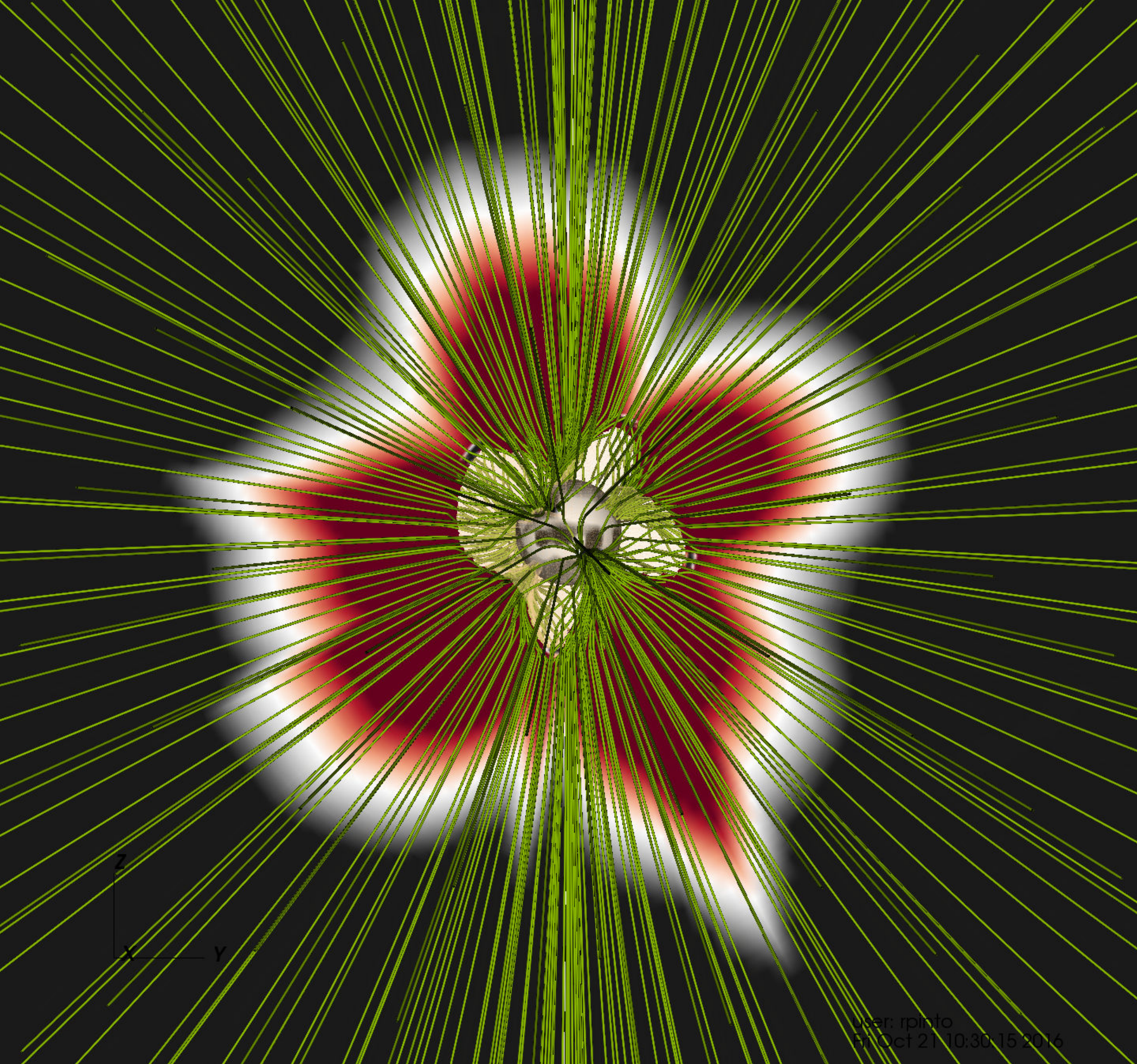}
  
  \caption{
    Alfvén Mach number profiles in the meridional plane (cut parallel to the plane-of-sky at mid Carrington rotation).
    Red and black represent the sub and super-alfvénic parts of the domain, respectively.
    The Alfvén surface is delineated in white.
    The top panel corresponds to CR 2055 (during the 2008 minimum) and the bottom panel to CR 2133 (close to the maximum of solar cycle 24).
  }
  \label{fig:alfven_speed}
\end{figure}

\prop{The baseline numerical model.}
The baseline model used to compute the solar wind profiles follows closely the strategy described in \citet{pinto_time-dependent_2009} and \citet{grappin_search_2010}, albeit with a number of modifications.
The numerical code solves the system of equations describing the heating and acceleration of a wind stream along a given magnetic flux-tube
\begin{eqnarray}
  \partial_t \rho  &+& \nabla\cdot\left(\rho\mathsf{u}\right)=0\ , \label{eq:rho} \\
  \partial_t \mathsf{u}&+&\left(\mathsf{u}\cdot\nabla_s \right)\mathsf{u} = -\frac{\nabla_s P}{\rho}  \nonumber \\
  &-& \frac{GM}{r^2}\cos\left(\alpha\right) + \nu\nabla_s^2u\ \label{eq:u}, \\
  \partial_t T &+& \mathsf{u}\cdot\nabla_s T + \left(\gamma-1\right)T\nabla\cdot\mathsf{u} =  \nonumber \\
  &-& \frac{\gamma-1}{\rho}\left[\nabla\cdot F_h + \nabla\cdot F_c +
    \rho^2\Lambda\left(T\right) \right] \label{eq:t}\ , 
\end{eqnarray}
where $\rho$ is the mass density, $\mathsf{u}$ is the wind speed, and $T$ is the plasma temperature.
The wind profiles are computed on a grid of points aligned with the magnetic field (with curvilinear coordinate $s$), $\alpha$ is the angle between the magnetic field and the vertical direction \citep[\emph{cf.}][]{li_solar_2011,lionello_validating_2014},
and $r$ represents the radial coordinate (distance to the center of the Sun).
The divergence operator is defined as 
\begin{equation}
  \nabla\cdot\left(*\right) =  \frac{1}{A\left(s\right)}\frac{\partial}{\partial s}\left(A\left(s\right) * \right) =
  B \frac{\partial}{\partial s}\left(\frac{*}{B}\right) \ ,
\end{equation}
where $B\left(s\right) \propto 1/A\left(s\right)$, $A\left(s\right)$ being the fluxtube's cross sectional area and $B\left(s\right)$ the magnetic field amplitude.
The ratio of specific heats is $\gamma = 5/3$. 
The terms $F_{\rm h}$, $F_{\rm c}$ denote the mechanical heating flux and the
Spitzer-Härm conductive heat flux, which are both field-aligned.
The radiative loss rate is $\Lambda\left(T\right)$.
For simplicity, we define here the mechanical heating flux $F_{\rm h}$ as a function which parametrizes the effects of the coronal heating processes \citep[rather than being the result of small-scale turbulent dissipation; \emph{cf. e.g.}][]{sokolov_magnetohydrodynamic_2013}.
We assigned it a phenomenological form inspired by those discussed by \citet{withbroe_temperature_1988}, \citet{mckenzie_acceleration_1995} and \citet{habbal_flow_1995}, but depending here on the basal magnetic field amplitude $\left|B_0\right|$, on the flux-tube expansion ratio $f$, and on the curvilinear coordinate $s$
\begin{equation}
  \label{eq_fluxp}
  F_{\rm h} = F_{\rm B0} \left(\frac{A_0}{A}\right)  \exp\left[-\frac{s-R_\odot}{H_{\rm f}}\right] \ .
\end{equation}
The heating coefficient $F_{\rm B0}$ is proportional $\left|B_0\right|$, and $H_{\rm f}$ represents an arbitrary damping length which is anti-correlated with the expansion ratio $f_{SS}$ in the low corona \citep[which is expected to lead to higher temperature peaks occurring at lower coronal altitudes and higher mass fluxes in the slower wind flows; see discussions by][]{cranmer_self-consistent_2007,wang_slow_2009}.
We ran a series of trial runs to calibrate the free parameters in Eq. (\ref{eq_fluxp}), and retained the forms $F_{\rm B0} = 12\e{5} \left|B_0\right| \un{erg\ cm^{-2}\ s^{-1}\ G}$ and $H_{\rm f} = 5\ f_{SS}^{-1.1} \un{\rsun}$.
The actual heating processes are still under debate, with physical mechanisms driven or triggered by MHD waves that propagate along open flux-tubes in the corona being favoured by the community.
Some of these invoke turbulent dissipation, resulting from the interaction of the upward propagating wave fronts with their reflected counterparts that feeds a turbulent cascade channeling energy from the larger injection scales to the smaller dissipative scales  \citep[\emph{cf.}][]{tu_mhd_1995,matthaeus_coronal_1999,verdini_alfven_2007,verdini_origin_2012,cranmer_self-consistent_2007}.
At the small-scale end of the turbulence spectrum, kinetic effects are likely to be responsible for the actual plasma heating, and specially for the selective heating of different species \citep{axford_origin_1992,kohl_uvcs/soho_1998,maneva_relative_2015}.
Another proposed scenario is that low frequency Alfvén waves, which are non-compressible modes, non-linearly convert energy to compressible modes, which dissipate much faster in the corona than their mother waves \citep{suzuki_making_2005,matsumoto_connecting_2014}.
In all the cases, the injected energy flux density and heat dissipation height are related to the geometry of each flux-tube (\emph{e.g} steepness of the stratification, gradients of wave phase speeds) as well as to the amplitude of the magnetic field (\emph{e.g} Poynting flux associated with transverse waves).
The formulation for $F_{\rm h}$ in Eq. (\ref{eq_fluxp}) provides a simple description of these phenomena that we can apply generally (for all flux tubes simulated), but that can be easily substituted in future work by a more precise phenomenology.

The radiative loss function is given by
\begin{equation}
  \Lambda(T) = 10^{-21}10^{[\log_{10}(T/T_{\rm M})]^2}\chi(T)\ ,
\end{equation}
where $\chi(T)$ equals unity for $T>0.02$~MK and varies linearly from $0$ to $1$ for $0.01<T<0.02$~MK. 
$T_{\rm M}$ equals $0.2$~MK.
We employ a nonuniform grid of $640$ points between the solar surface (where $\Delta r = 10^{-4}$~$R_\odot$) and  $31.5\rsun$ (where $\Delta r = 0.3\rsun$).
Time integration is done with a Runge-Kutta scheme of order 3, while an implicit finite-difference scheme of order 6 is used for the spatial dimension, except when computing temperature gradients in the conductive term, for which an explicit scheme of order 2 is applied. 
Numerical filtering is employed to increase the stability of the schemes \citep{lele_compact_1992}.
The top and bottom boundary conditions are transparent to perturbations propagating away from the domain, and are set in terms of the characteristic form of the hyperbolic part of equations (\ref{eq:rho})--(\ref{eq:t}), as in \citet{grappin_acoustic_1997}.
The non-hyperbolic terms in eqs. (\ref{eq:u}) and (\ref{eq:t}) are negligible at both numerical boundaries, except for $F_h$, which is prescribed \emph{a priori}. 
Placing the lower boundary at $r\approx 1\un{\rsun}$ lets us avoid imposing strong constraints on the mass and heat fluxes at the source region of the solar wind.
The plasma temperature is practically uniform and constant in the lowest layers of our domain, remaining close to $6000\un{K}$ and making both conduction and optically-thin radiative cooling vanish there.
Density variations are also negligible there, unlike at the top of chromosphere (see Fig. \ref{fig:wind_profiles} and Sect. \ref{sec:wind_maps}).
Each individual wind solution is obtained by a relaxation process during which the time-dependent system converges to a unique stationary final state,  as in \citet{pinto_time-dependent_2009} and \citet{grappin_search_2010} \citep[\emph{cf.} also background wind solution of][]{verdini_origin_2012}. 
To speedup the relaxation process, we start off from typical slow and fast wind solutions calculated in advance rather than performing full \emph{ab initio} computations (\emph{i.e,} the initial state already has a stable chromosphere, transition region and transsonic wind flow).

\prop{Performance}
As the core computing task consists of an ensemble of independent calculations, the parallel speedup scales linearly with the number of processes. 
We currently compute full synoptic maps (at a $5\degree$ angular resolution) in about $6\un{hr}$ using $320$ parallel computing cores.
In other words, the model can run almost in real-time for source magnetic field maps with a cadence larger or equal to $6\un{hr}$, or of four maps per day (which is the cadence of the standard PFSS maps currently available via SolarSoft).
Other geometrical setups are possible, using for example different angular resolutions or covering smaller fractions of the synoptic map.
The total computing load will, in all cases, depend mostly on the number of uni-dimensional samples used and on the typical wind speeds on the regions of the solar atmosphere considered (due to the CFL condition on the integration time-step; wind solutions on regions dominated by slow wind converge faster than those on regions dominated by fast wind).

%%%%%%%%%%%%%%%%%%%%%%%%%%%%%%%%%%%%%%%%%
\section{Results}
\label{sec:results}

\begin{figure*}

  \centering

  \includegraphics[clip, trim = 130 0 0 120, width=0.75\linewidth]{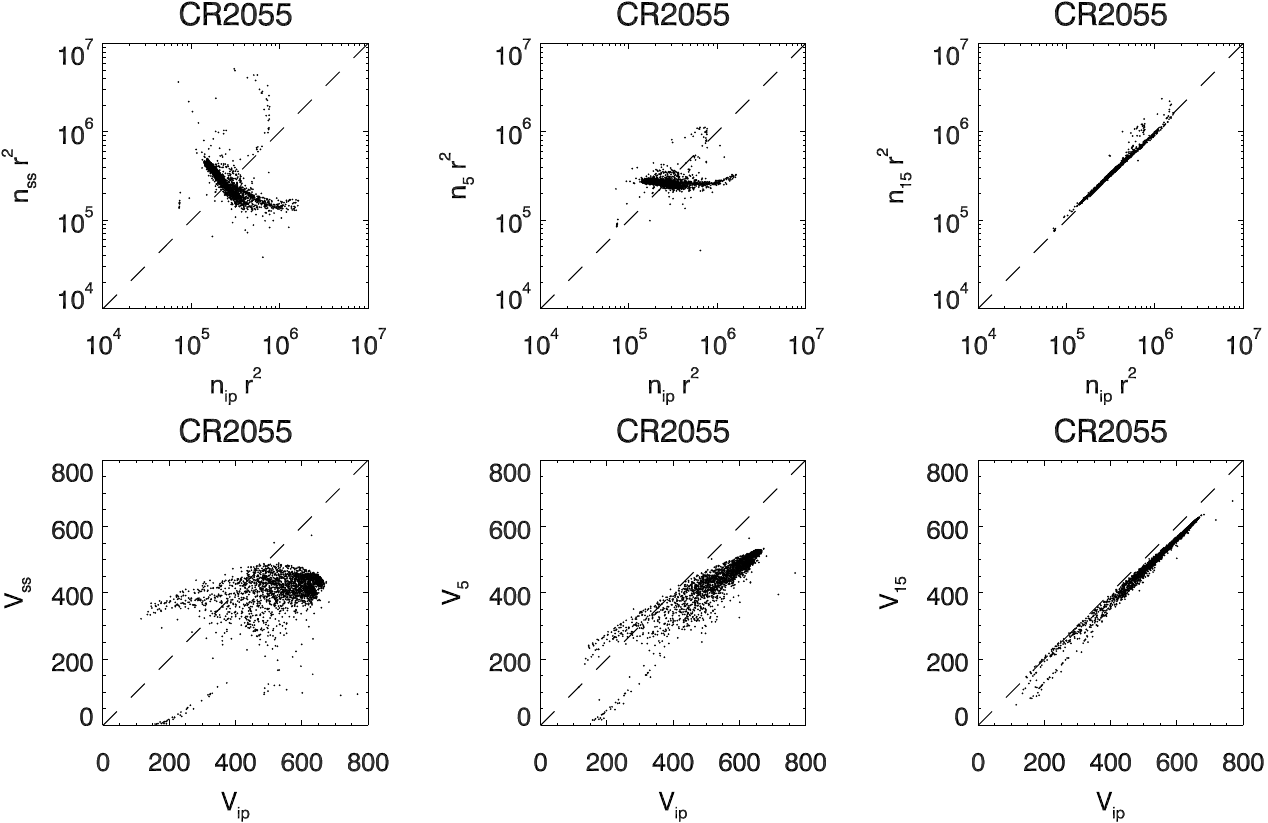}  

  \medskip \hrule \medskip

  \includegraphics[clip, trim = 130 0 0 120, width=0.75\linewidth]{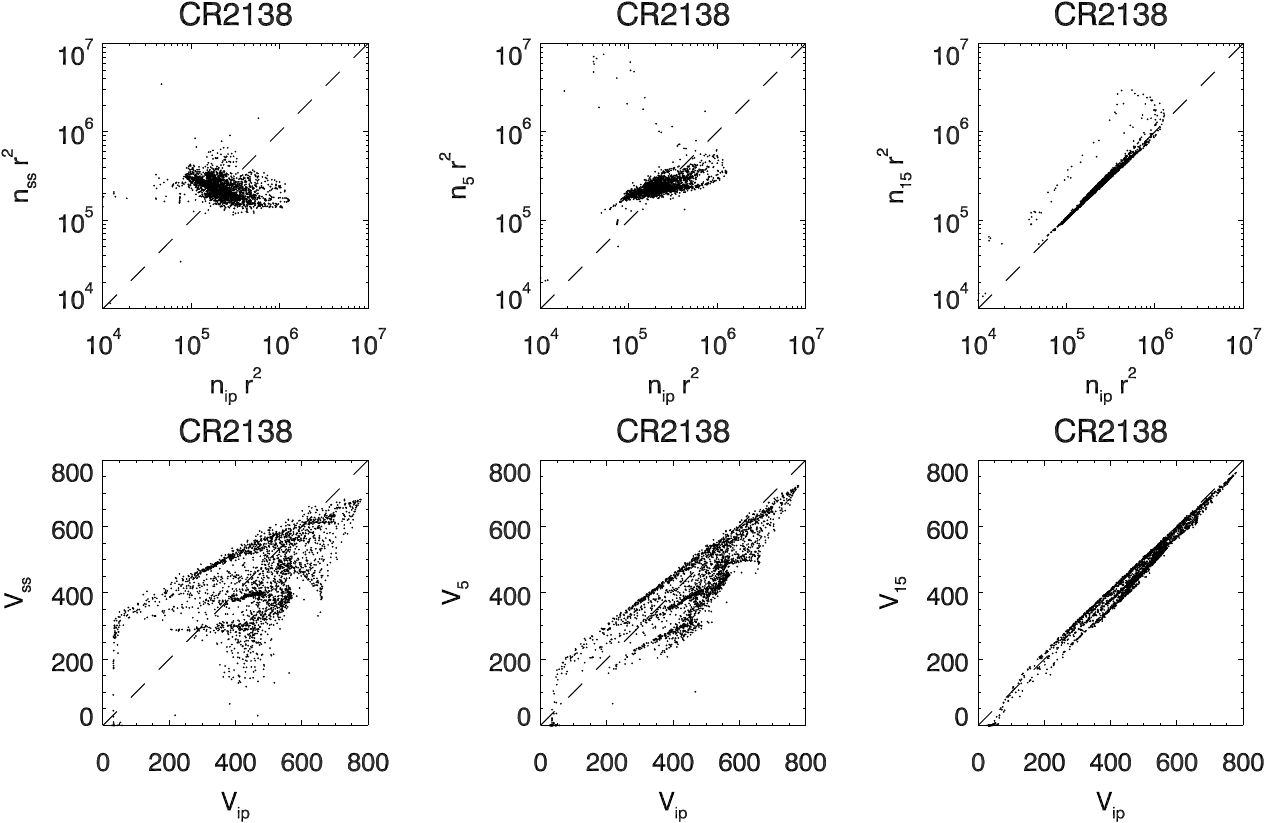} 
  \caption{
    Evolution the numerical density $n$ and of the wind speed $V$ along each flux-tube with height.
    Each scatter-plots show the normalized density ($n r^2$) and the wind speed ($V$) at a given height as a function of their values at the outer boundary at two different Carrington rotations (2055 and 2138) close to solar minimum and to solar maximum.
    Each dot represents one given flux-tube.
    The dashed lines mark the positions of the flux-tubes for which no change occurs between the height represented and the outer boundary.
    Points above or below this line indicate that the quantity represented decreases or increases, respectively, during the propagation between the two heights.
    \rev{The indices $ss$, $5$, $15$ and $ip$ in the axis labels correspond respectively to the height of the source-surface ($2.5\un{\rsun}$), $5$, $15$ and $30\un{\rsun}$}.
  }
  \label{fig:propag_diagrams}
\end{figure*}

\subsection{Magnetic field}
\label{sec:mag_field}

The three-dimensional geometry of the coronal magnetic field is given directly by PFSS extrapolations and is represented in Fig. \ref{fig:mvp_scheme} for Carrington rotations 2056 (at the end of solar cycle 23, during the solar minimum of 2008) and 2121 ($\sim 1.5$ years before the peak of cycle 24).
The former is characterized by equatorial streamers covering all longitudes, large coronal holes rooted at the poles, and a well-defined heliospheric current sheet (HCS) which remains close to the ecliptic plane (albeit with a noticeable warp).
The latter shows, in contrast, the presence of high-latitude streamers and of large coronal holes rooted close to the equator and a double-lobed HCS.
The figure also shows that the resulting slow and fast wind distribution relate to the large-scale geometry of the magnetic field, with slow wind confined to low latitudes during solar minimum and with slow and fast wind streams mixed up in latitude during solar maximum.
Figure \ref{fig:maps_inital_bfield} shows full Carrington maps of the magnetic field obtained after extrapolating the WSO magnetograms via PFSS and applying the corrections described in Sect. \ref{sec:methods} at different heights.
Each row represents a different Carrington rotation (repectively CRs 2055, 2068, 2079 and 2136).
The first three cover a period of time close to solar minimum (years 2007 and 2008).
The fourth column corresponds to the time of reversal of the global magnetic polarity at about the sunspot number maximum.
In this particular example, the HCS splits up into two separate structures with an almost cylindrical shape \citep[as discussed by]{wang_evidence_2014}.
The last row corresponds roughly to the beginning of the decaying phase of cycle 24.
The first column shows the magnetic field at the surface of the Sun (WSO data), the second column shows the magnetic field amplitude just below the source surface, and the last column at about $21.5\rsun$.
The orange dots show the positions of the foot-points of all the open flux-tubes considered and mark the coronal holes (the orange dots are not plotted on the 2nd and 3rd rows as they would sample the whole plane uniformly there).
As expected, the mix of polarities at the surface arrange into two (or into a few) magnetic sectors higher up in the corona, with one well defined polarity inversion line (PIL) marking the position of the heliospheric current sheet (HCS).
The open magnetic field amplitude becomes progressively uniform within each sector with height, but only in the upper part of the corona, well above the source-surface (\emph{cf.} second and third columns of Fig. \ref{fig:maps_inital_bfield}).
This is in good agreement with the measures of the radial magnetic fields in the interplanetary medium by space probes such as ULYSSES.
However, the PFSS coronal field reconstruction does not on its own generate such uniform open magnetic fields, and a straight-forward radial extension of the field would propagate the large scale non-uniformities in the magnetic flux up to all heliospheric heights.
This justifies the correction applied to the magnetic field geometry discussed in the beginning of Sect. \ref{sec:methods}, which consists of smooth adjustments to the flux-tube expansion ratios in the higher corona (between $2.5$ and $12\rsun$) such that the open magnetic flux becomes uniform (while keeping the total open flux invariant).
The HCS remains everywhere thin, in opposition to global MHD models which require and enhancement of cross-field diffusion on scales of the order of the transverse grid size, which are well above those characteristic of cross-field turbulence and reconnection processes \citep[see e.g][]{lazarian_reconnection_1999}.
Figures \ref{fig:profiles_inital_bfield} and \ref{fig:profiles_inital_expans} show the radial profiles of the unsigned magnetic field amplitude and of the expansion ratio for a small sample of open flux-tubes at two different Carrington rotations.
The blue dotted-lines show the same profiles without the correction to the magnetic flux-tubes in the high corona.
These adjustments to the flux-tube expansion ratios in the high corona have a moderate effect on the terminal wind speed distribution, a negligible effect on the wind temperature, but a strong effect on the wind density, and in particular on the correlation between density and terminal wind speed (see Sects. \ref{sec:wind_maps} and \ref{sec:correlations}).

\begin{figure}
  \centering
  \includegraphics[width=0.85\linewidth]{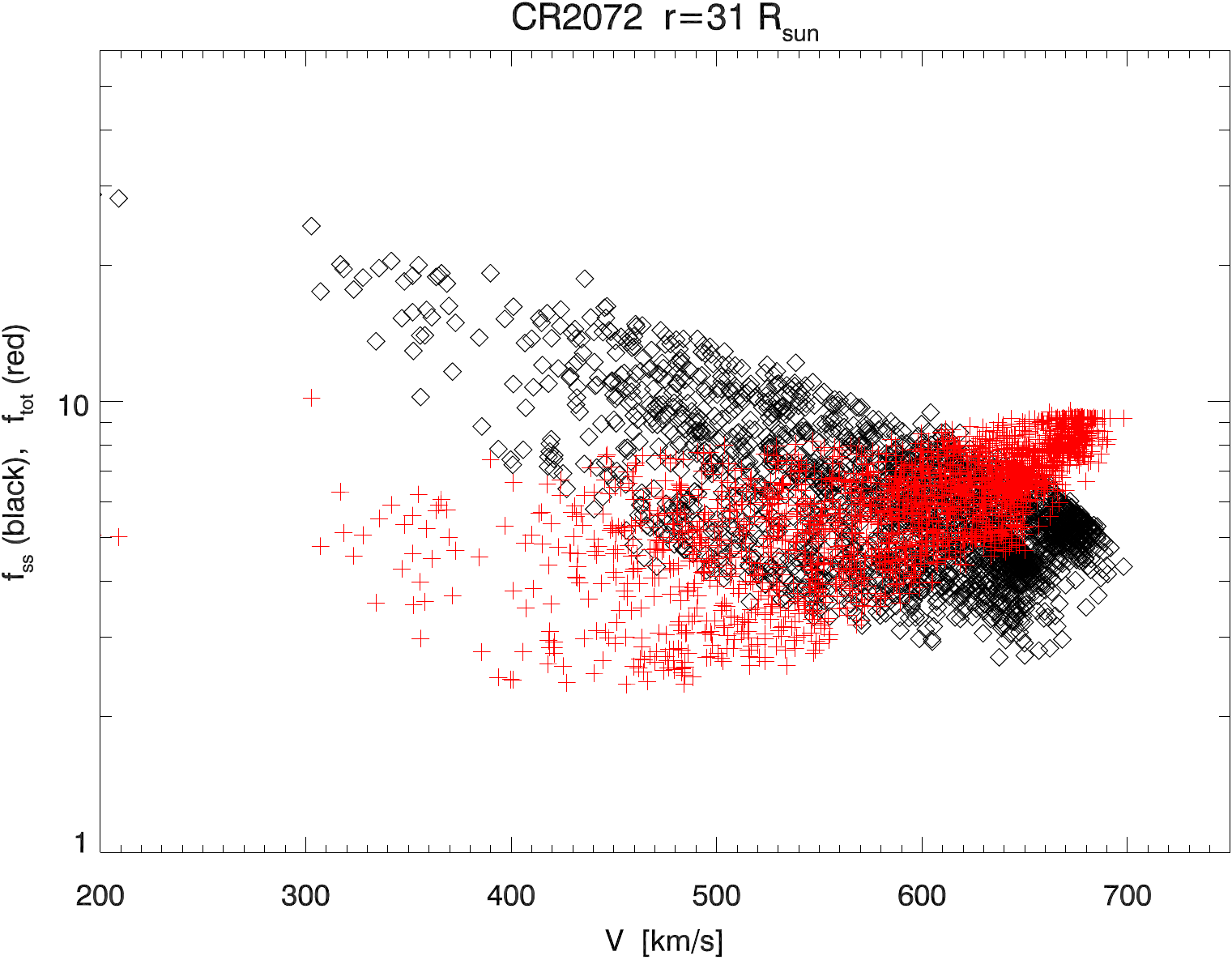}
  \caption{
    Dependence of the wind speed $V$ on the flux-tube expansion factor $f$ evaluated at the source-surface (black symbols) and at the outer boundary (red symbols).
  }
  \label{fig:v_vs_f}
\end{figure}

%\clearpage
\subsection{Solar wind properties}
\label{sec:wind_maps}

\rev{Multiple} density, temperature and wind speed profiles we obtained are shown in Fig. \ref{fig:wind_profiles} as a function of atmospheric height ($h=r-\rsun$) across the whole computational domain.
The log-log scale lets us visualize the full atmospheric stratification, especially in its lower layers.
The curves represented in the plot correspond to a sample of about 300 individual wind streams selected randomly from all the CR's in Figs. \ref{fig:maps_inital_bfield} and \ref{fig:crmaps_21rsun} that have asymptotic speeds falling close to $270,\ 360,\ 500$ and $650\un{km/s}$.
The curves are coloured according to the values of their asymptotic speeds using the same colour-scale as the two last columns in Fig. \ref{fig:mvp_scheme} and the first column in Fig. \ref{fig:crmaps_21rsun} (that is, from dark blue at $250\un{km/s}$ to dark red at $650\un{km/s}$).
The slower wind flows converge more slowly to their asymptotic state and exhibit a higher degree of variability than their faster counterparts.
The slowest of these streams (dark blue lines) can go through one or more zones of deceleration, unlike the fastest wind flows (dark red lines).
These can occur either below the source-surface on flux-tubes with sharp variations of their cross-sections (over-expansion, reconvergence) and inclination \citep[\emph{cf.}][]{pinto_flux-tube_2016}, and also above due to the smoother variations of their expansion factors in the high corona.
Slow wind flows are usually denser than fast flows across all coronal heights, but a few of them can be under-dense between $h=0.01$ and $\approx 2\un{\rsun}$.
Plasma temperature remains invariable in the bottom half of the chromosphere (to machine precision), and density shows only very small variations between all the computed solutions.
However, variations in basal magnetic field amplitude and flux-tube expansion alter the heating amplitudes and scale-heights (see Eq. \ref{eq_fluxp}), and lead to significant differences in the density of the various wind streams in the higher part of the chromosphere and in the low corona.
\rev{Typically}, slow wind flows are associated with stronger over-expansions in the low corona that lead to reduced $H_{\rm f}$, to shallower than average density fall-offs with height across the transition region (``evaporation''), and to higher mass fluxes.
This effect can be offset or reinforced by variations of $\left|B_0\right|$, which span a higher range for slow than for fast wind flows.
The balance between these variations in heating rate, density stratification, thermal conduction and radiative cooling determines the position of the TR and the temperature profile across the corona.

\prop{CR wind maps}
Figure \ref{fig:crmaps_21rsun} shows a sequence of maps of the wind speed, plasma temperature and density at $21.5\rsun$ for several Carrington rotations (at the same as in Fig. \ref{fig:maps_inital_bfield}).
The vast majority of the wind streams are close to their assymptotic state, the exceptions being those on the lower tail of the wind speed distribution, such as the $200 - 250 \un{km/s}$ flows appearing on the 4th row of the figure \citep[\emph{cf.}][]{sanchez-diaz_very_2016}.
Slow wind flows tend to concentrate on the close vicinity of the HCS, but also occur in the regions surrounding pseudo-streamers.
One such example is the feature on the western part of the maps on the second and third lines of the figure.
The fastest wind flows always occur within large coronal hole, which are rooted at the polar region during minimum, and sometimes at equatorial latitudes during maxima.
The last two rows of the figure show two examples of this situation at the the peak and decay phase of cycle 24 \citep[\emph{cf.}][]{wang_evidence_2014}.
The plasma temperature is generally well correlated with wind speed, while the plasma density anti-correlates (see Sect. \ref{sec:correlations} for a quantitative analysis).
The velocity and density maps shown in Figure \ref{fig:crmaps_21rsun} for solar minimum show a close resemblance to those in \citep{yang_time-dependent_2012}, but with the band of slow wind being much thinner (its thickness is closer to that in the WSA model maps).

The combination of the magnetic field amplitude of the corona with the obtained wind speeds and densities lets us deduce the distribution of the characteristic magneto-hydrodynamical phase speeds.
Figure \ref{fig:alfven_speed} represents the Alfvén Mach number $M_A$, which is the ratio between the wind speed and the Alfvén speed (measured in the direction parallel to the magnetic field, equal to the radial direction above the source-surface), in the meridional plane parallel to the plane-of-sky in the middle of Carrington rotations 2055 and 2133 (or equivalently, the meridional planes which cross longitudes $90\degree$ and $270\degree$ in the Carrington map).
The white shades indicate the positions for which $M_A = 1$, delineating the Alfvén surface.
During solar minimum (top panel), the Alfvén surface assumes typically a prolate double-lobed ellipsoidal shape, with the Alfvén radius $r_A$ reaching values close $10-12\un{\rsun}$ over the poles, $5 - 6\un{\rsun}$ at mid-latitudes, and showing strong incursions at low latitudes.
Thin outward ``spikes'' reaching heights above $15\un{\rsun}$ can sometimes appear in the close vicinity of the HCS, where the magnetic field amplitude and the local Alfvén speed quickly approach zero.
The Alfvén surface becomes closer to spherical during solar maxima (bottom panel of the figure), albeit with some irregularities, with an average Alfvén radius in the range $5-7\un{\rsun}$.
These variations are comparable to those found by \citet{pinto_coupling_2011} using idealized solar dynamo and wind models, but with a smaller contrast between the values of $r_A$ at minimum and maximum.
Overall, the values of $r_A$ we obtain are overall lower than those reported by \citet{deforest_inbound_2014} from the analysis of inbound propagation of density perturbations in coronograph data \citep[see also][]{tenerani_inward_2016,sanchez-diaz_observational_2017}.
The effects of setting a fixed source-surface radius $R_{SS}$ on the open magnetic flux amplitude discussed by \citet{reville_solar_2015} probably contribute to these disparities \citep[see also][]{lee_coronal_2011,arden_breathing_2014}.

Figure \ref{fig:propag_diagrams} compares the plasma density and wind speed along each individual flux-tube at different heights with their asymptotic values for two different Carrington rotations, in order to show how these quantities evolve during the propagation (see \citet{mcgregor_radial_2011} for similar scatter plots at higher heliospheric heights).
The dashed lines mark the positions for which no net evolution occurs during the propagation between the two heights considered.
Points placed above this line indicate that the plotted quantity has decreased during the trajectory, while points below the line indicate an increase.
\rev{The indices $ss$, $5$, $15$ and $ip$ in the axis labels correspond respectively to the height of the source-surface ($2.5\un{\rsun}$), $5$, $15$ and $30\un{\rsun}$}.
%The panel corresponding to CR 2070 also shows, over-plotted in blue, the solutions found without correcting for the flux-tube expansion in the high corona.
The figure shows that all of the fast wind streams are already close to their asymptotic state at $15\rsun$ (to within $8\%$ of the asymptotic speed).
The slow streams are accelerated more progressively, with a fraction of them falling within an envelope $40\%$ below the value of their asymptotic speed at the same height.
Note that the plots show the absolute difference to the asymptotic speeds, which can be of the same order for both fast and slow wind flows, unlike the relative differences (see also Fig. \ref{fig:wind_profiles}).
The general trend in our simulations is that the slow wind is accelerated more progressively than the fast wind, in accordance to, \emph{e.g.}, \citet{habbal_flow_1995} and \citet{sheeley_measurements_1997}.
Interestingly, a non-negligible fraction of the wind streams are decelerated in the intervals $2.5 - 21.5\rsun$ and $5 - 21.5\rsun$ (but not above $15\rsun$).
The decelerations are more frequent and more significant in the slow wind regime, for which the most extreme cases drop from about $400$ to $200\un{km/s}$ between $2.5$ and $15\rsun$.
%The blue scatter points for CR 2070 show that this not a spurious consequence of the geometric corrections applied to the magnetic field in the high corona. 
The wind speed diagrams show a more complex structure during phases of high solar activity, sometimes indicating the presence of two populations of wind flows in the low corona which converge higher up into a smoother distribution (cf. CR 2138 in the figure).
The evolution of the plasma density (normalized for the radial cross-field expansion) shows a rather regular trend, with a progressive reversal of the slope of the diagram between $r=R_{SS}$ and $15\uns{\rsun}$.
Both the initially overdense and underdense streams converge progressively and monotonically to their asymptotic values below $15\rsun$.

\begin{figure}
  \centering
  \centering
  \includegraphics[width=0.85\linewidth]{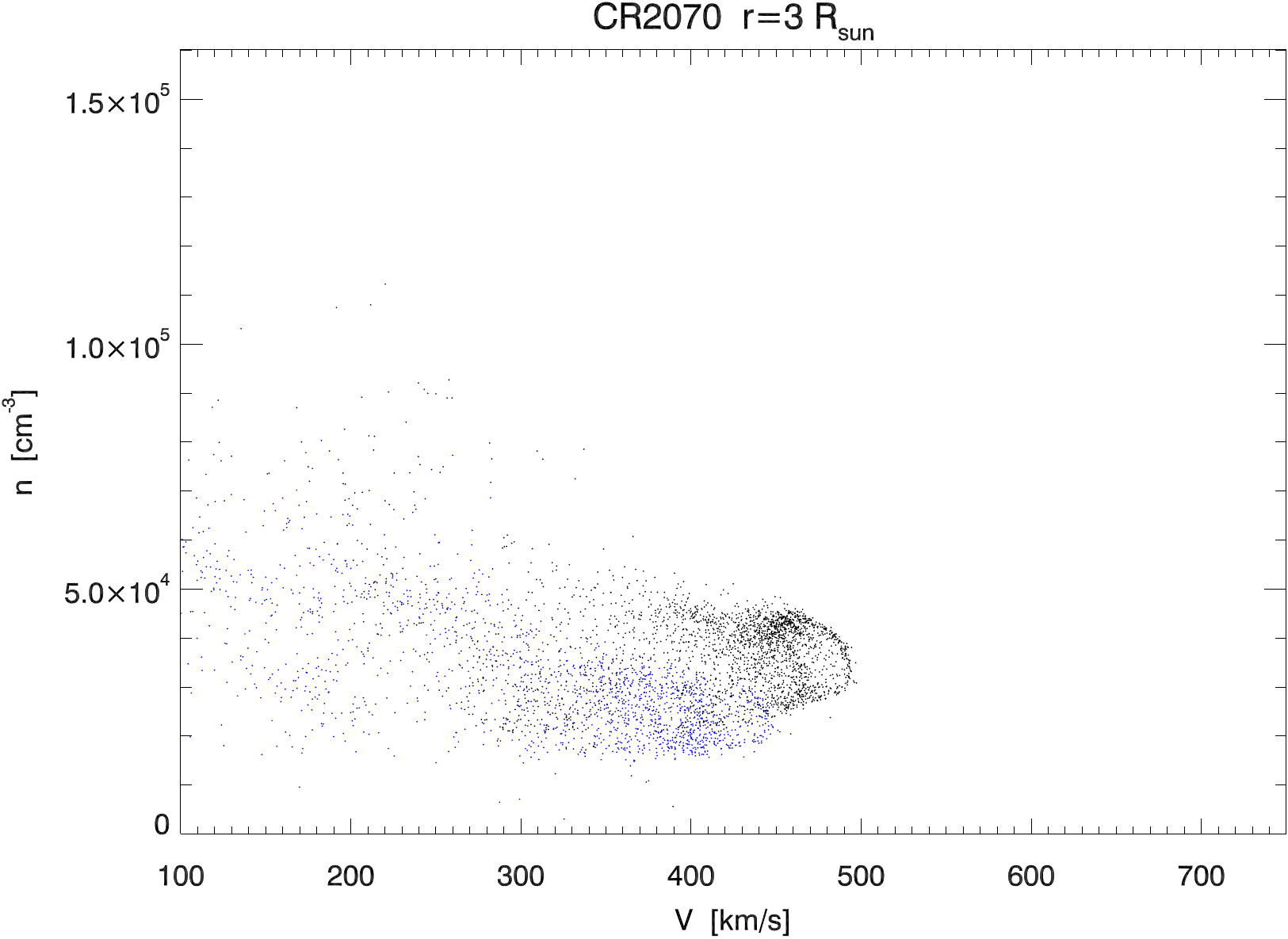}  \\ \medskip
  \includegraphics[width=0.85\linewidth]{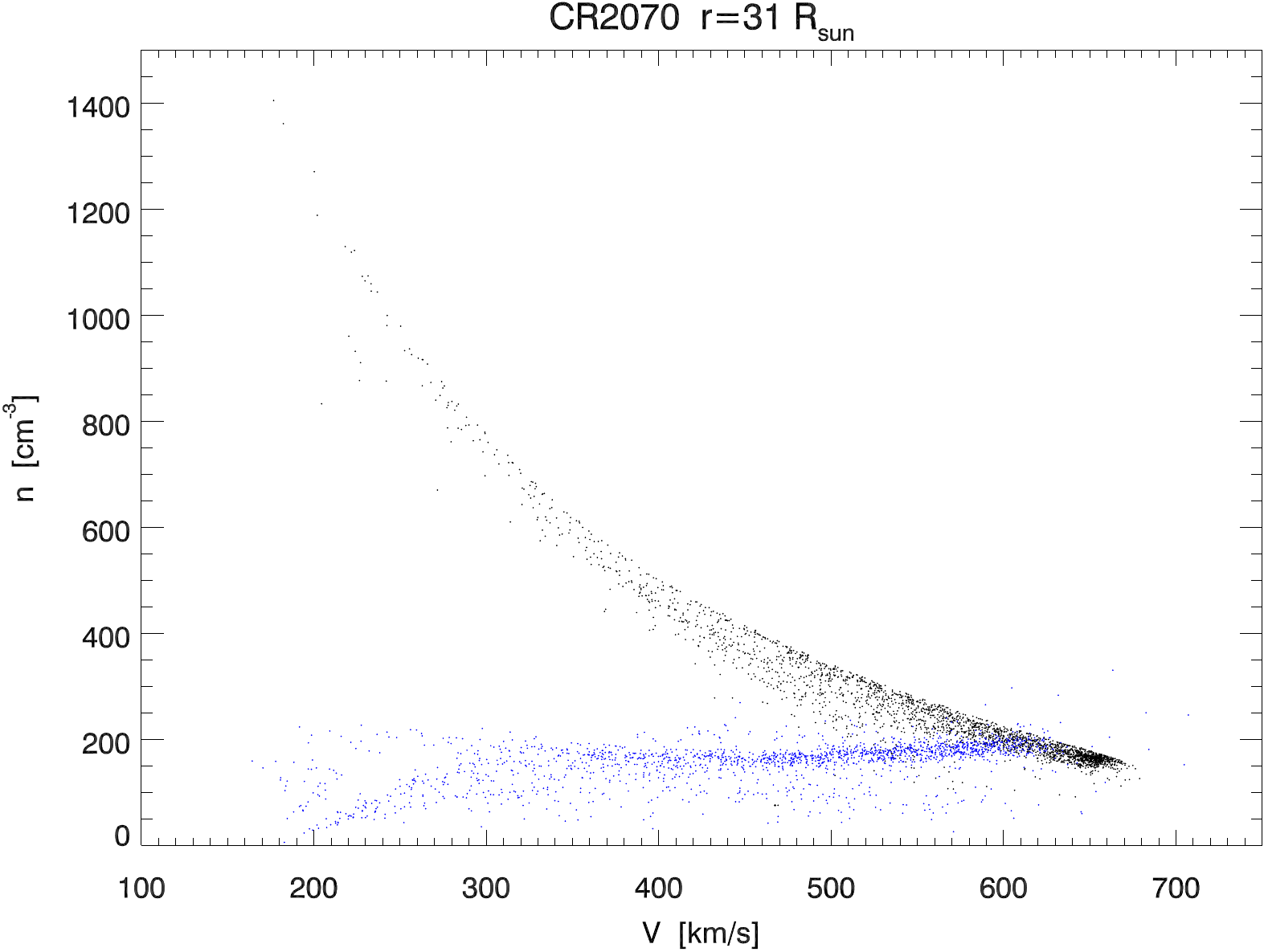} 
  \caption{
    Scatter-plot showing the correlation between density $n$ and wind speed $V$ at two different heights ($3$ and $31\un{\rsun}$) for CR 2070. 
    The black and blue dots correspond to the cases with and without expansion in the high corona. 
  }
  \label{fig:v_vs_n}
\end{figure}

\begin{figure}
  \centering
  \includegraphics[                width=0.85\linewidth]{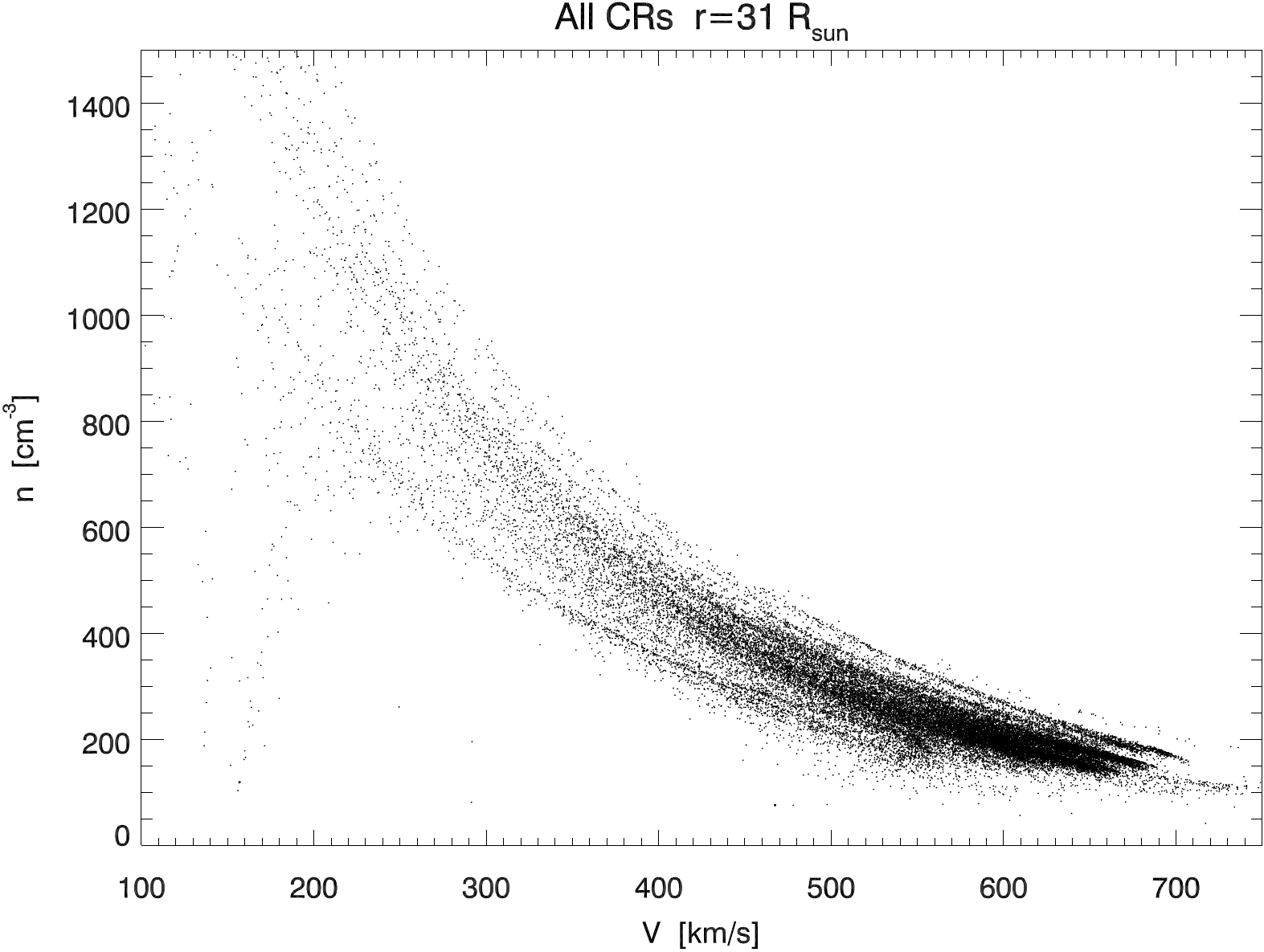}  \\ \medskip
  \includegraphics[                width=0.85\linewidth]{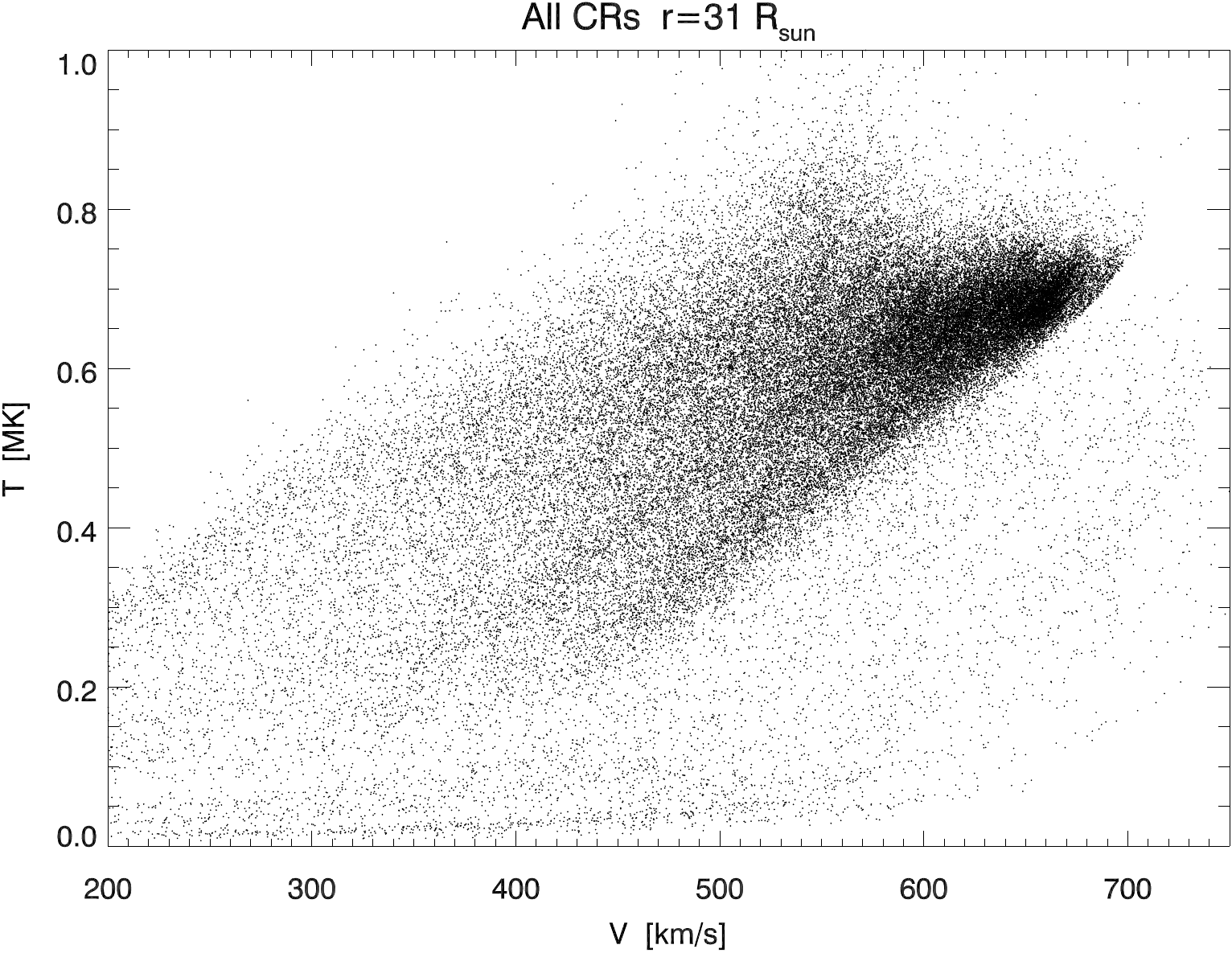}   
  \caption{Scatter-plots of the density $n$ (top panel) and of the plasma temperature $T$ (bottom panel) as a function of  the wind speed $V$ at $31\un{\rsun}$.}
  \label{fig:v_vs_T}
\end{figure}

\subsection{Correlations between speed, expansion, density and temperature}
\label{sec:correlations}

\prop{Speed and expansion}
The terminal wind speed we obtained correlate with the flux-tube expansion factors, but the degree and sign of the correlation depends on the reference height at which the latter are evaluated.
In particular, the well-know speed--expansion factor anti-correlation is only clearly observed in our solutions if the expansion factor is evaluated at (or close to) the source-surface.
Figure \ref{fig:v_vs_f} shows the dependence of the terminal wind speed ($V_{wind}$) on the flux-tube expansion factor measured at two different heights ($f_{SS}$ at $r=2.5\un{\rsun}$ represented as black diamonds, and $f_{tot}$ at $r=31.5\un{\rsun}$ represented as red crosses) for CR 2070, which illustrates this disparity particularly well (in general, $f_{tot}$ is simply not well correlated with $V_{wind}$, with the corresponding scatter-plot showing slopes which vary but remain close to flat on average).
The expansion that the flux-tubes suffer in the high corona contributes directly to this disparity ($f_{ss}$ would be equal to $f_{tot}$ otherwise), even though it remains smaller than a factor 2 for the vast majority of the flux-tubes considered.
The effect on the high-coronal expansion on the actual wind speeds is small.
This result is in agreement with the ideas of \citet{wang_solar_1990}, who suggested that $f_{SS}$ (expansion ratio measured at the height of the source-surface rather than the total expansion factor) is a good predictor for the terminal wind speed, in spite of the significant spread.
However, this particular height does not have a well-defined physical meaning, other than the fact that we are basing our computations on PFSS coronal field extrapolations.
It cannot be guaranteed that the anti-correlation between $V_{wind}$ and $f_{SS}$ remains valid for more general types of coronal field geometry.

\prop{Speed, density and temperature}
Figure \ref{fig:v_vs_n} shows scatter-plots of the plasma density and of the wind speed at $3$ and at $31\rsun$ for CR 2070.
The black dots correspond to the final wind solutions we obtained, and the blue dots to the solutions un-corrected for the expansion in the high corona.
In the final solutions, the $n$-$V$ relation shows a high scatter at $3\rsun$ and evolves into a well-defined anti-correlation, which is in agreement to the \emph{in-situ} measurements of HELIOS and ULYSSES.
The un-corrected solutions (which have a coronal magnetic field strictly aligned with the vertical direction above the source-surface) show a very different behaviour during the propagation of the wind flow into the high corona. 
The $n$-$V$ distributions start off with similar configuration at low coronal heights, but then evolve towards a state which can reach a flat or even positive slope.
In this incorrect scenario,  the slowest wind streams account for the largest disparities, remaining under-dense all the way through to the upper limit of the domain.
The correction applied to the magnetic field \rev{turns} the non-uniform magnetic field amplitude at $r_{ss}$ into uniform in the heliosphere and, while doing that, emulates to a certain extent the effects of cross-stream interactions on the wind density (see discussion on Sect. \ref{sec:justifications}).

Figure \ref{fig:v_vs_T} show scatter-plots of the plasma density (top panel) and of the temperature (bottom panel) against the wind speed at $31\rsun$ for the whole set of flux-tubes considered in the present study.
The terminal wind speed $V_{wind}$ is very clearly anti-correlated with density for all the data-set analyzed.
The plasma temperature is positively correlated with the terminal wind speed, in agreement with available \emph{in-situ} data, although with a higher scatter \citep[see][]{elliott_temporal_2012}.

% \subsection{Comparison with in-situ spacecraft data}
% \label{sec:in-situ}

\subsection{Comparison with coronograph observations}
\label{sec:coronograph}

\begin{figure}[!h]
  \centering
  
  \textsf{\textbf{CR 2079}} \\
  \includegraphics[width=.6\linewidth,clip,trim=0 18 0 35]{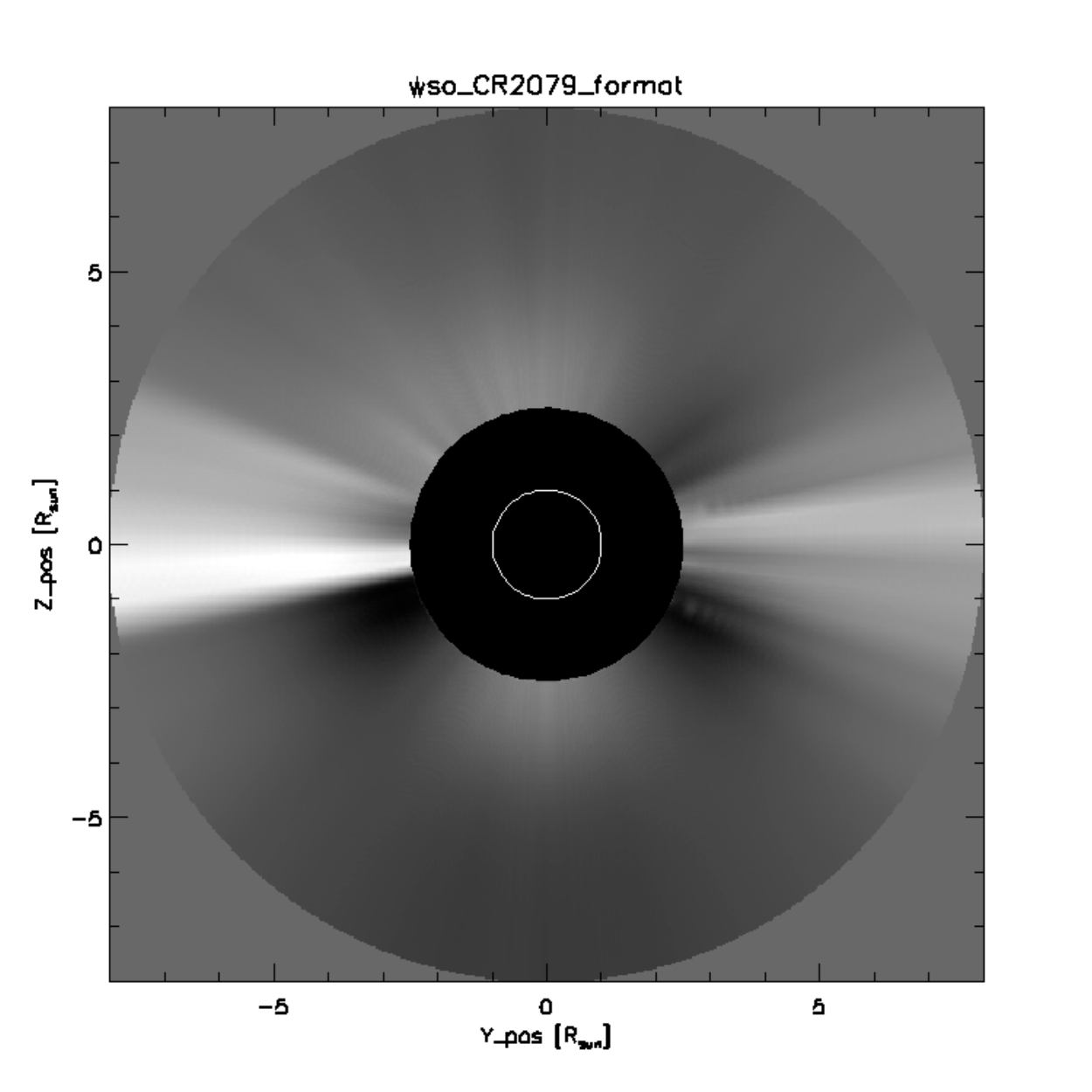}
  \raisebox{0.25\height}{\includegraphics[width=.38\linewidth]{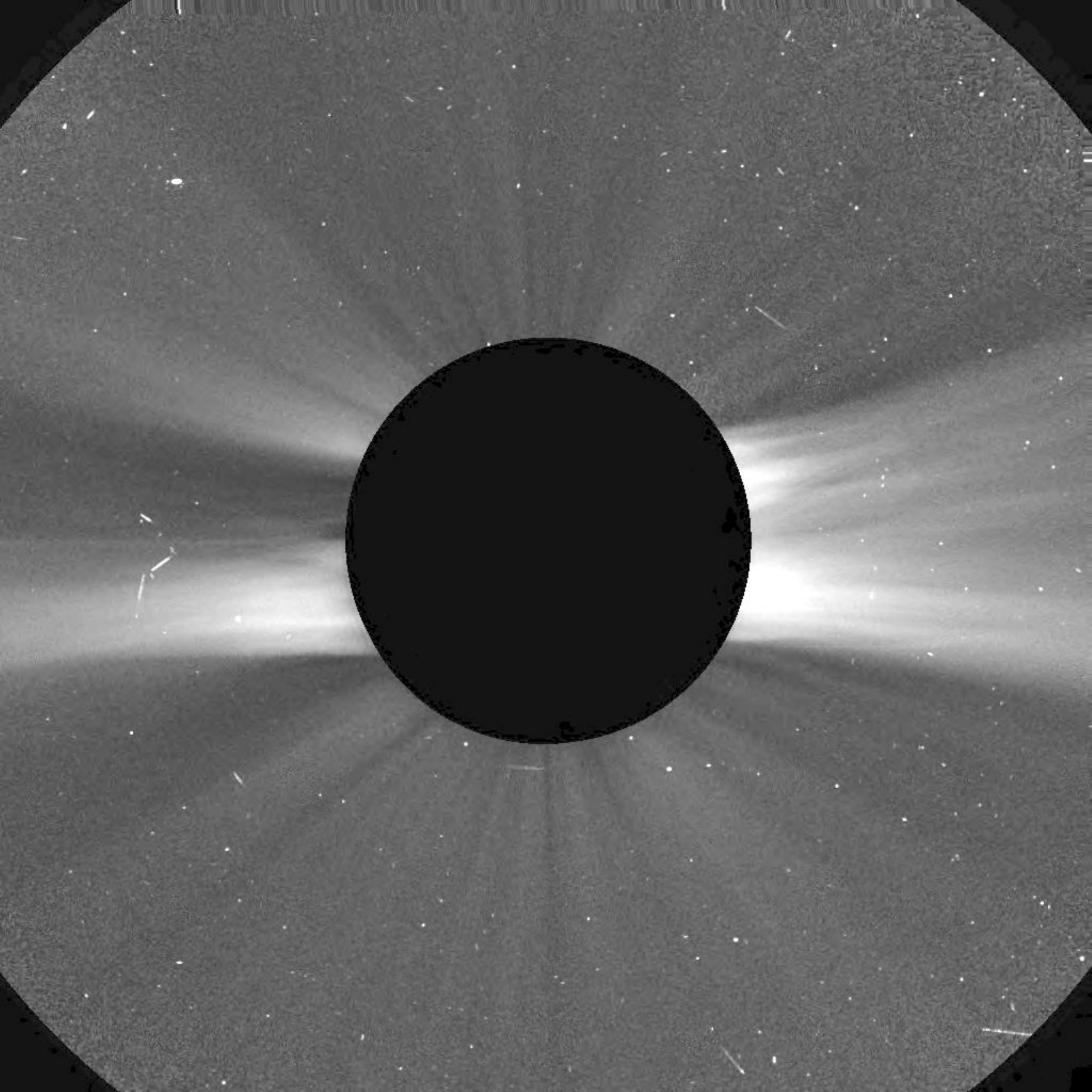}} \\

  \bigskip

  \textsf{\textbf{CR 2136}} \\
  \includegraphics[width=.6\linewidth,clip,trim=0 18 0 35]{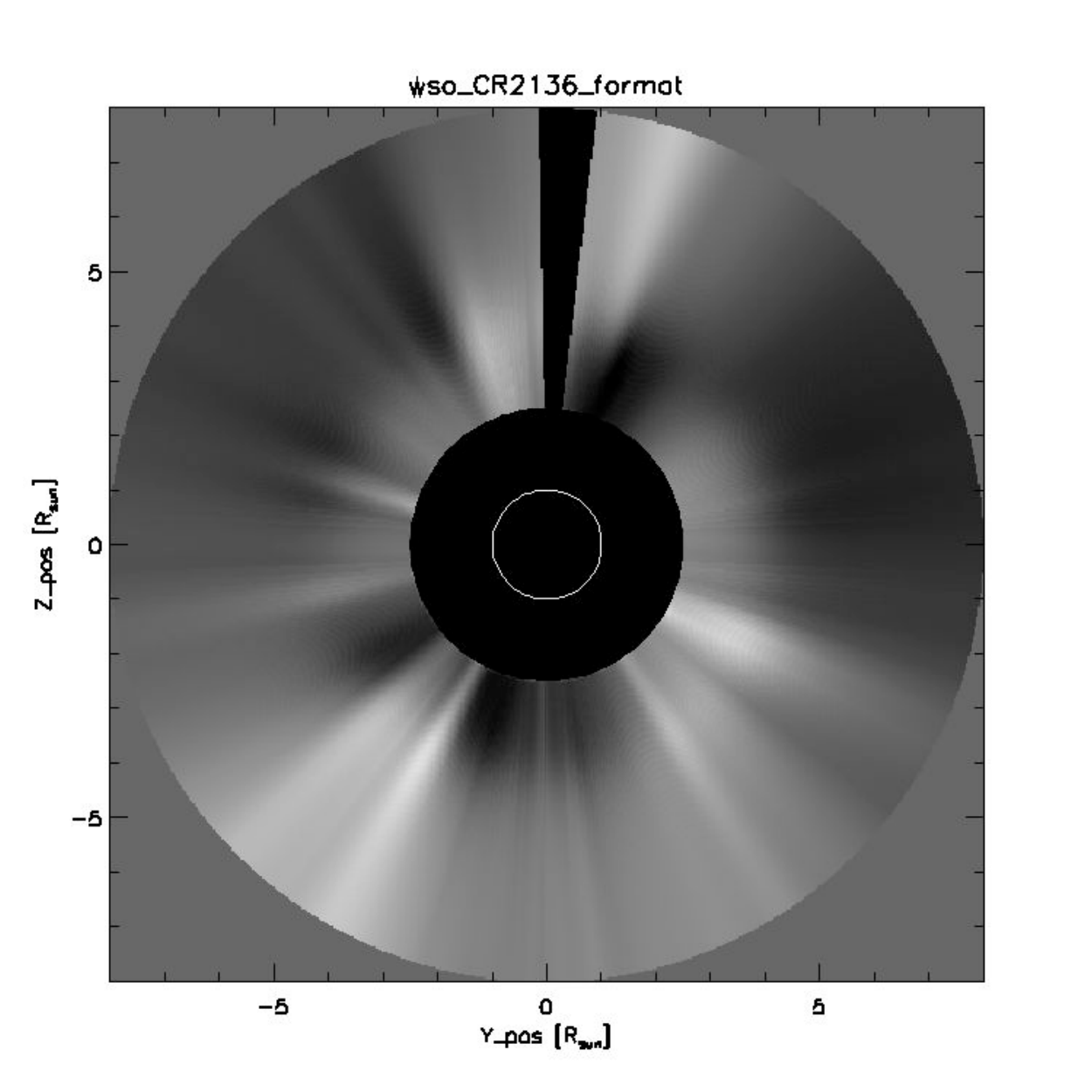} 
  \raisebox{0.25\height}{\includegraphics[width=.38\linewidth]{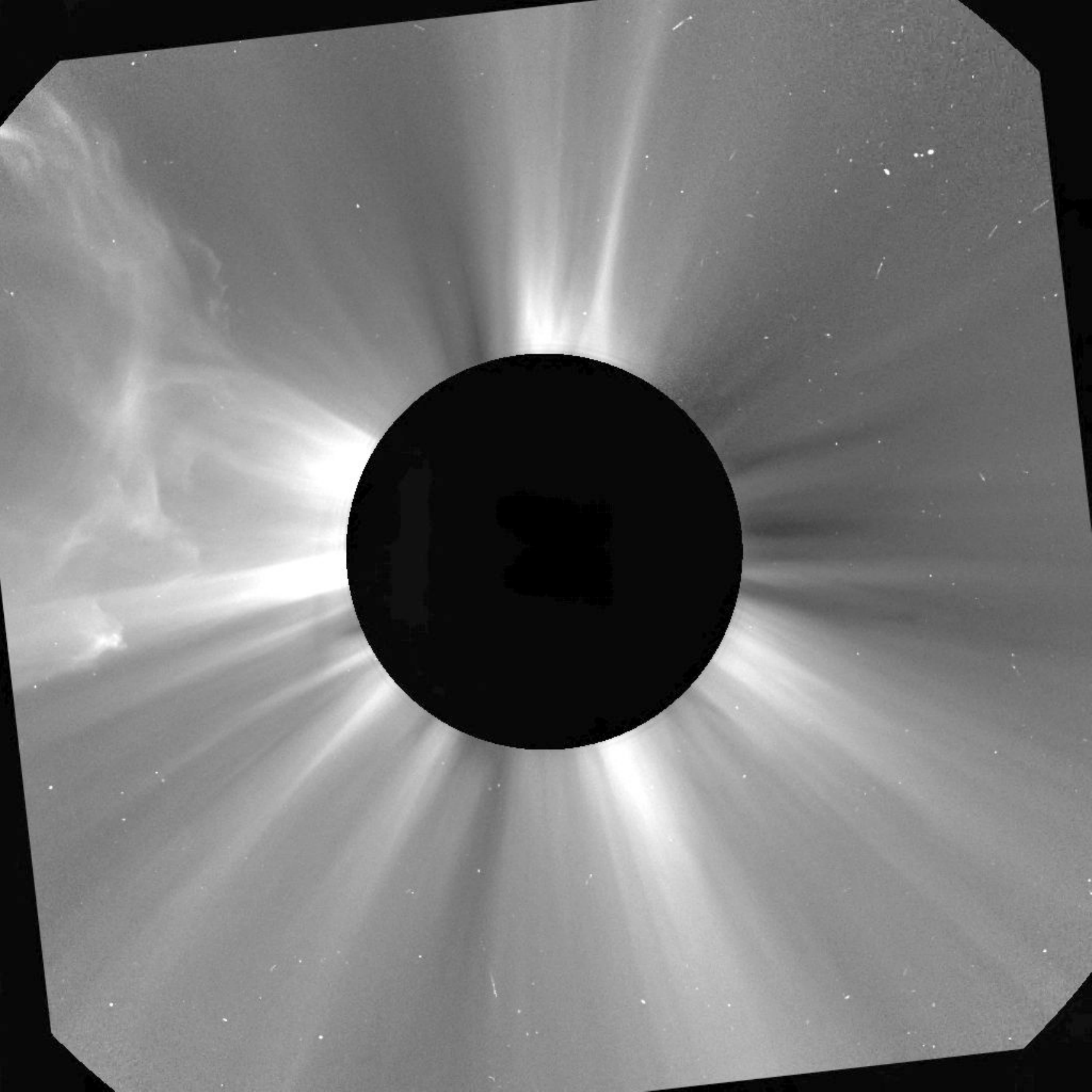}} \\
  \caption{Synthetic NRGF-filtered (left panel) and SoHO/LASCO-C2 (right) white-light images of the corona at solar minimum (top row) and solar maximum (bottom row).}
  \label{fig:forward_corona}
\end{figure}

\begin{figure*}[!h]
  \centering

  \textsf{\textbf{CR 2068 \hspace{1ex} $\mathbf{7\rsun}$ \hspace{1ex} West limb}} \\
  \includegraphics[width=.4\linewidth,clip,trim=0 18 0 33]{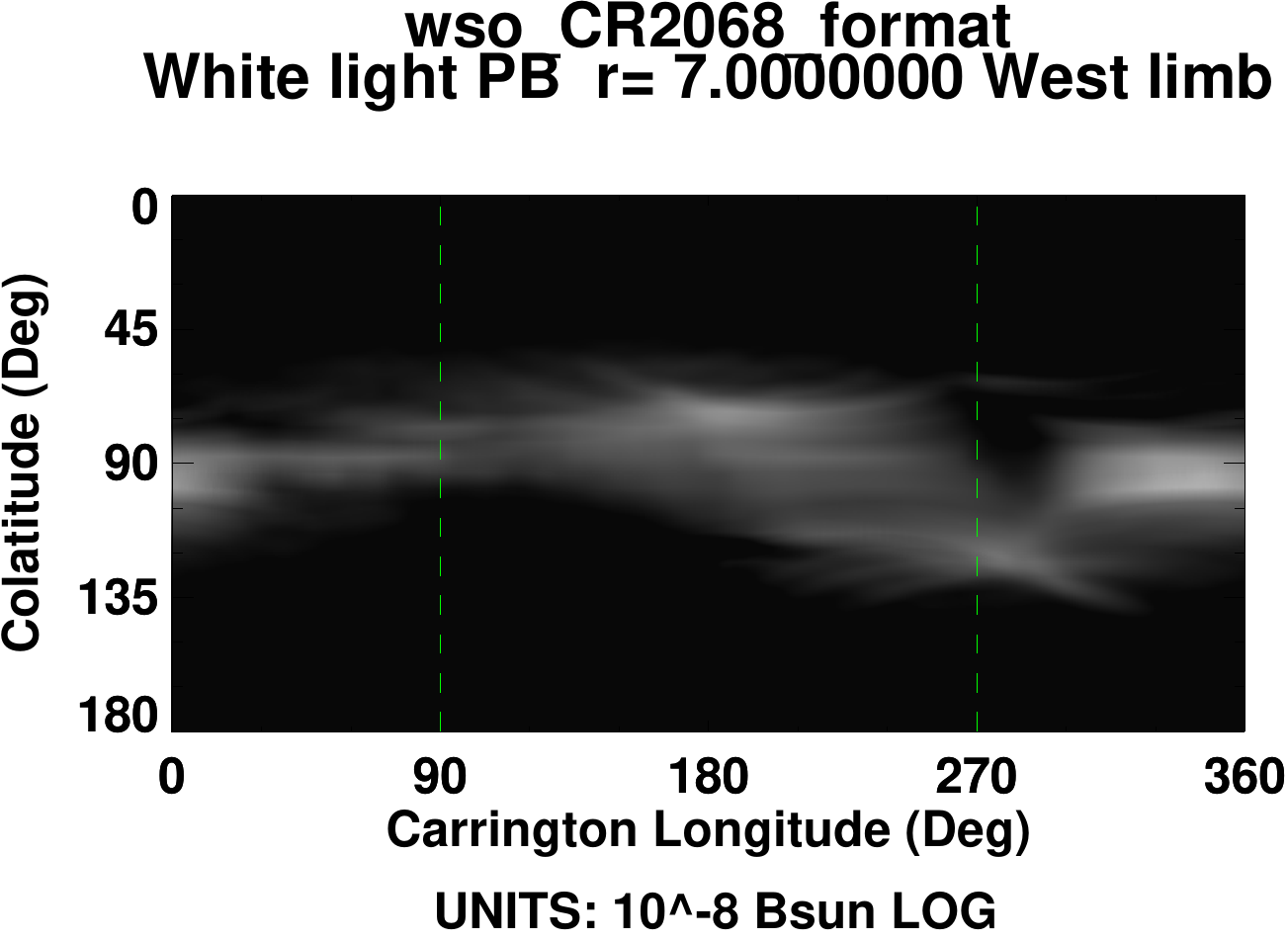}
  \raisebox{0.32\height}{\includegraphics[width=.4\linewidth,clip,trim=0 32 10 46]{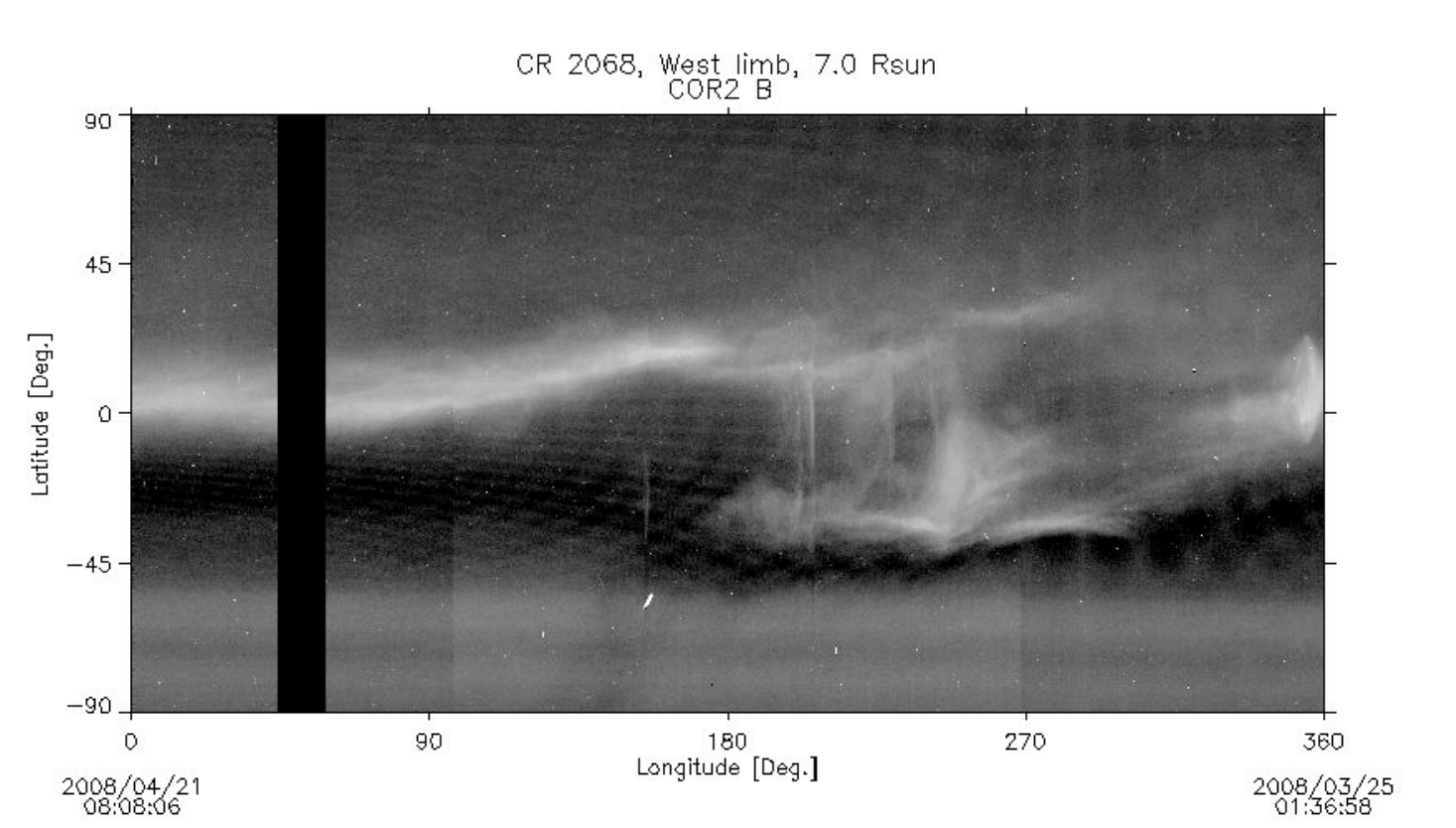}} \\

  \textsf{\textbf{CR 2122 \hspace{1ex} $\mathbf{7\rsun}$ \hspace{1ex} West limb}} \\
  \includegraphics[width=.4\linewidth,clip,trim=0 18 0 33]{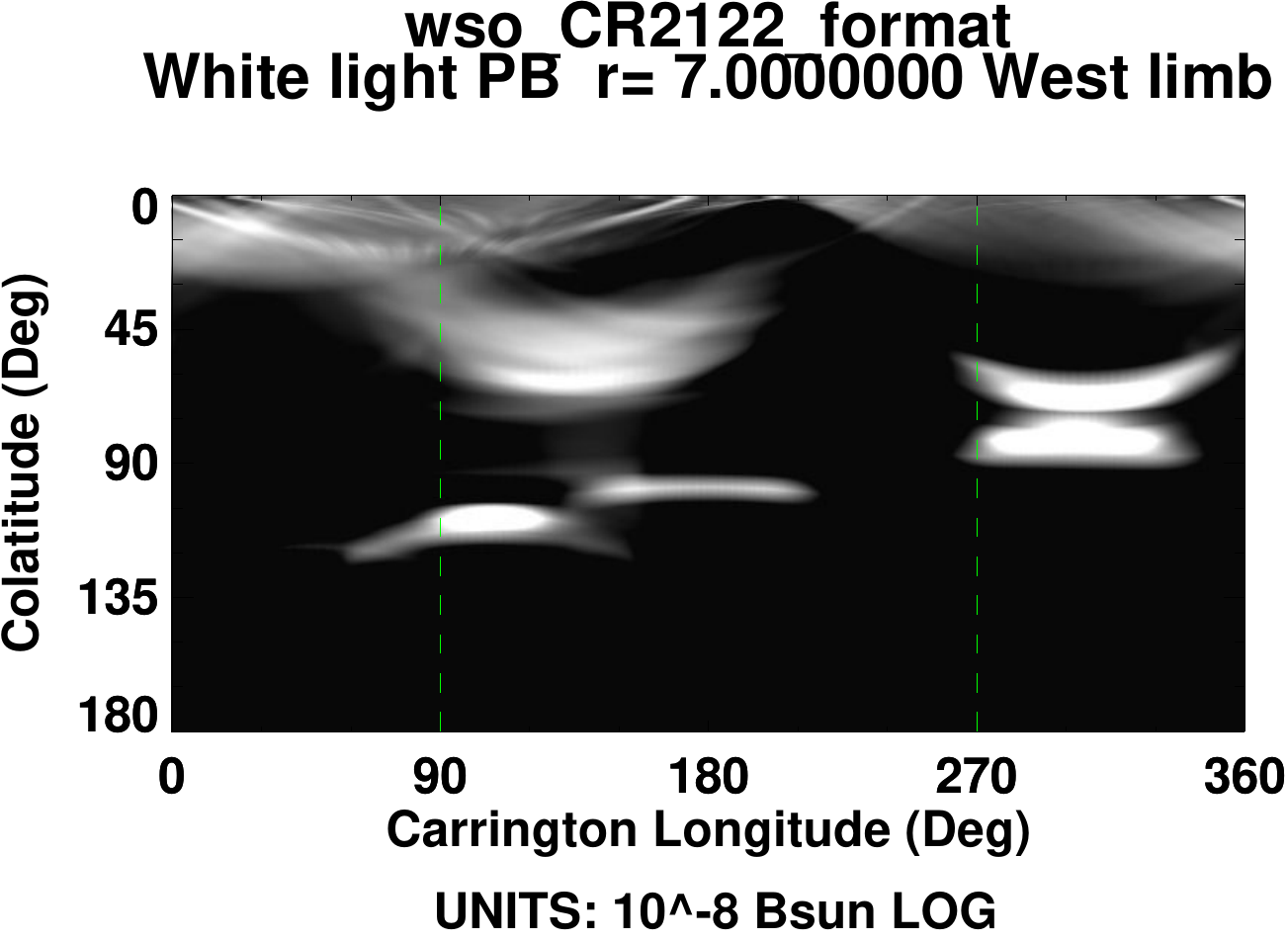}
  \raisebox{0.32\height}{\includegraphics[width=.4\linewidth,clip,trim=0 32 10 46]{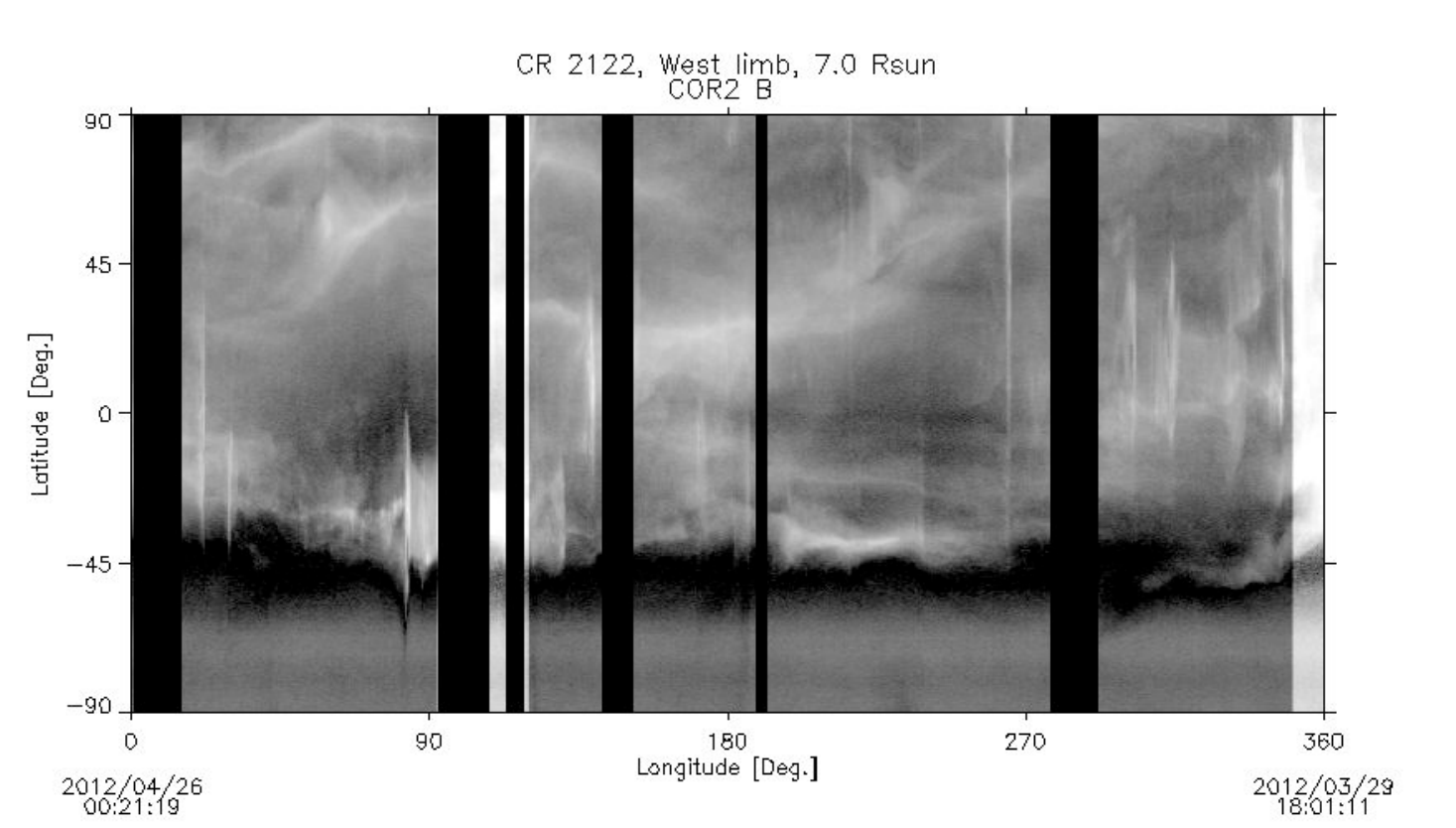}} 
  \caption{Synthetic white-light west-limb Carrington maps at $7\un{\rsun}$ (left panels) and SECCHI STEREO-A/B white-light images of the corona (right panels).}
  \label{fig:forward_carrmaps_7}
\end{figure*}

\begin{figure*}[!h]
  \centering

  \textsf{\textbf{CR 2068 \hspace{1ex} $\mathbf{13\rsun}$ \hspace{1ex} West limb}} \\
  \includegraphics[width=.4\linewidth,clip,trim=0 18 0 33]{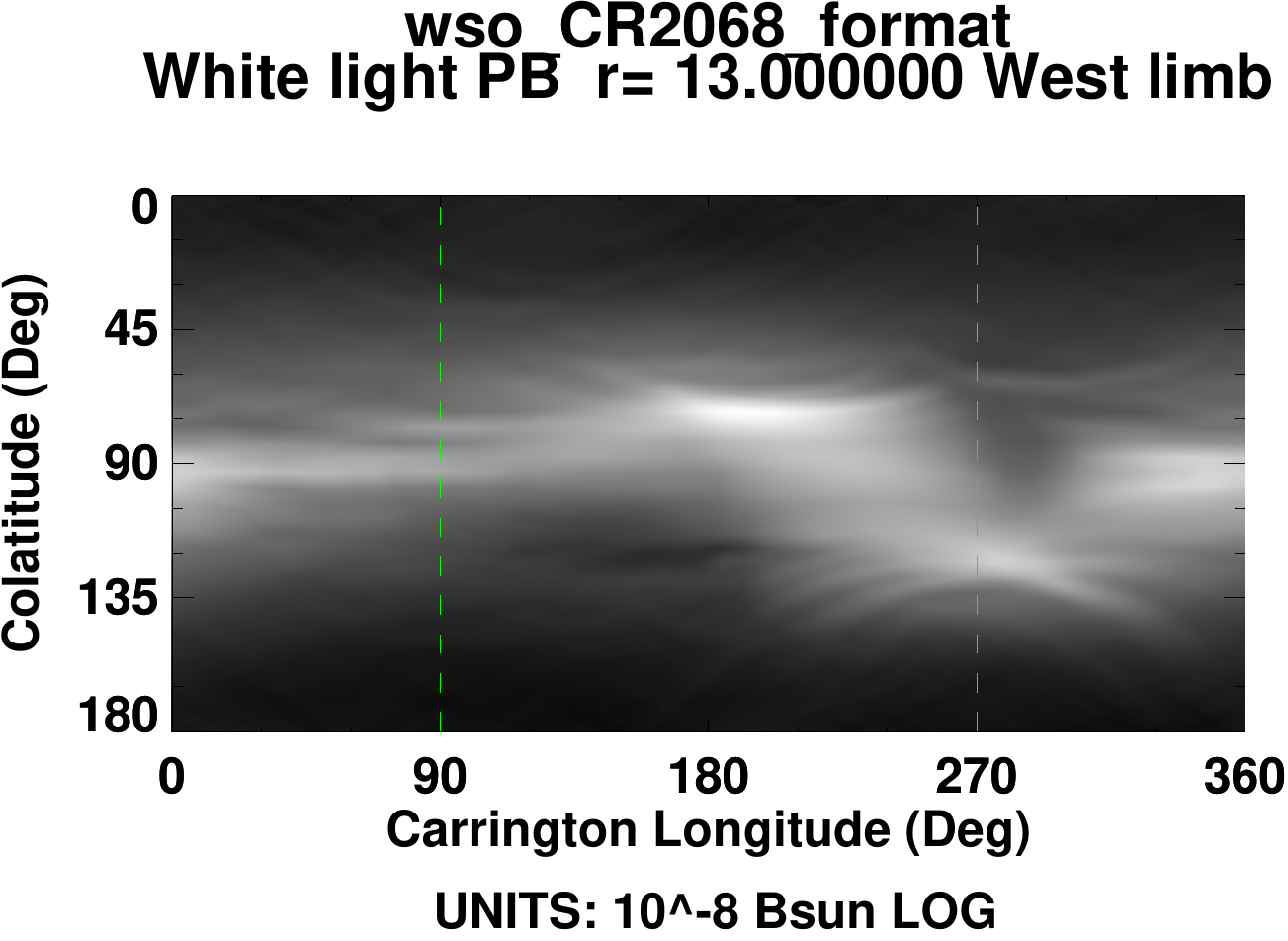}
  \raisebox{0.32\height}{\includegraphics[width=.4\linewidth,clip,trim=0 32 10 46]{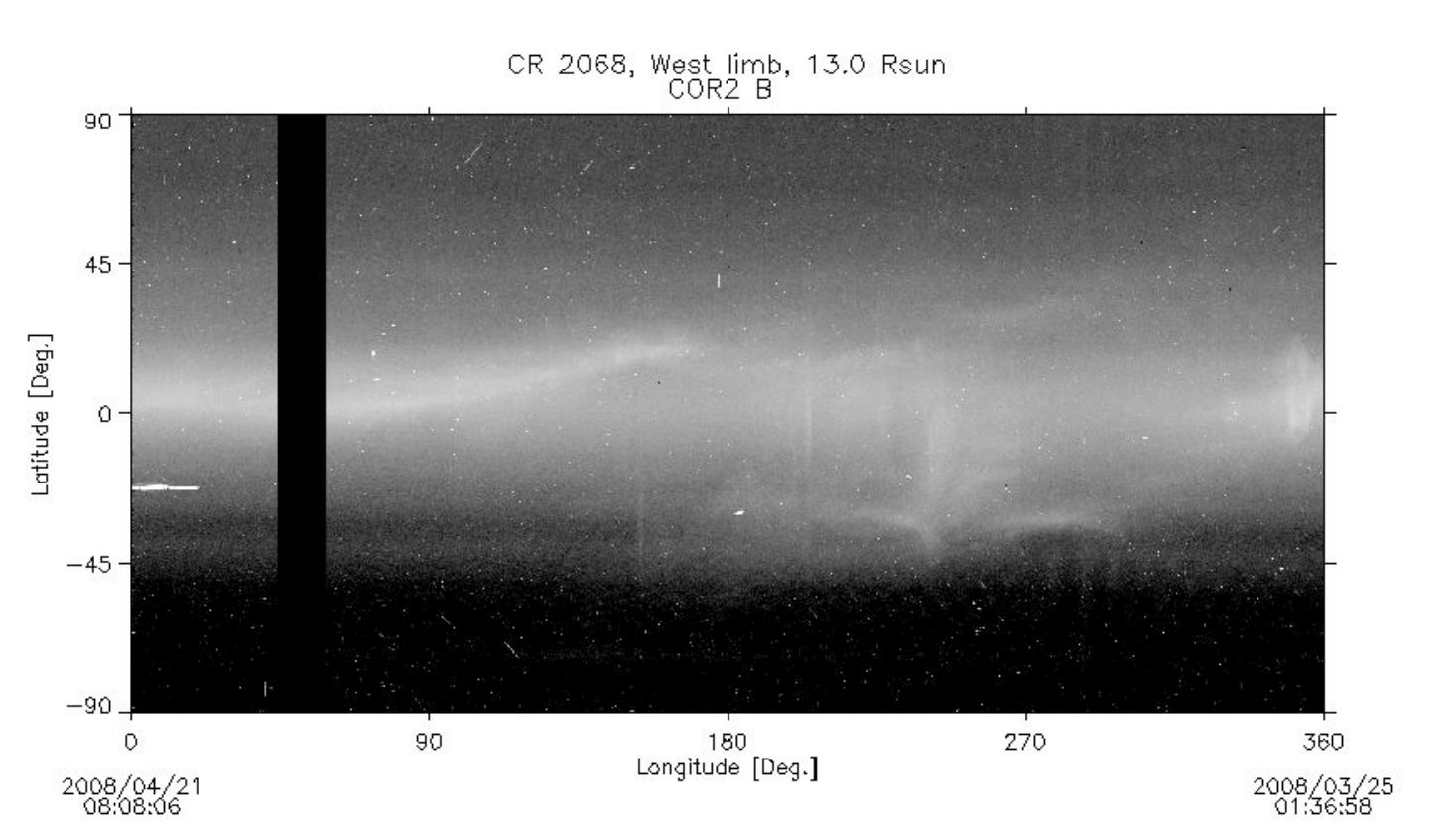}} \\

  \textsf{\textbf{CR 2122 \hspace{1ex} $\mathbf{13\rsun}$ \hspace{1ex} West limb}} \\
  \includegraphics[width=.4\linewidth,clip,trim=0 18 0 33]{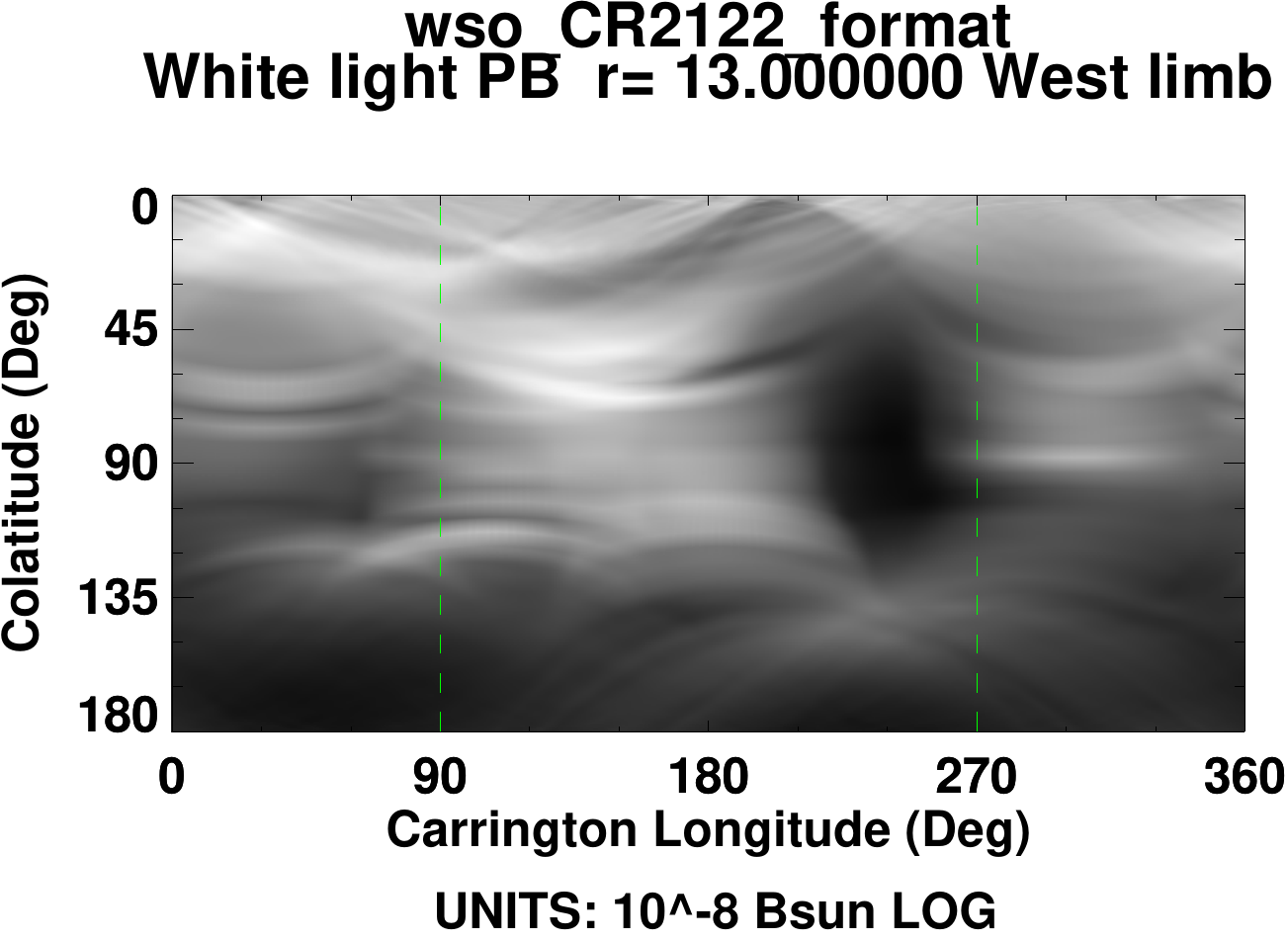}
  \raisebox{0.32\height}{\includegraphics[width=.4\linewidth,clip,trim=0 32 10 46]{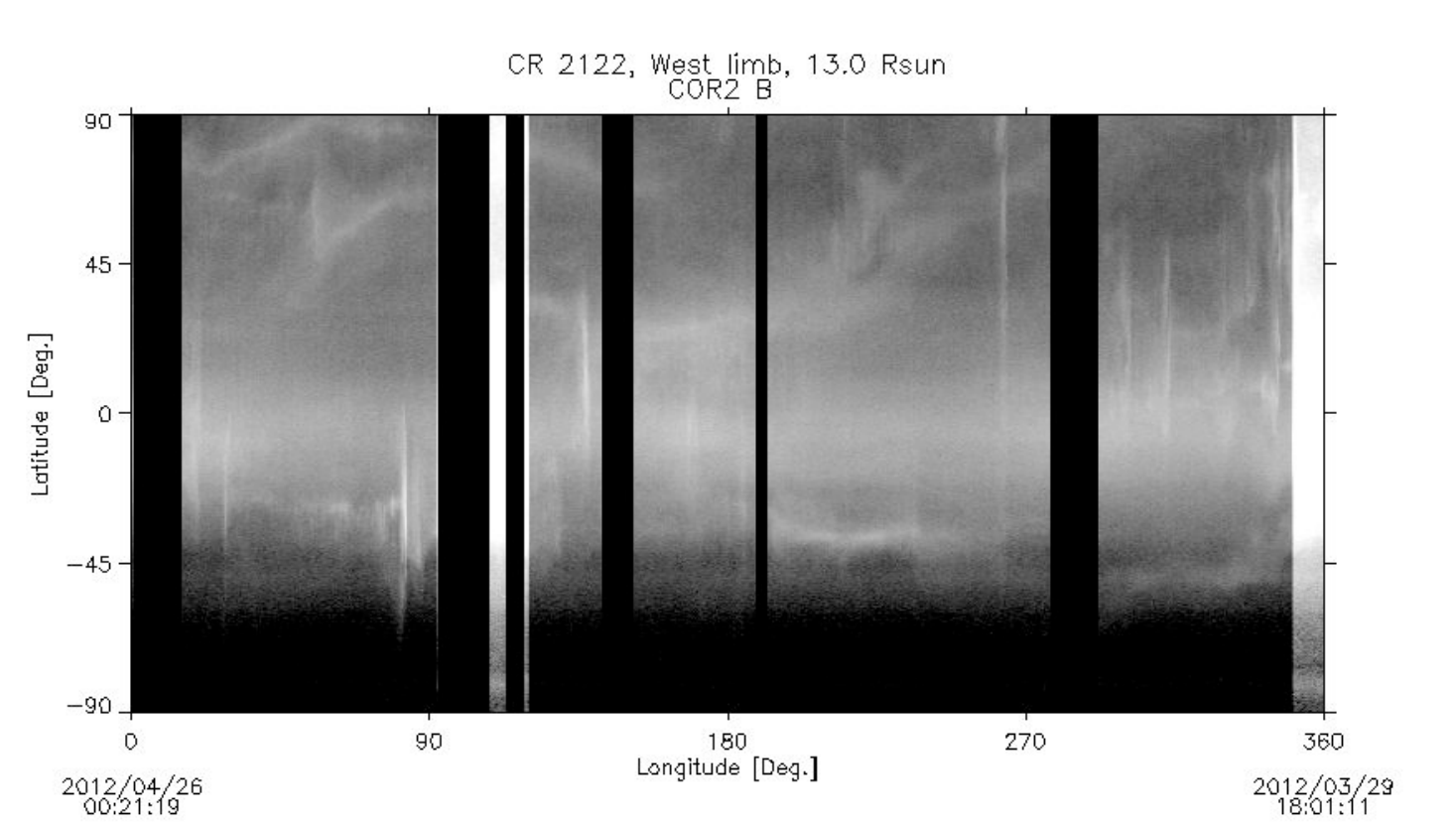}} 
  \caption{Synthetic white-light west-limb Carrington maps at $13\un{\rsun}$ (left panels) and SECCHI STEREO-A/B white-light images of the corona (right panels).}
  \label{fig:forward_carrmaps_13}
\end{figure*}

We used the FORWARD tool-set \citep{gibson_forward:_2016} to deduce the white-light emission from our wind model and to build synthetic images of the corona.
Figure \ref{fig:forward_corona} shows synthetic white-light polarised brightness (WLpB) images of the corona obtained from our simulations for Carrington rotations 2079 and 2136 sided by SoHO/LASCO-C2 images at the corresponding dates.
The synthetic images we filtered with a Normalizing Radial Graded Filter \citep[NRGF;][]{morgan_depiction_2006} to enhance the contrast of the coronal features and ease the qualitative comparison.
We found that the positions and widths of the main features are very well matched by our simulations for configurations typical both of solar minimum and solar maximum.
The main differences between the synthetic and the real coronagraph images relate to the low angular resolution of the magnetograms we have used ($5\degree \times 5\degree$), meaning that we cannot capture the finer structure of the streamers and pseudo-streamers, and the absence of transient events (the magnetic and wind models are stationary).
The lack of angular resolution is visible on the coronal features near the equator (both east and west) in the first set of images (for CR 2079).
The CME visible in the LASCO-C2 image on the bottom row is of course absent in the corresponding synthetic image.
We note furthermore that some of the streamers are not strictly aligned with the vertical direction in the C2 images, which probably corresponds to a temporary deflection due to CME activity \citep[see e.g.][]{rouillard_longitudinal_2012}.
The corresponding features in the synthetic images are perfectly vertically-aligned.

Figures \ref{fig:forward_carrmaps_7} and \ref{fig:forward_carrmaps_13} show Carrington maps of synthetic WLpB built using west limb cuts at two different heights ($r=7$ and $13\un{\rsun}$) with the corresponding real maps constructed using STEREO-B/COR2 data made available by the Naval Research Laboratory.
Once again we observe that the main features of the WL maps are very well reproduced (positions, slopes and widths) in our simulations, except for the signatures of coronal transients (CME) which appear as vertical traces in the COR2 maps.

%%%%%%%%%%%%%%%%%%%%%%%%%%%%%%%%%%%%%%%%%
\section{Discussion}
\label{sec:discussion}

\subsection{Strategy, strengths and caveats of the model}
\label{sec:validity}

MULTI-VP adopts a new approach that complements past and present efforts both on modelling the solar wind at global scales using full 3D MHD \citep[][among many others]{yang_time-dependent_2012,yang_self-consistent_2016,van_der_holst_alfven_2014,lee_solar_2009,gressl_comparative_2013,oran_global_2013-1} and on modelling the heating and transport processes occurring at smaller scales on the wind flow \citep[\emph{e.g}][]{verdini_alfven_2007,woolsey_turbulence-driven_2014,lionello_validating_2014,cranmer_self-consistent_2007,maneva_relative_2015,pinto_time-dependent_2009,grappin_two-temperature_2011}.
MULTI-VP computes detailed solutions of the background solar wind on a arbitrarily large bundle of open flux-tubes extending from the bottom of the chromosphere up to the high corona (typically up to $\sim 30\rsun$).
The model is able to sample large regions of the solar atmosphere (up to a full spherical domain) with more detailed thermodynamics and with significantly smaller computational requirements than the current full MHD global models.
MULTI-VP is furthermore unaffected by numercal resistive effects such as the spurious broadening of the HCS.
We currently compute the state of the whole corona in about $6\un{hrs}$ with moderate angular resolution ($5\degree\times5\degree$) and with a moderate number of allocated computing cores.
But the total execution can be significantly reduced, as the model is nearly perfectly scalable, and real-time operation can be envisaged.
The downsides of the MULTI-VP strategy are that it relies on coronal field reconstruction methods (or any other more or less realistic magnetic field model), it neglects cross-stream effects on the wind, and is only well defined for stationary flows.
The underlying numerical model is in fact fully time-dependent, but the setup used is however not well adapted to the study of large perturbations to the background flow, particularly in the direction transverse to the magnetic field.
In this manuscript, the magnetic field geometry is obtained externally by means of PFSS extrapolations, which imply the coronal magnetic field to be stationary.
We trace out open field-lines, and build an ensemble of magnetic flux-tubes fully taking into account the geometry of each one of them.
This translates into prescribing the magnetic field amplitude and polarity, areal expansion rate, field-line inclination and radial height as a function of the distance to the foot-point of each flux-tube.
The geometrical description is as general as possible, allowing for any type of field-line bend, kink, over-expansion or re-convergence profile.
This is of utmost importance, as both the location and spatial extent of field-line bends and regions with over/under-expansion play a crucial role on the properties of the wind flow which propagate along them \citep[see discussion by][]{pinto_flux-tube_2016}.
The model was tested against extreme scenarios which included sequences of exponential expansion and re-convergence, and field-line switchbacks.

\subsection{Other model parameters}
\label{sec:justifications}

The PFSS extrapolations produce coronal magnetic field of varying complexity up to the height of the source-surface (placed at $2.5\rsun$).
If one assumes radial and spherical expansion of the open magnetic field lines from the source surface outwards, the resulting magnetic fields at $30\un{\rsun}$ will be highly non uniform. 
This is contrary to in-situ measurements of the interplanetary magnetic field made by the Ulysses mission which showed that the latitudinal distribution of the radial field component is uniform \citep{balogh_heliospheric_1995}.
Furthermore, the plasma density distributions we calculated using radially expanding magnetic fields above the source-surface are incorrect. 
The correlation between wind speed and density is, in particular, very different from the expected one, with a tendency for the simulated solar wind speed to be correlated with density, in contrast with the well-known anti-correlation measured at $1\un{AU}$.
After careful analysis, we realized that both issues are related.
Flux-tubes with higher than average magnetic field amplitude tend to be over-dense, while flux-tubes with lower than average field strength tend to be under-dense.
Also, flux-tubes with higher than average magnetic field amplitude at the source-surface are bound to suffer some additional expansion in the high corona, while flux-tubes with lower than average field strength must undergo the inverse process.
It is not possible otherwise to generate an approximately uniform magnetic field (apart from polarity inversion at sector boundaries), as measured consistently in the interplanetary space.
We then proposed a correction to the radial variation of the magnetic field which consists of switching from spherical expansion to a smooth flux-tube expansion in the high corona (from a little below the source-surface and up to $12\rsun$).
The individual expansion factors are defined such that the total unsigned open magnetic flux is conserved.
In practice, the required additional expansions factors between $r=2.5\rsun$ and $12\rsun$ are small in respect to the ones for  the lower corona (they remain smaller than $2$, while the expansion factors between the surface and the source-surface can be as high as several hundreds, see Fig. \ref{fig:profiles_inital_bfield} and \ref{fig:profiles_inital_expans}).
This simple correction to the flux-tube profiles in the high corona was enough to produce the correct anti-correlation between the wind speed and density (see Fig. \ref{fig:v_vs_n}).

%\nota{(explain the physics, evoke transverse pressure balance.)}
%\nota{(Say future work should test other types of extrapolations methods which would not required this correction (PFSS+SCS, NLFFF, MHD models).)}
In more physical terms, the uniformisation of the magnetic field with height in the high corona and heliosphere is most likely a consequence of magnetic pressure balance, with neighbouring flux-tube adjusting their cross-sections in order to eliminate transverse pressure gradients.
As the magnetic pressure still dominates the transverse pressure budget at these heights, eliminating the pressure gradients implies that the magnetic field becomes close to uniform, with the density of the channeled flows accommodating for the subtle variations in cross-section with height.
As described in Sect. \ref{sec:results}, these variations in the expansion rate of the high coronal part of the flux-tubes produces a significant effect on the density, but a more moderate effect on the terminal wind speeds obtained.
The response of the wind flow speed in respect to variations in tube cross-section $A\left(s\right)$ is, to first approximation, proportional to its logarithmic gradient $\partial_s \log{A}\left(s\right)$ and to the ratio $v/\left(1 - M^2\right)$, where $v$ is the wind speed and $M$ is the sonic Mach number \citep{wang_two_1994,pinto_flux-tube_2016}.
Most of the additional expansion occurs well above the sonic point, such that $v/\left(1 - M^2\right) \ll 1$, and the log-gradient of $A\left(s\right)$ remains small.

Having fixed the magnetic field geometry, the only free parameters left in the model relate to the coronal heating functions.
The heating functions are empirical parametrizations of the actual coronal heating processes, as the small-scale dissipation mechanisms are still under debate and cannot be accounted for self-consistently in the model.
We chose to use a simple formulation based on commonly used phenomenological forms for the heating function with a few extra parameters.
More specifically, we used a heating formulation similar to that of \citet{withbroe_temperature_1988} with two main modifications (see eq. \ref{eq_fluxp}).
We made the coefficient proportional to the amplitude of the basal magnetic field $B_0$, to account for the energy input to the solar wind resulting from horizontal surface motions at the surface of the Sun. The corresponding energy flux density (the Poynting flux) is proportional to $B_0 v_\perp \sqrt{\rho}$ at the surface.
We also made the dissipation scale-height depend on the flux-tube expansion ratio, following ideas from \citet{wang_slow_2009}.
This varying dissipation scale-height increased the contrast between the slow and fast wind speed, making the solutions match \emph{in-situ} measurements more closely.
%We have calibrated this parameter against ACE data, and 
We have retained the same final form (see Eq. \ref{eq_fluxp}) for all calculations to keep our analysis as general as possible.
We did not make any other adjustement to the heating phenomenology, amplitude or scale-height depending on latitude, moment of the cycle or terminal wind speeds obtained.
Discriminating between different heating phenomenologies currently under debate is beyond the scope of this manuscript and will be subject of future work.

%%%%%%%%%%%%%%%%%%%%%%%%%%%%%%%%%%%%%%%%%
\section{Summary and perspectives}
\label{sec:conclusion}

We present and discuss the design, implementation and testing of a new solar wind model, called MULTI-VP.
The model calculates the dynamical and thermal properties of the solar wind from $1\un{\rsun}$ up to about $30\un{\rsun}$, and can cover the totality or a fraction of a spherical domain representing the three-dimensional open-field corona.
The model is initiated using an externally prescribed magnetic field geometry.
In the current study, we used magnetograms from the Wilcox Solar Observatory and performed standard PFSS extrapolations to derive the structure of the coronal magnetic field.
We defined simple phenomenological forms for the heating flux which result in correct angular distributions and amplitudes of slow and fast wind flows.
The model provides estimates of the wind speed, density and temperature (as well as any derived quantity such as MHD phase speeds) at a very moderate computational cost when compared to global 3D MHD simulations, and without resorting to semi-empirical hypothesis.
The model was designed to be as flexible and modular as possible, such that other sources of magnetogram or coronal field data, different heating scenarii (theoretical or data driven), and other types of domain can be easily setup. 

We produced of a series of Carrington maps of the solar wind speed, density, temperature and magnetic field amplitude, which display a correct distribution of slow to fast wind flows at different moments of the solar cycle.
The model solutions were used to derive synthetic white-light images of the corona, which compare very well with LASCO-C2 and STEREO COR2 data.
The model reproduces correctly the well-know correlations between wind speed, density and temperature measured \emph{in-situ} by spacecraft in the interplanetary medium.
We found that the inverse correlation between the density and wind speed is established in the high corona and that it is a consequence of the small adjustments that neighboring open flux-tubes undergo in order to maintain pressure balance in the transverse direction. 
This effect is intrinsically related to the uniformisation of the magnetic flux amplitude which takes place between the corona and the interplanetary medium.

%\prop{Future work.}
Future work will study the inclusion of alternative magnetogram data sources and coronal field reconstruction methods and of more sophisticated heating scenarios (theoretical or data-driven).
We plan on performing detailed comparisons between the results of our model with other well-established methods on the near future.
We aim at increasing the integration with other solar and heliospheric data and models, and at progressively approaching the requirements of real-time forecasting.

%%%%%%%%%%%%%%%%%%%%%%%%%%%%%%%%%%%%%%%%%
\acknowledgments

%We acknowledge \ldots\\
R. F. P. and A. P. R. acknowledge funding by the FP7 project \#606692 (HELCATS).
The numerical simulations of this study were performed using HPC resources from CALMIP (Grant 2016-P1504).
We are grateful to Y.-M. Wang for enlightening discussions and comments on the manuscript, and to R. Grappin for his commitment to the development of the baseline 1D wind model VP.
%%%%%%%%%%%%%%%%%%%%%%%%%%%%%%%%%%%%%%%%%
\bibliographystyle{aasjournal}
\bibliography{/data/BIBTEX/refs}

%% Show the entire author+affilation list when the collaboration and author truncation commands are used.  It has to

%% Include if \added, \replaced, \deleted  commands to see a summary list of all changes at the end of the article.

\end{document}